\documentclass[journal,twoside,onecolumn]{IEEEtran}
\usepackage{cite}

\ifCLASSINFOpdf
\else
   \usepackage[dvips]{graphicx}
\fi
\usepackage[cmex10]{amsmath}
\usepackage{amssymb,amsthm}
\usepackage{multirow,bigdelim}

\newtheorem{example}{Example}
\newtheorem{lemma}{Lemma}
\newtheorem{remark}{Remark}
\newtheorem{assumption}{Assumption}
\newtheorem{proposition}{Proposition}
\newtheorem{corollary}{Corollary}
\newtheorem{property}{Property}

\newcommand{\argmin}{\mathop{\mathrm{argmin}}\limits}

\begin{document}
%
\title{Large-System Analysis of Joint User Selection and Vector Precoding for Multiuser MIMO Downlink}
%
%
%

\author{Keigo~Takeuchi,~\IEEEmembership{Member,~IEEE,}
        Ralf~R.~M\"uller,~\IEEEmembership{Senior Member,~IEEE,}
        and~Tsutomu~Kawabata,~\IEEEmembership{Member,~IEEE,}
\thanks{Manuscript received, 2012. 
The work of K.~Takeuchi was in part supported by the JSPS Institutional 
Program for Young Researcher Overseas Visits and by the Grant-in-Aid for 
Young Scientists (B) (No.\ 23760329) from MEXT, Japan. 
The material in this paper was submitted to 2012 International Symposium on 
Information Theory and its Applications, Honolulu, Hawaii, USA, Oct.\ 2012.}
\thanks{K.~Takeuchi and T.~Kawabata are with the Department of Communication 
Engineering and Informatics, the University of Electro-Communications, 
Tokyo 182-8585, Japan (e-mail: ktakeuchi@uec.ac.jp, kawabata@uec.ac.jp).}
\thanks{R.~R.~M\"uller is with the Department of 
Electronics and Telecommunications, the Norwegian University of Science and 
Technology (NTNU), NO--7491 Trondheim, Norway (e-mail: ralf@iet.ntnu.no).}
}

%
%

\markboth{IEEE transactions on information theory,~Vol.~, No.~, 2012}%
{Takeuchi \MakeLowercase{\textit{et al.}}:Performance Improvement of Iterative 
Multiuser Detection for Large Sparse CDMA by Spatial Coupling}
%

\IEEEpubid{0000--0000/00\$00.00~\copyright~2012 IEEE}


\maketitle

\begin{abstract}
Joint user selection (US) and vector precoding (US-VP) is proposed for 
multiuser multiple-input multiple-output (MU-MIMO) downlink. The main 
difference between joint US-VP and conventional US is that US depends on 
data symbols for joint US-VP, whereas conventional US is independent of data 
symbols. The replica method is used to analyze the performance of joint 
US-VP in the large-system limit, where the numbers of transmit antennas, 
users, and selected users tend to infinity while their ratios are kept 
constant. The analysis under the assumptions of replica symmetry (RS) 
and 1-step replica symmetry breaking (1RSB) implies that optimal 
data-independent US provides nothing but the same performance as random US 
in the large-system limit, whereas data-independent US is capacity-achieving as 
only the number of users tends to infinity. It is shown that joint US-VP can 
provide a substantial reduction of the energy penalty in the large-system 
limit. Consequently, joint US-VP outperforms separate US-VP in terms of 
the achievable sum rate, which consists of a combination of vector precoding 
(VP) and data-independent US. In particular, data-dependent US can be applied 
to general modulation, and implemented with a greedy algorithm.  
\end{abstract}


%
\IEEEpeerreviewmaketitle

\section{Introduction}
\IEEEPARstart{M}{ultiple}-input multiple-output (MIMO) systems use multiple 
transmit and receive antennas to increase the spectral 
efficiency~\cite{Foschini98,Telatar99}. In early work, point-to-point 
MIMO or multiuser MIMO (MU-MIMO) uplink was 
investigated~\cite{Marzetta99,Zheng02,Biglieri02,Tse99,Verdu99,Mueller03}. 
In these MIMO systems, the receiver can utilize all received signals to 
detect the transmitted data. Recent research activities have been shifted to 
MU-MIMO downlink, in which one base station (BS) communicates with 
non-cooperative users. In the MU-MIMO downlink the main part of signal 
processing is at the transmitter side, whereas it is at the receiver side 
for the MU-MIMO uplink. 

Transmit strategies used for the MU-MIMO downlink depend on {\em duplexing}.  
For the MU-MIMO downlink with frequency-division duplexing (FDD), channel 
state information (CSI) is not available at the transmitter side. Instead, 
the BS may utilize limited feedback information about channel quality, 
transmitted through the uplink channels~\cite{Viswanath02,Chung03}. 
For the MU-MIMO downlink with time-division duplexing (TDD), on the other 
hand, channel state information (CSI) is used to pre-cancel inter-user 
interference (IUI) at the transmitter side. The CSI may be estimated by 
utilizing the fact that fading coefficients in both links are identical 
for TDD~\cite{Marzetta06,Marzetta10}. In particular, it is possible for the 
BS to attain accurate CSI when the coherence time is sufficiently long. 
In this paper, the MU-MIMO downlink with TDD is considered under the 
assumption that the coherence time is sufficiently long. For simplicity, 
we assume that perfect CSI is available at the transmitter and that 
the number of receive antenna for each user is one.  

The MU-MIMO downlink we consider is mathematically modeled as the MIMO 
broadcast channel (MIMO-BC) with perfect CSI at the transmitter. Recent 
excellent papers~\cite{Caire03,Viswanath03,Yu04,Weingarten06} have proved that 
the capacity region of the MIMO-BC with perfect CSI at the transmitter is 
achieved by dirty-paper coding (DPC)~\cite{Costa83}, which is a sophisticated 
scheme that pre-cancels IUI at the transmitter side. Since DPC is infeasible 
in terms of the computational complexity, however, it is an active research 
area and the target in this paper to construct a suboptimal scheme that 
achieves an acceptable tradeoff between performance and complexity. 

Zero-forcing transmit beamforming (ZFBF)~\cite{Spencer04,Choi04,Wiesel08} is 
a simple approach for pre-cancelling IUI at the transmitter side. The ZFBF 
decomposes the MIMO-BC into per-user interference-free channels. A drawback 
of the ZFBF is that {\em energy penalty}, which is the energy 
required for the pre-cancellation of IUI, increases rapidly as the number of 
(supported) users gets closer to the number of transmit antennas. An increase 
of the energy penalty results in a degradation of the receive signal-to-noise 
ratio (SNR). 

\IEEEpubidadjcol

The number of users is commonly larger than the number of transmit antennas 
for MU-MIMO downlink.  
In order to reduce the energy penalty, user selection (US) has been 
proposed~\cite{Tu03,Dimic05,Yoo06,Shen06}: A subset of users is selected 
to mitigate the increase of the energy penalty. Interestingly, it has been 
shown that a greedy algorithm of US can achieve the sum capacity of the 
MIMO-BC when {\em only} the number of users tends to 
infinity~\cite{Yoo06,Wang08}. This result can be understood as follows: 
If the channel vectors for all users were orthogonal to each other, the ZFBF  
would be optimal. However, there are dependencies between the channel 
vectors in general. In US, the BS attempts to select a subset of users with 
almost orthogonal channel vectors. It is possible to pick a finite number of 
almost orthogonal channel vectors from an infinite number of channel vectors 
under proper conditions. Thus, the ZFBF with US can achieve the sum capacity 
of the MIMO-BC when the number of users tends to infinity. Since the number 
of selected users should be comparable to the number of transmit antennas, 
the interpretation above implies that the performance of US degrades 
significantly as the number of transmit antennas gets closer to the number of 
users. 

The situation in which the number of transmit antennas is comparable to the 
number of users is becoming practical~\cite{Marzetta10}. As an alternative 
limit representing this situation, we consider the large-system limit 
in which the number of transmit antennas and the number of users tend to 
infinity while their ratio is kept constant. 

Vector perturbation or vector precoding (VP) is an effective pre-coding 
scheme that works well in the large-system 
limit~\cite{Hochwald05,Mueller08,Razi10}. In VP, the data symbols are 
modified to take values in relaxed alphabets to reduce the energy penalty. 
As relaxed alphabets, lattice-type alphabets~\cite{Hochwald05} and 
a continuous alphabet~\cite{Mueller08} have been proposed. In this paper, 
VP schemes with lattice-type and continuous alphabets are referred to as 
``lattice VP (LVP)'' and ``continuous VP (CVP),'' respectively. 
The search for a modified data symbol vector to minimize the energy penalty 
is NP-hard for LVP, so that LVP is infeasible for large alphabets or 
a large number of users. On the other hand, the search for CVP reduces to 
a quadratic optimization problem~\cite{Boyd04}, which may be solved by using 
an efficient algorithm. The large-system analysis in \cite{Mueller08,Zaidel12} 
has been shown that the performance of CVP is comparable to that of 
LVP in the large-system limit. In this paper, we only focus on CVP. 

A drawback of CVP is that the modulation is restricted to quadrature 
phase shift keying (QPSK). This restriction results in poor performance 
especially for the high SNR regime. In this paper, we propose a novel 
precoding scheme that is applicable for any modulation. The basic idea 
is to combine US and VP (US-VP). Joint US-VP we propose should not be 
confused with separate US-VP~\cite{Razi10}, in which a subset of users is 
first selected on the basis of CSI and subsequently VP is performed 
for the selected users. The crucial difference between the two schemes is 
that US depends on the data symbols for joint US-VP, whereas it is independent 
of the data symbols for separate US-VP. In this paper, joint US-VP is simply 
referred to as US-VP. 

Data-dependent US (DD-US) proposed in our previous work~\cite{Takeuchi121} 
can be regarded as a special example of US-VP: It is equivalent to 
US-VP with the original alphabet as the relaxed alphabet. DD-US allows us 
to use any modulation, as conventional US does. Furthermore, DD-US can be 
easily implemented with a suboptimal greedy algorithm for 
DD-US~\cite{Takeuchi121}. 

The goal of this paper is to assess the performance of US-VP. 
For that purpose, we consider the large-system limit 
in which the number of transmit antennas, the number of users, and the 
number of selected users tend to infinity while their ratios are kept 
constant. The replica method is used to analyze the performance of US-VP in 
the large-system limit. The replica method was originally developed in 
statistical physics~\cite{Sherrington75,Mezard87,Nishimori01}, and has been 
used to analyze the performance of MIMO systems~\cite{Moustakas03,Mueller04,Guo05,Takeda06,Wen07,Mueller08,Takeuchi08,Takeuchi122,Zaidel12} since 
Tanaka's pioneering work~\cite{Tanaka02}. 

The weakness of the replica method is that the method is based on several 
{\em non-rigorous} assumptions, such as the commutativity between the 
large-system limit and the other limits, replica continuity, and 
replica symmetry (RS). The commutativity was justified for a spin glass 
model~\cite{Hemmen79}, called Sherrington-Kirkpatrick (SK) model.   
The validity of replica continuity is open. The RS assumption 
may be broken for several models. In this case, the assumption of replica 
symmetry breaking (RSB) should be considered~\cite{Parisi80}. 
The RS assumption corresponds to the situation under which an energy function, 
called free energy, is unimodal. On the other hand, the RSB assumption 
corresponds to the situation under which the free energy has many local 
minima~\cite{Mezard87}. The simplest (strongest) assumption for RSB is 
referred to as 1-step RSB (1RSB). The most complex (weakest) assumption for 
RSB is called full-step RSB (full-RSB), which includes the RS assumption and 
the other lower-step RSB assumptions. In this paper, only the RS and 
1RSB assumptions are considered, since the assumption of higher-step RSB 
yields {\em numerically} unsolvable results for our problem. Thus, the 
results presented in this paper should be regarded as an approximation for 
the true ones. 

Recently, the validity of several results obtained from the replica method has 
been investigated. Korada and Montanari~\cite{Korada11} proved Tanaka's formula 
based on the RS assumption. Guerra and Talagrand's excellent 
works~\cite{Guerra03,Talagrand06} proved that the replica analysis under the 
full-RSB assumption provides the correct result for the SK model. The latter 
methodology might be applicable for our problem.  

This paper is organized as follows: After summarizing the notation used in 
this paper, in Section~\ref{sec2} we introduce the MIMO-BC and US-VP.  
Section~\ref{sec3} summarizes the main results of this paper. 
In Section~\ref{sec4}, we present numerical results based on the main results. 
Section~\ref{sec5} concludes this paper. The derivations of the 
main results are summarized in the appendices. 

\subsection{Notation}
For a complex number $z\in\mathbb{C}$, the real and imaginary parts of $z$ are 
denoted by $\Re[z]$ and $\Im[z]$, respectively. Furthermore, $z^{*}$ stands 
for the complex conjugate of $z$. For a matrix $\boldsymbol{A}$, the 
transpose, conjugate transpose, trace, and the determinant of $\boldsymbol{A}$ 
are denoted by $\boldsymbol{A}^{\mathrm{T}}$, $\boldsymbol{A}^{\mathrm{H}}$, 
$\mathrm{Tr}\boldsymbol{A}$, and $\det\boldsymbol{A}$, respectively. 
$\boldsymbol{I}_{N}$ stands for the $N\times N$ identity matrix.
$\boldsymbol{1}_{N}$ represents the $N$-dimensional vector whose elements are 
all one. $\mathrm{diag}\{a_{1},\ldots,a_{N}\}$ stands for the $N\times N$ 
diagonal matrix with $a_{n}$ as the $n$th diagonal element. 
The Kronecker product operator between two matrices is denoted by $\otimes$.  

For a set $\mathcal{A}=\{a_{i}:i=1,\ldots,N\}$, $\backslash a_{i}$ stands for 
the set $\{a_{i'}:\hbox{for all $i'\neq i$}\}$ obtained by eliminating $a_{i}$ 
from $\mathcal{A}$. Similarly, $\backslash\mathcal{A}_{i}$ denotes the union 
$\cup_{i'\neq i}\mathcal{A}_{i'}$ for sets $\{\mathcal{A}_{i}\}_{i=1}^{N}$. 
The direct product $\mathcal{A}_{1}\times\cdots\times\mathcal{A}_{N}$ is 
denoted by $\prod_{i=1}^{N}\mathcal{A}_{i}$. 

For a random variable $X$, $\mathbb{E}[X]$ and $\mathbb{V}[X]$ stand for 
the mean and variance of $X$, respectively. For the sequence of real random 
variables $\{X_{i}\}_{i=1}^{N}$, $X_{(i)}$ denotes the $i$th order statistic 
of $\{X_{i}\}$, i.e.\ $X_{(1)}\leq\cdots\leq X_{(N)}$~\cite{David03}.  
$\mathcal{N}(\boldsymbol{m},\boldsymbol{\Sigma})$ 
represents a real Gaussian distribution with mean $\boldsymbol{m}$ and 
covariance matrix $\boldsymbol{\Sigma}$. Similarly, 
$\mathcal{CN}(\boldsymbol{m},\boldsymbol{\Sigma})$ stands for 
a proper complex Gaussian distribution with mean 
$\boldsymbol{m}$ and covariance matrix $\boldsymbol{\Sigma}$~\cite{Neeser93}. 

For a discrete random variable $X$, the entropy of $X$ is denoted by $H(X)$. 
If $X$ is a continuous random variable, $h(X)$ represents the differential 
entropy. For two random variables $X$ and $Y$, the mutual information between 
$X$ and $Y$ is denoted by $I(X;Y)$. Throughout this paper, all logarithms are 
taken to base $2$, while the natural logarithm is denoted by $\ln$.  

Finally, we summarize several functions used in this paper. The function 
$\delta(\cdot)$ represents the Dirac delta function. For a proposition~$P$, 
the indicator function $1(P)$ is defined as 
\begin{equation}
1(P) 
= \left\{
\begin{array}{cl}
1 & \hbox{$P$ is true} \\ 
0 & \hbox{$P$ is false.} 
\end{array}
\right. 
\end{equation}
The probability density function (pdf) of a circularly symmetric complex 
Gaussian random variable with variance $\sigma^{2}$ is denoted by  
\begin{equation} \label{complex_Gauss}
p_{\mathrm{CG}}(z;\sigma^{2}) 
= \frac{1}{\pi\sigma^{2}}\mathrm{e}^{-\frac{|z|^{2}}{\sigma^{2}}} 
\quad \hbox{for $z\in\mathbb{C}$}. 
\end{equation}
Similarly, the pdf of a zero-mean real Gaussian random 
variable with variance $\sigma^{2}$ is written as 
\begin{equation} \label{Gauss} 
p_{\mathrm{G}}(x;\sigma^{2}) 
= \frac{1}{\sqrt{2\pi\sigma^{2}}}\mathrm{e}^{
 -\frac{x^{2}}{2\sigma^{2}} 
} \quad \hbox{for $x\in\mathbb{R}$.} 
\end{equation}
The standard Gaussian measure $Dx$ is defined as 
\begin{equation} \label{Gaussian_measure} 
Dx = p_{\mathrm{G}}(x;1) dx.  
\end{equation}
Furthermore, the function $Q(x)$ is given by 
\begin{equation} \label{Q_function}   
Q(x) = \int_{x}^{\infty}Dy. 
\end{equation}

\section{System Models} \label{sec2} 
\subsection{MIMO Broadcast Channel} 
We consider the MIMO-BC which consists of one BS with $N$ transmit antennas 
and $K$ users with one receive antenna. For simplicity, Rayleigh block-fading 
channels are assumed: The channel gains between the BS and each user are 
fixed during $T_{\mathrm{c}}$ time slots, and at the beginning of the next 
fading block the channel gains are independently sampled from a circularly 
symmetric complex Gaussian distribution. Let $y_{k,t}\in\mathbb{C}$ denote 
the received signal for user~$k$ in time slot~$t$. The receive vector 
$\boldsymbol{y}_{t}=(y_{1,t},\ldots,y_{K,t})^{\mathrm{T}}\in\mathbb{C}^{K}$ 
consisting of all received signals in time slot~$t$ is given by 
\begin{equation} \label{MIMO-BC} 
\boldsymbol{y}_{t} 
= \boldsymbol{H}\boldsymbol{u}_{t} + \boldsymbol{n}_{t}, 
\quad t=0,\ldots,T_{\mathrm{c}}-1.  
\end{equation}
In (\ref{MIMO-BC}), $\boldsymbol{n}_{t}=(n_{1,t},\ldots,n_{K,t})^{\mathrm{T}}
\sim\mathcal{CN}(\boldsymbol{0},N_{0}\boldsymbol{I}_{K})$ denotes the additive 
white Gaussian noise (AWGN) vector. The vector 
$\boldsymbol{u}_{t}\in\mathbb{C}^{N}$ is the transmit 
vector in time slot~$t$, which will be defined shortly. 
Each row vector of the channel matrix $\boldsymbol{H}\in\mathbb{C}^{K\times N}$ 
corresponds to the channel gains between the BS and each user. 
It is natural for the MIMO-BC to assume that each element of the channel 
matrix is $O(1)$ and that the time-average transmit power is constrained 
to below $P$. For convenience in analysis, we make an equivalent assumption: 
We assume that $\boldsymbol{H}$ has independent circularly symmetric 
complex Gaussian elements with variance $1/N$, and that the time-average  
transmit power is constrained to below $NP$, i.e.\ 
\begin{equation} \label{power_constraint} 
\frac{1}{T_{\mathrm{c}}}\sum_{t=0}^{T_{\mathrm{c}}-1}\|\boldsymbol{u}_{t}\|^{2} 
\leq NP. 
\end{equation}
Under these assumptions, the transmit SNR is defined as $P/N_{0}$. 

Slow fading is considered in this paper, i.e.\ 
$T_{\mathrm{c}}\rightarrow\infty$. Note that the channel matrix in 
(\ref{MIMO-BC}) is fixed during $T_{\mathrm{c}}$ time slots. In this situation, 
we can assume that the channel matrix $\boldsymbol{H}$ is known to the 
transmitter, since the transmitter can estimate the channel matrix on the 
basis of pilot signals transmitted from each user in a negligibly small 
portion of one fading block.  

\subsection{Zero-Forcing Transmit Beamforming} 
Let $\boldsymbol{x}_{t}=(x_{1,t},\ldots,x_{K,t})^{\mathrm{T}}\in\mathbb{C}^{K}$
denote the data symbol vector in time slot~$t$. The set   
$\mathcal{X}_{k}^{(\mathrm{all})}=\{x_{k,t}:t=0,\ldots,T_{\mathrm{c}}-1\}$ 
corresponds to the data symbols sent to user~$k$, and is assumed to be 
independent for different~$k$. Throughout this paper, power allocation is 
not considered: The data symbols $\{x_{k,t}\}$ are assumed to be independent 
and identically distributed (i.i.d.) zero-mean complex random variables with 
unit variance. 

The BS uses the information about the channel matrix to pre-cancel IUI. 
For $K\leq N$, the simplest method for pre-cancellation is 
ZFBF~\cite{Wiesel08}, in which the transmit vector $\boldsymbol{u}_{t}$ is 
given by 
\begin{equation} \label{transmit_vector_ZF} 
\boldsymbol{u}_{t} 
= \sqrt{\frac{NP}{\mathcal{E}(\boldsymbol{H},\{\boldsymbol{x}_{t}\})}}
\boldsymbol{u}_{t}^{(\mathrm{ZF})}(\boldsymbol{H},\boldsymbol{x}_{t}), 
\end{equation}
with 
\begin{equation} \label{ZFBF} 
\boldsymbol{u}_{t}^{(\mathrm{ZF})}(\boldsymbol{H},\boldsymbol{x}_{t}) 
= \boldsymbol{H}^{\mathrm{H}}
\left(
 \boldsymbol{H}\boldsymbol{H}^{\mathrm{H}}
\right)^{-1}\boldsymbol{x}_{t}. 
\end{equation}
In (\ref{transmit_vector_ZF}), the energy penalty 
$\mathcal{E}(\boldsymbol{H},\{\boldsymbol{x}_{t}\})$ is defined as the 
time-average power of the ZFBF vectors~(\ref{ZFBF}),  
\begin{equation} \label{energy_penalty} 
\mathcal{E}(\boldsymbol{H},\{\boldsymbol{x}_{t}\}) 
= \frac{1}{T_{\mathrm{c}}}\sum_{t=0}^{T_{\mathrm{c}}-1} 
\left\|
 \boldsymbol{u}_{t}^{(\mathrm{ZF})}(\boldsymbol{H},\boldsymbol{x}_{t}) 
\right\|^{2}. 
\end{equation}
It is straightforward to confirm that the transmit 
vector~(\ref{transmit_vector_ZF}) satisfies the power 
constraint~(\ref{power_constraint}). 

The ZFBF~(\ref{transmit_vector_ZF}) decomposes the MIMO-BC~(\ref{MIMO-BC}) 
into per-user channels 
\begin{equation} \label{interference_free_channel} 
y_{k,t} 
= \sqrt{\frac{NP}{\mathcal{E}(\boldsymbol{H},\{\boldsymbol{x}_{t}\})}}x_{k,t} 
+ n_{k,t}, 
\end{equation} 
for all $k$. This implies that the receive SNR is given by 
$NP/(\mathcal{E}(\boldsymbol{H},\{\boldsymbol{x}_{t}\})N_{0})$. 
The drawback of the ZFBF is an increase of the energy 
penalty~(\ref{energy_penalty}). Substituting the ZFBF vector~(\ref{ZFBF}) 
into (\ref{energy_penalty}) yields  
\begin{IEEEeqnarray}{rl} 
\mathcal{E}(\boldsymbol{H},\{\boldsymbol{x}_{t}\}) 
=& \frac{1}{T_{\mathrm{c}}}\sum_{t=0}^{T_{\mathrm{c}}-1} 
\boldsymbol{x}_{t}^{\mathrm{H}}\left(
 \boldsymbol{H}\boldsymbol{H}^{\mathrm{H}} 
\right)^{-1}\boldsymbol{x}_{t} \label{energy_penalty_ZF} \\ 
\rightarrow&\mathrm{Tr}\left\{
 \left(
  \boldsymbol{H}\boldsymbol{H}^{\mathrm{H}} 
 \right)^{-1}
\right\}, \label{energy_penalty_ZF_inf}
\end{IEEEeqnarray}
in $T_{\mathrm{c}}\rightarrow\infty$. The Mar$\breve{\rm c}$enko-Pastur 
law~\cite{Tulino04} 
implies that the energy penalty~(\ref{energy_penalty_ZF_inf}) per user 
converges almost surely to 
\begin{equation} \label{asymptotic_energy_penalty_ZF}
\frac{1}{K}\mathcal{E}(\boldsymbol{H},\{\boldsymbol{x}_{t}\})
\rightarrow \frac{1}{1-\alpha}, 
\end{equation}
in the large-system limit, where both $K$ and $N$ tend to infinity with their 
ratio $\alpha=K/N$ kept constant. The asymptotic energy 
penalty~(\ref{asymptotic_energy_penalty_ZF}) diverges as $\alpha$ gets 
closer to $1$ from below. Since the receive SNR 
$NP/(\mathcal{E}(\boldsymbol{H},\{\boldsymbol{x}_{t}\})N_{0})$ is inversely 
proportional to the energy penalty, this divergence results in a fatal 
degradation of the receive SNR.  

\subsection{Vector Precoding} 
As a method for improving the drawback of ZFBF, VP with ZFBF was 
proposed~\cite{Hochwald05,Mueller08}. In VP, each data symbol 
$x_{k,t}$ is modified to take values in a relaxed alphabet 
$\mathcal{M}_{x_{k,t}}\subset\mathbb{C}$, depending on the original data 
symbol $x_{k,t}$, to reduce the energy penalty. 
The modified data symbol vector 
$\tilde{\boldsymbol{x}}_{t}\in\prod_{k=1}^{K}\mathcal{M}_{x_{k,t}}$ 
based on the minimization of the energy 
penalty~(\ref{energy_penalty_ZF}) is given by 
\begin{equation} \label{VP} 
\tilde{\boldsymbol{x}}_{t} 
= \argmin_{\tilde{\boldsymbol{x}}_{t}\in\mathcal{M}_{x_{1,t}}\times
\cdots\times\mathcal{M}_{x_{K,t}}}
\tilde{\boldsymbol{x}}_{t}^{\mathrm{H}}\left(
 \boldsymbol{H}\boldsymbol{H}^{\mathrm{H}} 
\right)^{-1}\tilde{\boldsymbol{x}}_{t}. 
\end{equation}  
Note that the modified vector~(\ref{VP}) to minimize each instantaneous 
power $\|\boldsymbol{u}_{t}^{(\mathrm{ZF})}
(\boldsymbol{H},\boldsymbol{x}_{t})\|^{2}$ for the ZFBF~(\ref{ZFBF}) 
minimizes the energy penalty~(\ref{energy_penalty_ZF}) for the ZFBF. 

\begin{example}[CVP]
Suppose that QPSK is used. For CVP~\cite{Mueller08}, 
the relaxed alphabet $\mathcal{M}_{x}$ for a QPSK data symbol $x$ is given by 
\begin{equation} \label{CVP} 
\mathcal{M}_{x} = \tilde{\mathcal{M}}_{\Re[x]} 
+ \mathrm{i}\tilde{\mathcal{M}}_{\Im[x]}, 
\end{equation}
with 
\begin{equation} \label{each_CVP} 
\tilde{\mathcal{M}}_{b} 
= \left\{
\begin{array}{ll}
[b,\infty) & \hbox{for $b=1/\sqrt{2}$} \\ 
(-\infty,b] & \hbox{for $b=-1/\sqrt{2}$}. 
\end{array}
\right.
\end{equation}
M\"uller et al.~\cite{Mueller08} showed that the CVP results in a significant 
reduction of the energy penalty, compared to the conventional ZFBF. 
The minimization problem (\ref{VP}) with (\ref{CVP}) reduces to a quadratic 
optimization problem~\cite{Boyd04}, so that one can use an efficient 
algorithm to solve (\ref{VP}). 
\end{example}

The point of the CVP is that the modified data symbol vector 
$\tilde{\boldsymbol{x}}_{t}$ depends on the channel matrix $\boldsymbol{H}$. 
Consequently, the energy penalty 
$\mathcal{E}(\boldsymbol{H},\{\tilde{\boldsymbol{x}}_{t}\})$, given by 
(\ref{energy_penalty}), for the CVP never tends to 
(\ref{energy_penalty_ZF_inf}) in $T_{\mathrm{c}}\rightarrow\infty$. In fact, 
the energy penalty for the CVP was shown to be bounded in the limit 
$\alpha\rightarrow1$ after taking the large-system limit~\cite{Zaidel12}.

\subsection{Joint User Selection and Vector Precoding} 
We propose US-VP based on the combination of US and VP. US-VP is performed 
every $T$ ($\ll T_{\mathrm{c}}$) time slots. Let 
$\mathcal{K}_{i}\subset\mathcal{K}_{\mathrm{all}}=\{1,\ldots,K\}$, with size 
$\tilde{K}=|\mathcal{K}_{i}|$ ($\leq N$), denote the set of selected users in 
the $i$th block of US ($i=0,\ldots,T_{\mathrm{c}}/T-1$). The corresponding 
modified data symbol vectors are denoted 
by $\tilde{\boldsymbol{x}}_{\mathcal{K}_{i},t}\in\prod_{k\in\mathcal{K}_{i}}
\mathcal{M}_{x_{k,t}}$ for $t=iT,\ldots,(i+1)T-1$. 
The set of selected users $\mathcal{K}_{i}$ and the corresponding modified 
vectors $\{\tilde{\boldsymbol{x}}_{\mathcal{K}_{i},t}:t=iT,\ldots,(i+1)T-1\}$ 
are selected\footnote{
If there are multiple solutions, one solution is selected randomly and 
uniformly.} to minimize the energy penalty~(\ref{energy_penalty}): 
\begin{equation}  
(\mathcal{K}_{i}, \{\tilde{\boldsymbol{x}}_{\mathcal{K}_{i},t}\}) 
= \argmin_{\mathcal{K}_{i},\{\tilde{\boldsymbol{x}}_{\mathcal{K}_{i},t}\}}
\mathcal{E}_{i}(\boldsymbol{H}_{\mathcal{K}_{i}},
\{\tilde{\boldsymbol{x}}_{\mathcal{K}_{i},t}\}), \label{US-VP}
\end{equation}
where the minimization is taken over 
$\{\mathcal{K}_{i}\subset\mathcal{K}_{\mathrm{all}}:|\mathcal{K}_{i}|=\tilde{K}\}$ 
and $\{\tilde{\boldsymbol{x}}_{\mathcal{K}_{i},t}\in\prod_{k\in\mathcal{K}_{i}}
\mathcal{M}_{x_{k,t}}:t=iT,\ldots,(i+1)T-1\}$, 
with 
\begin{equation} \label{energy_penalty_US-VP} 
\mathcal{E}_{i}(\boldsymbol{H}_{\mathcal{K}_{i}},
\{\tilde{\boldsymbol{x}}_{\mathcal{K}_{i},t}\}) 
= \frac{1}{T}\sum_{t=iT}^{(i+1)T-1}
\left\|
 \boldsymbol{u}_{t}^{(\mathrm{ZF})}(\boldsymbol{H}_{\mathcal{K}_{i}},
 \tilde{\boldsymbol{x}}_{\mathcal{K}_{i},t})
\right\|^{2}. 
\end{equation}
In (\ref{energy_penalty_US-VP}), the ZFBF vector 
$\boldsymbol{u}_{t}^{(\mathrm{ZF})}(\boldsymbol{H}_{\mathcal{K}_{i}},
\tilde{\boldsymbol{x}}_{\mathcal{K}_{i},t})$ is given by (\ref{ZFBF}). 
Furthermore, $\boldsymbol{H}_{\mathcal{K}_{i}}\in\mathbb{C}^{\tilde{K}\times N}$ 
denotes the channel matrix corresponding to the selected users 
$\mathcal{K}_{i}$, which is obtained by collecting the row vectors for the 
selected users $\mathcal{K}_{i}$ from the channel matrix $\boldsymbol{H}$. 

\begin{example}[DD-US]
DD-US is defined as the US-VP~(\ref{US-VP}) with the 
original alphabet as the relaxed alphabet, i.e.\ 
$\mathcal{M}_{x_{k,t}}=\{x_{k,t}\}$. Thus, the modified data symbol vector 
$\tilde{\boldsymbol{x}}_{\mathcal{K}_{i},t}$ is equal to the original data 
symbol vector $\boldsymbol{x}_{\mathcal{K}_{i},t}\in\mathbb{C}^{\tilde{K}}$, 
obtained by stacking the data symbols $\{x_{k,t}\}$ for the selected 
users $\mathcal{K}_{i}$. The minimization problem~(\ref{US-VP}) for the 
DD-US can be approximately solved by using a greedy algorithm proposed 
in \cite{Takeuchi121}. 
\end{example}

\begin{example}[US-CVP]
Suppose that QPSK is used. Joint US and CVP (US-CVP) is defined as the 
US-VP~(\ref{US-VP}) with the CVP~(\ref{CVP}). Unfortunately, the minimization 
problem~(\ref{US-VP}) is not convex. It may be possible to extend the 
greedy algorithm for the DD-US~\cite{Takeuchi121} to the US-CVP. 
Obviously, the obtained algorithm should be more complex than the greedy 
algorithm for the DD-US.  
\end{example}

The transmit vector $\boldsymbol{u}_{t}$ for  
US-VP~(\ref{US-VP}) in time slot~$t$ is given by 
\begin{equation} \label{transmit_vector_US-VP} 
\boldsymbol{u}_{t} 
= \sqrt{\frac{NP}{\mathcal{E}(\{\boldsymbol{H}_{\mathcal{K}_{i}}\},
\{\tilde{\boldsymbol{x}}_{\mathcal{K}_{i},t}\})}}
\boldsymbol{u}_{t}^{(\mathrm{ZF})}(\boldsymbol{H}_{\mathcal{K}_{i}},
\tilde{\boldsymbol{x}}_{\mathcal{K}_{i},t}), 
\end{equation}
where the energy penalty $\mathcal{E}(\{\boldsymbol{H}_{\mathcal{K}_{i}}\},
\{\tilde{\boldsymbol{x}}_{\mathcal{K}_{i},t}\})$ for the US-VP is defined as 
\begin{equation} \label{time_averaged_energy_penalty} 
\mathcal{E}(\{\boldsymbol{H}_{\mathcal{K}_{i}}\},
\{\tilde{\boldsymbol{x}}_{\mathcal{K}_{i},t}\}) 
= \frac{1}{T_{\mathrm{c}}/T}\sum_{i=0}^{T_{\mathrm{c}}/T-1}
\mathcal{E}_{i}(\boldsymbol{H}_{\mathcal{K}_{i}},
\{\tilde{\boldsymbol{x}}_{\mathcal{K}_{i},t}\}),  
\end{equation}
with (\ref{energy_penalty_US-VP}). 
In order to simplify detection in each user, the data symbols for 
non-selected users are discarded at the {\em transmitter} 
side~\cite{Takeuchi121}. This implies that the 
ZFBF~(\ref{transmit_vector_US-VP}) with US-VP~(\ref{US-VP}) 
decomposes the MIMO-BC~(\ref{MIMO-BC}) into per-user channels 
\begin{equation} \label{equivalent_channel} 
y_{k,t} 
= \sqrt{\frac{NP}{\mathcal{E}(\{\boldsymbol{H}_{\mathcal{K}_{i}}\},
\{\tilde{\boldsymbol{x}}_{\mathcal{K}_{i},t}\})}}\left\{
 s_{k,i}\tilde{x}_{k,t} + (1-s_{k,i})I_{k,t}
\right\} + n_{k,t}, 
\end{equation}
for all $k$. In (\ref{equivalent_channel}), 
$\tilde{x}_{k,t}\in\mathcal{M}_{x_{k,t}}$ denotes the modified data symbol 
corresponding to the original data symbol $x_{k,t}$. The variable 
$s_{k,i}\in\{0,1\}$ indicating whether user~$k$ has been selected in block~$i$ 
is defined as 
\begin{equation} \label{indicator}
s_{k,i} = \left\{
\begin{array}{ll}
1 & k\in\mathcal{K}_{i} \\ 
0 & k\notin\mathcal{K}_{i}.  
\end{array}
\right.
\end{equation}
Furthermore, $I_{k,t}\in\mathbb{C}$ denotes the interference to the 
non-selected user~$k\notin\mathcal{K}_{i}$, given by 
\begin{equation} \label{interference} 
I_{k,t} 
= \vec{\boldsymbol{h}}_{k}\boldsymbol{u}_{t}^{(\mathrm{ZF})}
(\boldsymbol{H}_{\mathcal{K}_{i}},\tilde{\boldsymbol{x}}_{\mathcal{K}_{i},t}),
\end{equation} 
where $\vec{\boldsymbol{h}}_{k}\in\mathbb{C}^{1\times N}$ denotes the $k$th row 
vector of the channel matrix $\boldsymbol{H}$. Note that the indices $t$ 
of $y_{k,t}$ and $x_{k,t}$ in (\ref{equivalent_channel}) are identical to 
each other, since the data symbols for the non-selected users 
$k\notin\mathcal{K}_{i}$ have been discarded. This simplifies the detection 
of (\ref{indicator}).   

It is easy for user~$k$ to blind-detect {\em one} variable $s_{k,i}$ from 
the $T$ observations $\{y_{k,t}\}$ in each block. Using the decision-feedback 
of $\tilde{x}_{k,t}$ from the decoder improve the accuracy of 
detection~\cite{Takeuchi121}. In order to reduce the energy penalty, 
small $T$ should be used. On the other hand, too small $T$ makes it 
difficult to detect. As one option, dozens of time slots should be used as 
the block length $T$. For example, the energy loss due to detection errors is 
at most 0.2--0.5~dB for $T=16$~\cite{Takeuchi121}.

\begin{remark} \label{remark1} 
Let us discuss the relationship between the DD-US and conventional US. 
The set of selected users $\mathcal{K}_{0}$ in the first block for the DD-US 
is given by 
\begin{equation} \label{DD-US} 
\mathcal{K}_{0} = \argmin_{\mathcal{K}_{0}\subset\mathcal{K}_{\mathrm{all}}:
|\mathcal{K}|=\tilde{K}}
\mathcal{E}_{0}(\boldsymbol{H}_{\mathcal{K}_{0}},
\{\boldsymbol{x}_{\mathcal{K}_{0},t}\}), 
\end{equation}
with (\ref{energy_penalty_US-VP}). On the other hand, 
when the minimization of the energy penalty~(\ref{energy_penalty_ZF_inf}) 
for the ZFBF or equivalently of (\ref{energy_penalty_ZF}) in 
$T_{\mathrm{c}}\rightarrow\infty$ is used as the US criterion, 
the set of selected users $\mathcal{K}$ for conventional US is given by  
\begin{equation} \label{conventional_US} 
\mathcal{K} = \argmin_{\mathcal{K}\subset\mathcal{K}_{\mathrm{all}}:
|\mathcal{K}|=\tilde{K}}\lim_{T\rightarrow\infty}
\mathcal{E}_{0}(\boldsymbol{H}_{\mathcal{K}},\{\boldsymbol{x}_{\mathcal{K},t}\}),  
\end{equation} 
where we have re-written the coherence time $T_{\mathrm{c}}$ as $T$. 
Note that (\ref{conventional_US}) is independent of the data symbols, 
since the object function tends to $\mathrm{Tr}\{(\boldsymbol{H}_{\mathcal{K}}
\boldsymbol{H}_{\mathcal{K}}^{\mathrm{H}})^{-1}\}$. 

The minimization and the limit in (\ref{conventional_US}) is not 
commutative. It is straightforward to prove the inequality 
\begin{IEEEeqnarray}{rl}  
&\lim_{T\rightarrow\infty}\min_{\mathcal{K}_{0}\subset\mathcal{K}_{\mathrm{all}}:
|\mathcal{K}_{0}|=\tilde{K}}\mathcal{E}_{0}(\boldsymbol{H}_{\mathcal{K}_{0}},
\{\boldsymbol{x}_{\mathcal{K}_{0},t}\}) \nonumber \\ 
\leq& \min_{\mathcal{K}_{0}\subset\mathcal{K}_{\mathrm{all}}:
|\mathcal{K}_{0}|=\tilde{K}}\lim_{T\rightarrow\infty}
\mathcal{E}_{0}(\boldsymbol{H}_{\mathcal{K}_{0}},
\{\boldsymbol{x}_{\mathcal{K}_{0},t}\}). \label{relationship}
\end{IEEEeqnarray}
Comparing (\ref{DD-US}) and (\ref{conventional_US}), we find that the energy 
penalty of the DD-US in $T\rightarrow\infty$ provides a lower bound on that 
of the conventional US.  Let us prove the inequality~(\ref{relationship}). 
We start with a trivial inequality 
\begin{equation} \label{relationship1} 
\min_{\mathcal{K}_{0}\subset\mathcal{K}_{\mathrm{all}}:
|\mathcal{K}_{0}|=\tilde{K}}\mathcal{E}_{0}(\boldsymbol{H}_{\mathcal{K}_{0}},
\{\boldsymbol{x}_{\mathcal{K}_{0},t}\}) 
\leq \mathcal{E}_{0}(\boldsymbol{H}_{\mathcal{K}},
\{\boldsymbol{x}_{\mathcal{K},t}\}),  
\end{equation}
where $\mathcal{K}\subset\mathcal{K}_{\mathrm{all}}$ with 
$|\mathcal{K}|=\tilde{K}$ denotes the set of selected 
users~(\ref{conventional_US}) for the conventional US. We next take the 
limit $T\rightarrow\infty$. Since $\mathcal{K}$ is independent of the 
data symbols, we can use the weak law of large numbers for the right-hand 
side (RHS) of (\ref{relationship1}) to find that 
$\mathcal{E}_{0}(\boldsymbol{H}_{\mathcal{K}},
\{\boldsymbol{x}_{\mathcal{K},t}\})$ converges in probability to 
$\mathrm{Tr}\{(\boldsymbol{H}_{\mathcal{K}}
\boldsymbol{H}_{\mathcal{K}}^{\mathrm{H}})^{-1}\}$ or 
equivalently the RHS of (\ref{relationship}) in $T\rightarrow\infty$. 
Thus, we obtain the inequality~(\ref{relationship}). 
\end{remark}

\section{Main Results} \label{sec3} 
\subsection{Large-System Analysis}
We use the replica method to analyze the performance of US-VP 
in the large-system limit where the number of transmit 
antennas~$N$, the number of users~$K$, and the number of selected 
users~$\tilde{K}$ tend to infinity while their ratios $\alpha=K/N$ and 
$\kappa=\tilde{K}/K$ are kept constant. 
Without loss of generality, we focus on the first block $i=0$ of US-VP 
and drop the subscripts $i$ from 
$\mathcal{E}_{i}(\boldsymbol{H}_{\mathcal{K}_{i}},
\{\tilde{\boldsymbol{x}}_{\mathcal{K}_{i},t}\})$, $\mathcal{K}_{i}$, and 
$s_{k,i}$.  

The asymptotic performance of US-VP is characterized via 
a solvable US-VP problem. We first define the solvable problem. For a positive 
parameter~$q$, let us define a random variable 
$\tilde{E}_{k}(\{\tilde{x}_{k,t}\},q)$ as 
\begin{equation} \label{tilde_E_k} 
\tilde{E}_{k}(\{\tilde{x}_{k,t}\},q) = \frac{1}{T}\sum_{t=0}^{T-1}
|\tilde{x}_{k,t} - \sqrt{q}z_{k,t}|^{2}. 
\end{equation}
In (\ref{tilde_E_k}), $\{z_{k,t}\}$ are independent circularly symmetric 
complex Gaussian random variables with unit variance. 
The normalized parameter $q/(\alpha\kappa)$ will 
be shortly shown to be equal to the average energy penalty per selected user 
in the large-system limit. The solvable US-VP problem is the following 
minimization problem: 
\begin{equation} \label{equivalent_minimization} 
E_{K} 
= \min_{\mathcal{K}\subset\mathcal{K}_{\mathrm{all}}:|\mathcal{K}|=\tilde{K}}
\min_{\{\tilde{x}_{k,t}\in\mathcal{M}_{x_{k,t}}\}}
\frac{1}{K}\sum_{k\in\mathcal{K}}\tilde{E}_{k}(\{\tilde{x}_{k,t}\},q). 
\end{equation}

The asymptotic performance of US-VP is characterized via 
three quantities for (\ref{equivalent_minimization}) in the large-system 
limit. The minimization in (\ref{equivalent_minimization}) with respect to 
$\{\tilde{x}_{k,t}\}$ is straightforwardly solved to obtain 
\begin{equation} \label{equivalent_US} 
E_{K} 
= \min_{\mathcal{K}\subset\mathcal{K}_{\mathrm{all}}:|\mathcal{K}|=\tilde{K}}
\frac{1}{K}\sum_{k\in\mathcal{K}}E_{k}(q),  
\end{equation}
with 
\begin{equation} \label{E_k} 
E_{k}(q) 
= \frac{1}{T}\sum_{t=0}^{T-1}
|\tilde{x}_{k,t}^{(\mathrm{opt})}(q) - \sqrt{q}z_{k,t}|^{2},  
\end{equation}
where $\tilde{x}_{k,t}^{(\mathrm{opt})}(q)$ denotes the optimal modified 
data symbol, given by 
\begin{equation} \label{modified_symbol} 
\tilde{x}_{k,t}^{(\mathrm{opt})}(q) 
= \argmin_{\tilde{x}_{k,t}\in\mathcal{M}_{x_{k,t}}}
|\tilde{x}_{k,t} - \sqrt{q}z_{k,t}|^{2}. 
\end{equation}
In order to solve the minimization~(\ref{equivalent_US}) analytically, 
we write the order statistics for the random variables $\{E_{k}(q)\}$ as 
$\{E_{(k)}(q)\}$, i.e.\ $E_{(1)}(q)\leq \cdots \leq E_{(K)}(q)$~\cite{David03}. 
The minimization~(\ref{equivalent_US}) reduces to 
\begin{equation} \label{equivalent_US1} 
E_{K} = \frac{1}{K}\sum_{k=1}^{K}1\left(
 \frac{k}{K}\leq \kappa
\right)E_{(k)}(q). 
\end{equation}

The three quantities that characterizes the asymptotic performance of 
US-VP is the mean and variance for (\ref{equivalent_US1}) 
in the large-system limit, and the $\tilde{K}$th order statistic 
$E_{(\tilde{K})}(q)$ for (\ref{E_k}) in the large-system limit. 
The three quantities are given via the cumulative distribution function (cdf) 
of (\ref{E_k}), 
\begin{equation} \label{cdf} 
F_{T}(x;q) = \mathrm{Pr}(E_{k}(q)\leq x). 
\end{equation}
Note that the cdf~(\ref{cdf}) is monotonically increasing, 
because of $z_{k,t}\sim\mathcal{CN}(0,1)$. Thus, there exists 
the inverse function of (\ref{cdf}), denoted by $F_{T}^{-1}(x;q)$. 

\begin{lemma} \label{lemma1} 
Let $\xi_{\kappa,T}(q)$ denote the $\kappa$-quantile for the cdf~(\ref{cdf}), 
\begin{equation} \label{quantile} 
\xi_{\kappa,T}(q) = F_{T}^{-1}(\kappa;q). 
\end{equation}
Then, the $\tilde{K}$th order statistic $E_{(\tilde{K})}(q)$ for (\ref{E_k}) 
converges in probability to the $\kappa$-quantile $\xi_{\kappa,T}(q)$  
in the large-system limit. 
\end{lemma}
\begin{IEEEproof}[Proof of Lemma~\ref{lemma1}]
Since $E_{(\tilde{K})}(q)$ is a sample $\kappa$-quantile for 
the independent random variables~(\ref{E_k}) with the common 
cdf~(\ref{cdf}), Bahadur's theorem~\cite{Bahadur66} or its 
modification~\cite{Ghosh71} implies that 
\begin{equation} \label{Bahadur_theorem}
E_{(\tilde{K})}(q)  
= \xi_{\kappa,T}(q) - \frac{\hat{F}_{T}(\xi_{\kappa,T}(q);q) - \kappa}
{F_{T}'(\xi_{\kappa,T}(q);q)} + o(K^{-1/2}), 
\end{equation}
in the large-system limit, where $\hat{F}_{T}(x;q)$ is the 
empirical cdf for (\ref{E_k}), given by 
\begin{equation} \label{empirical_cdf} 
\hat{F}_{T}(x;q) 
= \frac{1}{K}\sum_{k=1}^{K}1(E_{k}(q)\leq x). 
\end{equation}
The mean and variance of the empirical cdf~(\ref{empirical_cdf}) 
at $x=\xi_{\kappa,T}(q)$ are given by 
\begin{equation}
\mathbb{E}[\hat{F}_{T}(\xi_{\kappa,T}(q);q)]
=\kappa, 
\end{equation}
\begin{equation}
\mathbb{V}[\hat{F}_{T}(\xi_{\kappa,T}(q);q)]
=\frac{\kappa(1-\kappa)}{K}, 
\end{equation}
respectively. Thus, the second term on the RHS of 
(\ref{Bahadur_theorem}) is a quantity of $O(K^{-1/2})$. This observation 
implies that (\ref{Bahadur_theorem}) converges in probability 
to the $\kappa$-quantile~(\ref{quantile}) in the large-system limit. 
\end{IEEEproof}

\begin{lemma}[Stigler 1974] \label{lemma2} 
Let $\mu_{\kappa,T}(q)$ and $\sigma_{\kappa,T}^{2}(q)$ denote the mean 
of $E_{K}$ and the variance of $\sqrt{K}E_{K}$, given by 
(\ref{equivalent_US1}), in the large-system limit. Then,   
\begin{equation} \label{mean}
\mu_{\kappa,T}(q) = \int_{0}^{\kappa}F_{T}^{-1}(x;q)dx, 
\end{equation}
\begin{IEEEeqnarray}{r}
\sigma_{\kappa,T}^{2}(q) = 
\int_{0}^{\xi_{\kappa,T}(q)}\int_{0}^{\xi_{\kappa,T}(q)} 
[F_{T}(\min(x,y);q) \nonumber \\ 
 - F_{T}(x;q)F_{T}(y;q) ]dxdy, \label{variance}  
\end{IEEEeqnarray}
where the $\kappa$-quantile $\xi_{\kappa,T}(q)$ is given by (\ref{quantile}).  
\end{lemma}
\begin{IEEEproof}[Proof of Lemma~\ref{lemma2}]
The function $1(k/K\leq\kappa)$ in (\ref{equivalent_US1}) is bounded and 
continuous almost everywhere (a.e.) $F_{T}^{-1}$, since the cdf~(\ref{cdf}) 
is monotonically increasing. Thus, we can use Stigler's 
theorems~\cite[Theorems~1 and 3]{Stigler74} to obtain (\ref{mean}) 
and (\ref{variance}). 
\end{IEEEproof}

We need calculate the cdf~(\ref{cdf}) to evaluate the three 
quantities~(\ref{quantile}), (\ref{mean}), and (\ref{variance}). 
See Appendix~\ref{calculation_cdf} for how to calculate the 
cdf~(\ref{cdf}). 

\subsection{Average Energy Penalty} \label{sec3_2}  
The average of the energy penalty~(\ref{time_averaged_energy_penalty}) for 
US-VP~(\ref{US-VP}), denoted by $\bar{\mathcal{E}}=\mathbb{E}[
\mathcal{E}(\{\boldsymbol{H}_{\mathcal{K}_{i}}\},
\{\tilde{\boldsymbol{x}}_{\mathcal{K}_{i},t}\}) ]$, 
is analyzed in the large-system limit. We use the replica method under 
the RS and 1RSB assumptions~\cite{Mezard87,Nishimori01}. Roughly speaking, 
the RS assumption corresponds to postulating that there are no local 
minimizers to the minimization~(\ref{US-VP}) in the large-system limit. 
On the other hand, the 1RSB assumption is the simplest assumption for 
the case where there are many local minimizers in the large-system limit. 

\begin{proposition} \label{proposition1} 
Suppose that $q_{0}$ is the solution to the fixed-point equation
\begin{equation} \label{fixed_point_RS} 
q_{0} = \alpha\mu_{\kappa,T}(q_{0}), 
\end{equation}
where $\mu_{\kappa,T}(q)$ is given by (\ref{mean}). If (\ref{fixed_point_RS}) 
has multiple solutions, the smallest solution $q_{0}$ is selected. 
Under the RS assumption, the average energy penalty per selected user 
$\bar{\mathcal{E}}/\tilde{K}$ converges to $q_{0}/(\alpha\kappa)$ 
in the large-system limit. 
\end{proposition}
\begin{IEEEproof}[Derivation of Proposition~\ref{proposition1}]
See Appendix~\ref{proposition1_proof}. 
\end{IEEEproof}
\begin{proposition} \label{proposition2}
Suppose that $q_{1}$ satisfies the coupled fixed-point equations 
\begin{equation} \label{fixed_point_1RSB1}
\ln\left(
 1 + \frac{q_{1}}{\chi}
\right)
= \frac{\alpha}{\chi}\left(
 \mu_{\kappa,T}(q_{1}) 
 - \frac{T\sigma_{\kappa,T}^{2}(q_{1})}{2\chi}
\right), 
\end{equation}
\begin{equation} \label{fixed_point_1RSB2} 
\frac{q_{1}}{\chi+q_{1}} = \frac{\alpha}{\chi}\left(
 \mu_{\kappa,T}(q_{1}) 
 - \frac{T\sigma_{\kappa,T}^{2}(q_{1})}{\chi}
\right), 
\end{equation}
for some $0<\chi<\infty$, 
in which $\mu_{\kappa,T}(q)$ and $\sigma_{\kappa,T}^{2}(q)$ 
are given by (\ref{mean}) and (\ref{variance}), respectively. 
If there are multiple solutions, the smallest solution $q_{1}$ is selected. 
Under the 1RSB assumption, the average energy penalty per selected user 
$\bar{\mathcal{E}}/\tilde{K}$ converges to $q_{1}/(\alpha\kappa)$ 
in the large-system limit. 
\end{proposition}
\begin{IEEEproof}[Derivation of Proposition~\ref{proposition2}] 
See Appendix~\ref{proposition2_proof}. 
\end{IEEEproof}

The asymptotic energy penalty for VP was calculated with the R-transform for 
the empirical eigenvalue distribution of 
$(\boldsymbol{H}\boldsymbol{H}^{\mathrm{H}})^{-1}$~\cite{Mueller08,Zaidel12}. 
Since it is difficult to apply this method to our case, another method is 
used in the calculation of the energy penalty, as presented in 
Appendix~\ref{proposition1_proof}. Note that the meanings of RSB are 
different for the two methods. The two methods should yield the same result 
under the full-RSB assumption, since the full-RSB assumption is expected to 
provide the correct solution~\cite{Guerra03,Talagrand06}. However, they may 
yield different results under the RS and 1RSB assumptions, since these   
assumptions are approximations. In fact, the two methods seem to yield 
different results under the 1RSB assumption, whereas the same result is 
obtained under the RS assumption. 

It is straightforward to show that the RS assumption provides a smaller  
prediction of the energy penalty than the 1RSB assumption. 

\begin{property} \label{property1} 
Let $q_{0}$ and $q_{1}$ denote the solutions defined in 
Proposition~\ref{proposition1} and Proposition~\ref{proposition2}, 
respectively. Then, 
\begin{equation}
q_{0} < q_{1}. 
\end{equation} 
\end{property}
\begin{IEEEproof}[Proof of Property~\ref{property1}] 
Eliminating $\sigma_{\kappa,T}^{2}(q_{1})$ with (\ref{fixed_point_1RSB1}) and 
(\ref{fixed_point_1RSB2}) yields 
\begin{equation} \label{reduced_fixed_point} 
2\chi\ln\left(
 1 + \frac{q_{1}}{\chi}
\right) - \frac{\chi q_{1}}{\chi+q_{1}} = \alpha\mu_{\kappa,T}(q_{1}). 
\end{equation}
We write the left-hand side (LHS) of (\ref{reduced_fixed_point}) as 
$f(q_{1},\chi)$. It is straightforward to prove $f(q,\chi)<q$ for 
any $q>0$ and $\chi>0$. Calculating the first and second derivatives of 
$f(q,\chi)$ with respect to $\chi$, we obtain 
\begin{equation} \label{deriv1}
\frac{\partial f}{\partial\chi} 
= 2\ln\left(
 1 + \frac{q}{\chi}
\right) - \frac{q(2\chi+3q)}{(\chi+q)^{2}}, 
\end{equation}
\begin{equation} \label{deriv2} 
\frac{\partial^{2}f}{\partial\chi^{2}} 
= - \frac{2q^{3}}{\chi(\chi+q)^{3}}<0. 
\end{equation}
Since $\lim_{\chi\rightarrow\infty}\partial f/\partial\chi=0$, (\ref{deriv1}) 
and (\ref{deriv2}) imply $\partial f/\partial\chi>0$ for any $q>0$. 
Furthermore, $\lim_{\chi\rightarrow\infty}f(q,\chi)=q$ indicates 
$f(q,\chi)<q$ for any $q>0$ and $\chi>0$.  

Let us prove $q_{0}<q_{1}$. Since (\ref{mean}) is positive in $q\rightarrow0$, 
from (\ref{fixed_point_RS}) we find $q\leq\alpha\mu_{\kappa,T}(q)$ for any 
$q\in(0,q_{0})$. Then,   
\begin{equation}
f(q,\chi) < q \leq \alpha\mu_{\kappa,T}(q), 
\end{equation}
for any $q\in(0,q_{0}]$ and $\chi>0$. This inequality implies that  
(\ref{reduced_fixed_point}) has no solutions for any $q_{1}\in(0,q_{0}]$. 
Thus, we obtain $q_{0}<q_{1}$.  
\end{IEEEproof}

We next calculate the average energy penalty in $T\rightarrow\infty$ to 
derive a performance bound for separate US-VP. 

\begin{corollary} \label{corollary1} 
Suppose that $q_{0}$ is the solution to the fixed-point equation
\begin{equation} \label{fixed_point_inf}  
q_{0} = \alpha\kappa\mathbb{E}\left[
 \min_{\tilde{x}_{k,t}\in\mathcal{M}_{x_{k,t}}}
 |\tilde{x}_{k,t} - \sqrt{q_{0}}z_{k,t}|^{2}
\right].  
\end{equation}
If (\ref{fixed_point_inf}) has multiple solutions, 
the smallest solution $q_{0}$ is selected. Under the RS assumption, 
the average energy penalty per selected user 
$\bar{\mathcal{E}}/\tilde{K}$ converges to $q_{0}/(\alpha\kappa)$ 
in $T\rightarrow\infty$ after taking the large-system limit. 
\end{corollary}
\begin{IEEEproof}[Proof of Corollary~\ref{corollary1}] 
See Appendix~\ref{proof_corollary1} 
\end{IEEEproof} 
An informal derivation of (\ref{fixed_point_inf}) is as follows: 
First of all, one should recall that $\mu_{\kappa,T}(q)$ have been defined 
as the mean of (\ref{equivalent_US}) in the large-system limit. The weak law 
of large numbers implies that each term~(\ref{E_k}) in (\ref{equivalent_US})  
converges in probability to the expectation $\mathbb{E}[E_{k}(q)]$ 
in $T\rightarrow\infty$, which is equal to the expectation on the RHS of 
(\ref{fixed_point_inf}) with $q_{0}=q$. Thus, the minimization in 
(\ref{equivalent_US}) should make no sense in $T\rightarrow\infty$, i.e., 
(\ref{equivalent_US}) should tend to $\kappa\mathbb{E}[E_{k}(q)]$ 
in $T\rightarrow\infty$. This implies that the fixed-point 
equation~(\ref{fixed_point_RS}) reduces to (\ref{fixed_point_inf}) in 
$T\rightarrow\infty$.  

As noted in Remark~\ref{remark1}, the energy penalty for the DD-US in 
$T\rightarrow\infty$ provides a lower bound on that for the conventional 
(data-independent) US~(\ref{conventional_US}). For the DD-US, it is 
possible to solve (\ref{fixed_point_inf}),  
\begin{equation} \label{fixed_point_DD-US_inf}
\frac{q_{0}}{\alpha\kappa} = \frac{1}{1 - \alpha\kappa}. 
\end{equation} 
Comparing (\ref{asymptotic_energy_penalty_ZF}) 
and (\ref{fixed_point_DD-US_inf}), we find that the energy penalty for 
the DD-US in $T\rightarrow\infty$ is achievable by the ZFBF with random 
US (RUS), referred to as ZFBF-RUS in this paper, in which a subset of 
users with size $\tilde{K}$ is selected randomly and uniformly.  
This observation under the RS assumption implies that the performance of 
the ZFBF with the optimal US is equal to that of the ZFBF-RUS in the 
large-system limit. Furthermore, from Property~\ref{property1} we can conclude 
that the 1RSB assumption yields a wrong result in $T\rightarrow\infty$. 
The same statements also hold for separate US-CVP: Under the RS assumption, 
the performance of separate US-CVP is equal to that of the CVP with 
RUS (CVP-RUS) in the large-system limit. Furthermore, 
the 1RSB assumption yields a wrong result in $T\rightarrow\infty$. 

One cannot conclude from the results under the RS and 1RSB assumptions that 
conventional (data-independent) US makes no sense in the large-system limit, 
since there is a possibility that the full-RSB assumption provides a smaller 
energy penalty than the RS assumption. Unfortunately, it is difficult to 
calculate the energy penalty under higher-step RSB assumptions,  
so that whether this statement is correct should be checked by using another 
methodology. We leave this issue as future work, since it is beyond the 
scope of this paper. 

\subsection{Sum Rate}
Before investigating the achievable sum rate of US-VP, the joint 
distribution of the indicator variable~(\ref{indicator}) and the modified data 
symbols $\tilde{\mathcal{X}}_{k}=\{\tilde{x}_{k,t}:t=0,\ldots,T-1\}$ 
given the data symbols $\mathcal{X}_{k}=\{x_{k,t}:t=0,\ldots,T-1\}$ 
is analyzed in the large-system limit. This joint distribution is used to 
calculate the achievable sum rate. 

Let $\{\mathcal{A}_{t}\}$ denote measurable subsets of $\mathbb{C}$ for 
$t=0,\ldots,T-1$. The joint distribution is shown to be characterized via 
the conditional probability 
\begin{equation} \label{joint_distribution} 
\mathrm{Pr}\left(
 \left.
  E_{k}(q)\leq \xi_{\kappa,T}(q), 
  \{\tilde{x}_{k,t}^{(\mathrm{opt})}(q)\in\mathcal{A}_{t}\} 
 \right| \mathcal{X}_{k}
\right), 
\end{equation}
where $E_{k}(q)$, $\tilde{x}_{k,t}^{(\mathrm{opt})}(q)$, and $\xi_{\kappa,T}(q)$ 
are given by (\ref{E_k}), (\ref{modified_symbol}), and (\ref{quantile}), 
respectively. 

\begin{proposition} \label{proposition3} 
Suppose that $q_{0}$ is the same solution as in 
Proposition~\ref{proposition1}. Under the RS assumption, the 
conditional joint probability 
$\mathrm{Pr}(s_{k}=1, \tilde{\mathcal{X}}_{k}\in\prod_{t=0}^{T-1}
\mathcal{A}_{t} | \mathcal{X}_{k})$ converges to (\ref{joint_distribution}) 
with $q=q_{0}$ in the large-system limit. 
\end{proposition}
\begin{IEEEproof}[Derivation of Proposition~\ref{proposition3}]
See Appendix~\ref{proposition3_proof}. 
\end{IEEEproof}

\begin{proposition} \label{proposition4} 
Suppose that $q_{1}$ is the same solution as in 
Proposition~\ref{proposition2}. Under the 1RSB assumption, the 
conditional joint probability 
$\mathrm{Pr}(s_{k}=1, \tilde{\mathcal{X}}_{k}\in\prod_{t=0}^{T-1}
\mathcal{A}_{t} | \mathcal{X}_{k})$ 
converges to (\ref{joint_distribution}) with $q=q_{1}$ 
in the large-system limit. 
\end{proposition}
\begin{IEEEproof}[Derivation of Proposition~\ref{proposition4}]
See Appendix~\ref{proposition4_proof}. 
\end{IEEEproof}

It is straightforward to find 
$\mathrm{Pr}(s_{k}=1)=\kappa$: Marginalizing (\ref{joint_distribution}) 
yields 
\begin{equation}
\mathrm{Pr}(s_{k}=1) 
= \mathrm{Pr}\left(
 E_{k}(q)\leq \xi_{\kappa,T}(q)
\right), 
\end{equation}
which is equal to the cdf $F_{T}(\xi_{\kappa,T}(q);q)$, given by (\ref{cdf}). 
The definition of the $\kappa$-quantile~(\ref{quantile}) implies 
$\mathrm{Pr}(s_{k}=1)=\kappa$ for any $q$. 

We next investigate the achievable sum rate of US-VP in the large-system 
limit. The achievable sum rate $R$ is given by 
\begin{equation} \label{sum_rate} 
R = \sum_{k=1}^{K}R_{k}, 
\end{equation}
where the achievable rate $R_{k}$ for user~$k$ is defined as the mutual 
information per time slot between all data symbols 
$\mathcal{X}_{k}^{(\mathrm{all})}$ for user~$k$ and all received signals 
$\mathcal{Y}_{k}^{(\mathrm{all})}=\{y_{k,t}:t=0,\ldots,T_{\mathrm{c}}-1\}$ for 
user~$k$~\cite{Cover06,Tse05}, 
\begin{equation} \label{user_rate} 
R_{k} = \lim_{T_{\mathrm{c}}\rightarrow\infty}\frac{1}{T_{\mathrm{c}}}
I\left(
 \mathcal{X}_{k}^{(\mathrm{all})};\mathcal{Y}_{k}^{(\mathrm{all})}
\right). 
\end{equation}
In (\ref{user_rate}), the received signal $y_{k,t}$ is transmitted through 
the equivalent channel~(\ref{equivalent_channel}), in which $y_{k,t}$ depends 
on the data symbol $x_{k,t}$ through the modified data symbol 
$\tilde{x}_{k,t}\in\mathcal{M}_{x_{k,t}}$. 

A crucial assumption in evaluating the achievable rate~(\ref{user_rate}) 
is the self-averaging property of the energy 
penalty~(\ref{time_averaged_energy_penalty}) in the large-system limit. 

\begin{assumption}[Self-Averaging Property] \label{self-averaging} 
The energy penalty per selected user for US-VP converges 
in probability to the expectation $\bar{\mathcal{E}}
=\mathbb{E}[\mathcal{E}(\{\boldsymbol{H}_{\mathcal{K}_{i}}\},
\{\tilde{\boldsymbol{x}}_{\mathcal{K}_{i},t}\})]$ in the large-system limit. 
\end{assumption}

The energy penalty corresponds to free energy in the low-temperature 
limit or equivalently ground state energy in statistical 
physics~\cite{Mezard87,Nishimori01} (See Appendix~\ref{proposition1_proof}). 
Normalized ground state energy is believed to be self-averaging for many 
disordered systems. In fact, the self-averaging property of ground state 
energy was proved for a generalized spin glass 
model~\cite{Pastur91,Guerra02,Guerra032} and for MIMO systems~\cite{Korada10}. 
Since the proof of Assumption~\ref{self-averaging} is beyond the scope of 
this paper, we postulate the self-averaging property of the energy penalty 
in the large-system limit. 

The following lemma provides a genie-aided upper bound on the achievable sum 
rate~(\ref{sum_rate}) in the large-system limit. 

\begin{lemma} \label{lemma3} 
Suppose that Assumption~\ref{self-averaging} holds. Then, the achievable sum 
rate~(\ref{sum_rate}) per transmit antenna $R/N$ is bounded from above by 
\begin{equation} \label{sum_rate_bound} 
\overline{C} 
= \frac{\alpha}{T}\left\{
 H(\kappa) + \kappa I(\mathcal{X}_{k};\mathcal{Y}_{k}|s_{k}=1)
\right\}
\end{equation}
in the large-system limit, with $\mathcal{X}_{k}=\{x_{k,t}:t=0,\ldots,T-1\}$ 
and $\mathcal{Y}_{k}=\{y_{k,t}:t=0,\ldots,T-1\}$ denoting the data symbols 
and the received signals in the first block, respectively. 
In (\ref{sum_rate_bound}), $H(\kappa)$ denotes the binary entropy function 
\begin{equation} \label{entropy_function} 
H(\kappa) = -\kappa\log\kappa - (1-\kappa)\log(1-\kappa). 
\end{equation}
\end{lemma}
\begin{IEEEproof}[Proof of Lemma~\ref{lemma3}]
The received signal $y_{k,t}$ given by (\ref{equivalent_channel}) depends on 
all data symbols for user~$k$ through the energy 
penalty~(\ref{time_averaged_energy_penalty}). Under 
Assumption~\ref{self-averaging}, this dependencies disappear in the 
large-system limit: The equivalent channel~(\ref{equivalent_channel}) 
reduces to 
\begin{equation} \label{asymptotic_channel} 
y_{k,t} 
= \sqrt{\frac{P}{q}}\left\{
 s_{k,i}\tilde{x}_{k,t} + (1-s_{k,i})I_{k,t}
\right\} + n_{k,t}  
\end{equation}
in the large-system limit, where $q$ is equal to $q_{0}$ or $q_{1}$ in 
Propositions~\ref{proposition1} and~\ref{proposition2}. 
Since the US-VP~(\ref{US-VP}) is performed block 
by block, the received signals~(\ref{asymptotic_channel}) are i.i.d.\ block 
by block. As a result, the achievable rate~(\ref{user_rate}) reduces to 
\begin{equation} \label{user_rate1} 
R_{k} = \frac{1}{T}I(\mathcal{X}_{k};\mathcal{Y}_{k}) 
\end{equation}
in the large-system limit. 

We next consider a genie-aided upper bound on (\ref{user_rate1}), in which 
a genie informs each user about whether he/she has been selected in the 
first block, 
\begin{equation} \label{genie-aided_bound} 
R_{k} < \frac{1}{T}I(\mathcal{X}_{k};\mathcal{Y}_{k},s_{k}), 
\end{equation}
where $s_{k}$ is the indicator variable~(\ref{indicator}) to represent 
whether user~$k$ has been selected. In (\ref{genie-aided_bound}), we have 
dropped the subscript~$i$ from $s_{k,i}$. The upper 
bound~(\ref{genie-aided_bound}) is formally obtained from the chain rule for 
mutual information~\cite{Cover06}, 
\begin{IEEEeqnarray}{rl}
I(\mathcal{X}_{k};\mathcal{Y}_{k},s_{k})
=& I(\mathcal{X}_{k};\mathcal{Y}_{k}) 
+ I(\mathcal{X}_{k};s_{k}|\mathcal{Y}_{k}) \nonumber \\ 
>& I(\mathcal{X}_{k};\mathcal{Y}_{k}).  
\end{IEEEeqnarray}

Applying the chain rule for mutual information to the RHS of 
(\ref{genie-aided_bound}) yields 
\begin{equation} \label{bound1} 
I(\mathcal{X}_{k};\mathcal{Y}_{k},s_{k}) 
= I(\mathcal{X}_{k};s_{k}) 
+ I(\mathcal{X}_{k};\mathcal{Y}_{k}|s_{k}). 
\end{equation}
Since $s_{k}$ is a binary variable, the conditional entropy 
$H(s_{k}|\mathcal{X}_{k})$ is non-negative~\cite{Cover06}, so that 
the first term on the RHS of (\ref{bound1}) 
is bounded from above by the entropy of $s_{k}$, 
\begin{equation} \label{first_term} 
I(\mathcal{X}_{k};s_{k}) = H(s_{k}) - H(s_{k}|\mathcal{X}_{k}) < H(s_{k}). 
\end{equation}
Since the prior probability $\mathrm{Pr}(s_{k}=1)$ is equal to $\kappa$, 
the entropy $H(s_{k})$ is equal to the binary entropy function $H(\kappa)$. 
On the other hand, the second term should be equal to 
\begin{equation} \label{second} 
I(\mathcal{X}_{k};\mathcal{Y}_{k}|s_{k}) 
= \kappa I(\mathcal{X}_{k};\mathcal{Y}_{k}|s_{k}=1), 
\end{equation}
where $\kappa$ is the probability with which $s_{k}$ takes $1$. 
This can be understood as follows: 
The equivalent channel~(\ref{asymptotic_channel}) 
implies that user~$k$ receives the interference~(\ref{interference}) when 
$s_{k}=0$. In this case, the interference $I_{k,t}$ does not contain 
the desired data symbols $\{\tilde{x}_{k,t}\}$ for user~$k$. Strictly speaking, 
there may be dependencies between the received signals for $s_{k}=0$ and 
the desired data symbols, since the set of selected users 
depends on the desired data symbols for user~$k$. However, this dependencies 
should be negligible in the large-system limit, so that we 
obtain (\ref{second}). 
Combining (\ref{sum_rate}), (\ref{genie-aided_bound}), (\ref{bound1}), 
(\ref{first_term}) with $H(s_{k})=H(\kappa)$, and (\ref{second}), 
we arrive at the upper bound~(\ref{sum_rate_bound}). 
\end{IEEEproof}

In the derivation of the upper bound~(\ref{sum_rate_bound}), we have used 
the two upper bounds~(\ref{genie-aided_bound}) and (\ref{first_term}). 
The {\em looseness} of (\ref{sum_rate_bound}) due to the latter bound is 
negligible as $T\rightarrow\infty$, since $T^{-1}H(\kappa)$ tends to zero.   
On the other hand, the genie-aided bound~(\ref{genie-aided_bound}) also 
becomes tight as $T\rightarrow\infty$, since the detection of $s_{k}$ becomes 
easy as $T$ increases. See \cite{Takeuchi121} for an iterative algorithm to 
detect $s_{k}$. For example, the SNR loss required for detecting $s_{k}$ is 
at most $0.5$~dB for a sum rate per transmit antenna of $0.5$~bps/Hz when 
$T=16$ and QPSK are used. Furthermore, the SNR loss is at most $0.2$~dB 
for $1$~bps/Hz. These observations may imply that the upper 
bound~(\ref{sum_rate_bound}) is reasonably tight for a few dozen $T$. 

As shown in Appendix~\ref{calculation_mutual_inf_DD-US}, $s_{k}$ is 
independent of $\mathcal{X}_{k}$ for the DD-US with QPSK. 
As a result, the mutual information $I(\mathcal{X}_{k};\mathcal{Y}_{k}|s_{k})$ 
in (\ref{sum_rate_bound}) reduces to 
\begin{equation}
I(\mathcal{X}_{k};\mathcal{Y}_{k}|s_{k}=1) 
= \kappa TI(x_{k,t};y_{k,t}|s_{k}=1)  
\end{equation}
for the DD-US with QPSK. However, it is still hard to calculate the upper 
bound~(\ref{sum_rate_bound}) for general modulation, since $s_{k}$ depends on 
$\mathcal{X}_{k}$. From Lemma~\ref{lemma3}, we obtain an upper bound for the 
DD-US that is possible to calculate for large $T$. 

\begin{corollary}[Bound for DD-US]  \label{corollary2} 
Suppose that Assumption~\ref{self-averaging} holds. Then, the achievable sum 
rate~(\ref{sum_rate}) per transmit antenna $R/N$ for the DD-US is bounded 
from above by 
\begin{equation} \label{sum_rate_bound_DD-US} 
\overline{C}_{\mathrm{DD-US}} 
= \alpha\left\{
 \frac{H(\kappa)}{T} + \kappa I(x_{k,t}; y_{k,t} | s_{k}=1)
\right\}, 
\end{equation}
where $H(\kappa)$ denotes the binary entropy function~(\ref{entropy_function}). 
\end{corollary}
\begin{IEEEproof}[Proof of Corollary~\ref{corollary2}]
It is sufficient from Lemma~\ref{lemma3} to prove the following upper 
bound: 
\begin{equation} \label{second_term} 
I(\mathcal{X}_{k};\mathcal{Y}_{k}|s_{k}=1) 
\leq TI(x_{k,t}; y_{k,t} | s_{k}=1). 
\end{equation}
By definition, 
\begin{IEEEeqnarray}{rl}
I(\mathcal{X}_{k};\mathcal{Y}_{k}|s_{k}=1) 
= h(\mathcal{Y}_{k}|s_{k}=1) - h(\mathcal{Y}_{k}|\mathcal{X}_{k},s_{k}=1). 
\label{second_term1} 
\end{IEEEeqnarray}
Since $\tilde{x}_{k,t}=x_{k,t}$ holds for the DD-US, the second term 
$h(\mathcal{Y}_{k}|\mathcal{X}_{k},s_{k}=1)$ is equal to 
the differential entropy $Th(n_{k,t})$ of the noise $\{n_{k,t}\}$ in 
(\ref{asymptotic_channel})\footnote{
This statement does not hold for the US-CVP. Consequently, we need to 
derive a tight lower bound of the second term to obtain an upper bound 
on (\ref{sum_rate_bound}). Unfortunately, we are unable to find such a tight 
lower bound that can be calculated easily.}. On the other hand, 
the conditional differential entropy $h(\mathcal{Y}_{k}|s_{k}=1)$ is bounded 
from above by the sum of the conditional differential entropy for each 
received signal~\cite{Cover06}, 
\begin{equation} \label{entropy_bound1} 
h(\mathcal{Y}_{k}|s_{k}=1) \leq \sum_{t=0}^{T-1}h(y_{k,t}|s_{k}=1). 
\end{equation}
These observations imply that the RHS of (\ref{second_term1}) is equal 
to the upper bound~(\ref{second_term}). 
\end{IEEEproof}

We shall explain how to calculate the conditional mutual information 
$I(x_{k,t}; y_{k,t} | s_{k}=1)$, which is given by 
\begin{equation} \label{mutual_information_DD-US} 
I(x_{k,t}; y_{k,t} | s_{k}=1) 
= \mathbb{E}\left[
 \log\frac{p(y_{k,t}|x_{k,t},s_{k}=1)}{p(y_{k,t}|s_{k}=1)}
\right], 
\end{equation}
with 
\begin{equation}
p(y_{k,t}|s_{k}=1) 
= \mathbb{E}_{x_{k,t}}\left[
 \left. 
  p(y_{k,t}|x_{k,t},s_{k}=1)
 \right| s_{k}=1 
\right], \label{conditional_pdf_y}
\end{equation} 
where the conditional pdf $p(y_{k,t}|x_{k,t},s_{k}=1)$ is given by 
\begin{IEEEeqnarray}{rl}
&p(y_{k,t}|x_{k,t},s_{k}=1) \nonumber \\ 
=& \mathbb{E}_{\tilde{x}_{k,t}}\left[
 \left. 
  p(y_{k,t}|\tilde{x}_{k,t},s_{k}=1)
 \right| x_{k,t},s_{k}=1
\right]. \label{conditional_pdf_yx}
\end{IEEEeqnarray}
In (\ref{conditional_pdf_yx}), the pdf $p(y_{k,t}|\tilde{x}_{k,t},s_{k}=1)$ 
characterizes the equivalent channel~(\ref{asymptotic_channel}), 
\begin{equation} \label{channel_pdf} 
p(y_{k,t}|\tilde{x}_{k,t},s_{k}=1) 
= p_{\mathrm{CG}}\left(
 y_{k,t} - \sqrt{\frac{P}{q}}\tilde{x}_{k,t}; N_{0} 
\right), 
\end{equation}
with (\ref{complex_Gauss}), 
where $q$ is equal to $q_{0}$ or $q_{1}$ in Propositions~\ref{proposition1} 
and~\ref{proposition2}. 

In order to calculate the expectations in (\ref{conditional_pdf_y}) 
and (\ref{conditional_pdf_yx}), we need the joint posterior probability of 
$\tilde{x}_{k,t}$ and $x_{k,t}$ given $s_{k}=1$, given by  
\begin{IEEEeqnarray}{rl}  
&\mathrm{Pr}(\tilde{x}_{k,t}\in\tilde{\mathcal{A}},
x_{k,t}\in\mathcal{A}|s_{k}=1) \nonumber \\ 
=& \frac{\mathrm{Pr}(s_{k}=1,\tilde{x}_{k,t}\in\tilde{\mathcal{A}}|
x_{k,t})\mathrm{Pr}(x_{k,t}\in\mathcal{A})}
{\mathrm{Pr}(s_{k}=1)}, \label{posterior_pdf}
\end{IEEEeqnarray}
with measurable sets $\tilde{\mathcal{A}}\subset\mathbb{C}$ and 
$\mathcal{A}\in\mathbb{C}$. 
In (\ref{posterior_pdf}), $\mathrm{Pr}(x_{k,t}\in\mathcal{A})$ 
denotes the prior probability of $x_{k,t}$. Furthermore, the conditional 
probability $\mathrm{Pr}(s_{k}=1,\tilde{x}_{k,t}\in\tilde{\mathcal{A}}|
x_{k,t})$ is characterized via Proposition~\ref{proposition3} 
or Proposition~\ref{proposition4}: 
\begin{IEEEeqnarray}{rl}  
&\mathrm{Pr}(s_{k}=1,\tilde{x}_{k,t}\in\tilde{\mathcal{A}}|x_{k,t}) 
\nonumber \\ 
=& \mathrm{Pr}\left(
 \left.
  E_{k}(q)\leq \xi_{\kappa,T}(q), 
  \tilde{x}_{k,t}^{(\mathrm{opt})}(q)\in\tilde{\mathcal{A}}  
 \right| x_{k,t}
\right), \label{selection_probability}
\end{IEEEeqnarray}
with $q=q_{0}$ and $q=q_{1}$ under the RS and 1RSB assumptions, 
respectively. In summary, it is possible to calculate the mutual 
information~(\ref{mutual_information_DD-US}) from 
(\ref{conditional_pdf_y})--(\ref{selection_probability}).  
See Appendix~\ref{calculation_mutual_inf_DD-US} for the details. 

\section{Numerical Results} \label{sec4} 
\subsection{Energy Penalty}
US-VP is compared to ZFBF-RUS and CVP-RUS in terms of the average energy 
penalty. As noted in Section~\ref{sec3_2}, ZFBF-RUS and CVP-RUS provide 
lower bounds on the energy penalties of conventional US with ZFBF and 
{\em separate} US-CVP, respectively. The block length $T$ is kept finite, 
while the coherence time $T_{\mathrm{c}}$ is implicitly assumed to tend to 
infinity. 

\begin{figure}[t]
\begin{center}
\includegraphics[width=0.5\hsize]{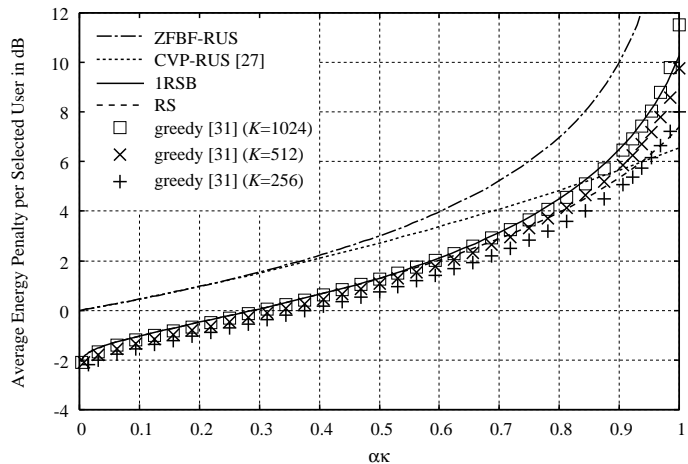}
\end{center}
\caption{
$\bar{\mathcal{E}}/\tilde{K}$ versus $\alpha\kappa=\tilde{K}/N$ 
for $\alpha=4$, $T=64$, and Gaussian signaling. 
}
\label{fig1} 
\end{figure}

We first focus on the DD-US with Gaussian signaling 
$x_{k,t}\sim\mathcal{CN}(0,1)$. Figure~\ref{fig1} shows the average energy 
penalty per selected users $\bar{\mathcal{E}}/\tilde{K}$ based on 
Propositions~\ref{proposition1} and~\ref{proposition2}. For comparison, 
the energy penalties of ZFBF-RUS and CVP-RUS are also shown on the basis of 
Corollary~\ref{corollary1}. The energy penalty of CVP-RUS was 
originally evaluated in \cite{Mueller08}. Furthermore, we plot the energy 
penalty of a greedy algorithm for the DD-US with Gaussian signaling proposed in 
\cite{Takeuchi121}. We obtain three observations: First, the RS solution is 
indistinguishable from the 1RSB solution for low-to-moderate 
$\alpha\kappa=\tilde{K}/N$, whereas there are a gap between the two solutions 
for large $\alpha\kappa$. Secondly, the energy penalty of the greedy algorithm 
is close to the RS and 1RSB solutions for low-to-moderate $\alpha\kappa$. 
This observation implies that the two solutions can provide acceptable 
approximations for the actual energy penalty in the same region. Finally, 
the DD-US outperforms ZFBF-RUS and CVP-RUS for low-to-moderate $\alpha\kappa$. 
Note that the energy penalty of CVP-RUS for finite-sized systems gets 
closer from {\em above} to the asymptotic one~\cite{Zaidel12}, whereas 
the energy penalty of the DD-US gets closer from {\em below} to the 
asymptotic one, as shown in Fig.~\ref{fig1}. This implies that the performance 
gap between the DD-US and CVP-RUS should be larger for finite-sized systems.   

\begin{figure}[t]
\begin{center}
\includegraphics[width=0.5\hsize]{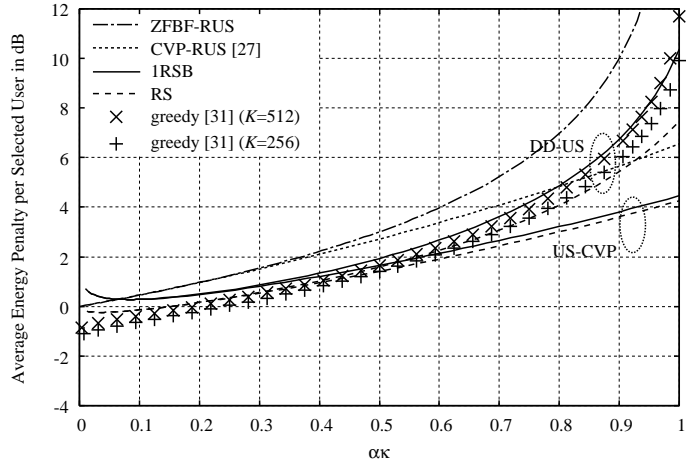}
\end{center}
\caption{
$\bar{\mathcal{E}}/\tilde{K}$ versus $\alpha\kappa=\tilde{K}/N$ 
for $\alpha=4$, $T=64$, and QPSK. 
}
\label{fig2} 
\end{figure}

We next focus on the average energy penalty of the DD-US with QPSK, shown 
in Fig.~\ref{fig2}. For comparison, the energy penalty of the US-CVP is also 
shown on the basis of Propositions~\ref{proposition1} and \ref{proposition2}. 
Furthermore, we plot the energy penalty of the greedy algorithm for the DD-US 
with QPSK~\cite{Takeuchi121}. Three observations are obtained: First, 
the RS and 1RSB solutions for the DD-US are indistinguishable from the 
respective solutions for the US-CVP in the low-to-moderate regime of 
$\alpha\kappa$. Secondly, the RS and 1RSB solutions for the US-CVP are close 
to each other for moderate-to-large $\alpha\kappa$, whereas there is a gap 
between the two solutions for small $\alpha\kappa$. Finally, the 1RSB 
solution provides an acceptable approximation for moderate-to-large 
$\alpha\kappa$, while the RS solution does for small $\alpha\kappa$. 

\begin{figure}[t]
\begin{center}
\includegraphics[width=0.5\hsize]{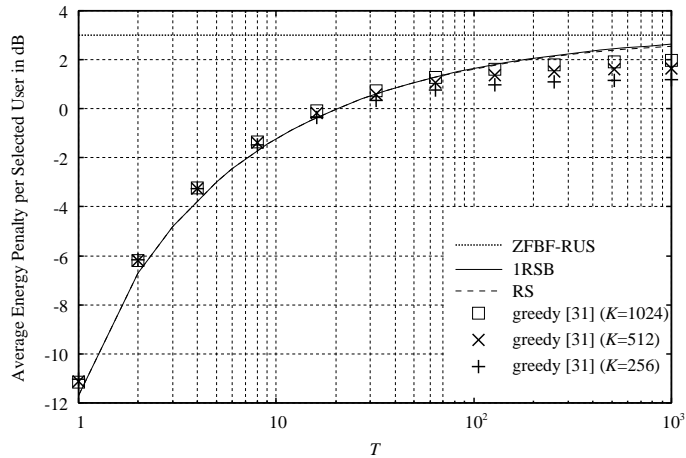}
\end{center}
\caption{
$\bar{\mathcal{E}}/\tilde{K}$ versus $T$ for $\alpha=4$, 
$\alpha\kappa=0.5$, and Gaussian signaling. 
} 
\label{fig3} 
\end{figure}

\begin{figure}[t]
\begin{center}
\includegraphics[width=0.5\hsize]{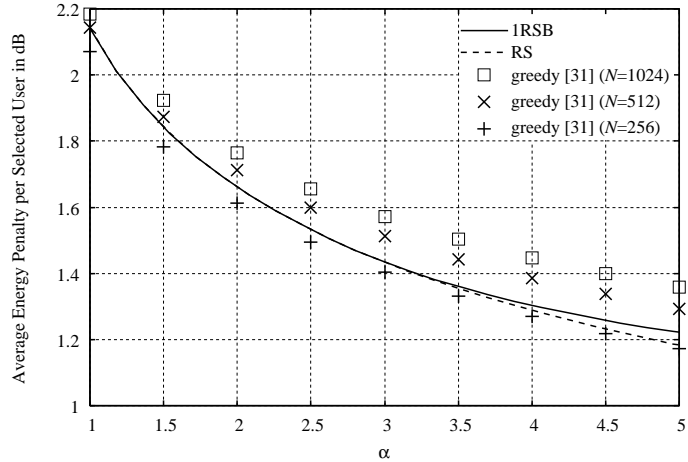}
\end{center}
\caption{
$\bar{\mathcal{E}}/\tilde{K}$ versus $\alpha$ for $\alpha\kappa=0.5$, 
$T=64$, and Gaussian signaling. 
} 
\label{fig4} 
\end{figure}

Finally, we investigate the impacts of $T$ and $\alpha$ on the average 
energy penalty. Figure~\ref{fig3} shows the energy penalty of the DD-US 
with Gaussian signaling versus $T$. For small $T$, the energy penalty of the 
DD-US increases quickly as $T$ grows. For moderate-to-large $T$, on the other 
hand, it increases slowly toward that for ZFBF-RUS. Figure~\ref{fig4} shows 
the energy penalty of the DD-US with Gaussian signaling versus $\alpha$. 
The RS solution is indistinguishable from the 1RSB solution, except for 
large $\alpha$. We find that the gap between the analytical predictions and 
the energy penalty of the greedy algorithm~\cite{Takeuchi121} becomes large 
as $\alpha$ increases. This may be due to the suboptimality of the greedy 
algorithm~\cite{Takeuchi121}.  

\subsection{Sum Rate} 
The DD-US is compared to ZFBF-RUS, CVP-RUS, and the DPC without power 
allocation in terms of the achievable sum rate. For the DD-US, we use the 
upper bound~(\ref{sum_rate_bound_DD-US}) on the achievable sum rate of the 
DD-US. The achievable sum rate of CVP-RUS was evaluated in \cite{Zaidel12}. 
The achievable sum rate of the DPC without power allocation is equal to the 
sum capacity of a dual MIMO uplink~\cite{Viswanath03} with no power 
allocation. The sum capacity of the dual MIMO uplink is possible to calculate 
in the large-system limit~\cite{Verdu99}. 

\begin{figure}[t]
\begin{center}
\includegraphics[width=0.5\hsize]{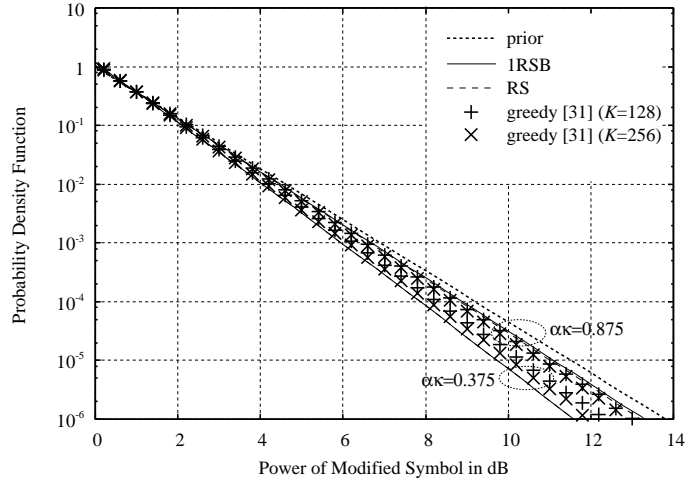}
\end{center}
\caption{
The pdf of $|\tilde{x}_{k,t}|^{2}$ given $s_{k}=1$ for $\alpha=4$, 
$T=64$, and Gaussian signaling. 
}
\label{fig5} 
\end{figure}

Before presenting the achievable rates, the distribution of the power of the 
modified symbol $\tilde{x}_{k,t}$ given $s_{k}=1$ is investigated for the 
DD-US with Gaussian signaling, which can be calculated 
via (\ref{posterior_pdf}). Figure~\ref{fig5} shows the pdf of 
$|\tilde{x}_{k,t}|^{2}$ given $s_{k}=1$. For comparison, we also plot the prior 
pdf of the original data symbol $x_{k,t}$. The DD-US selects the data symbols 
with smaller power to reduce the energy penalty. Consequently, the pdf of the 
power of the modified symbol $\tilde{x}_{k,t}$ has lighter tail than the 
prior pdf. This non-Gaussianity of the modified symbol results in a rate 
loss. 

\begin{figure}[t]
\begin{center}
\includegraphics[width=0.5\hsize]{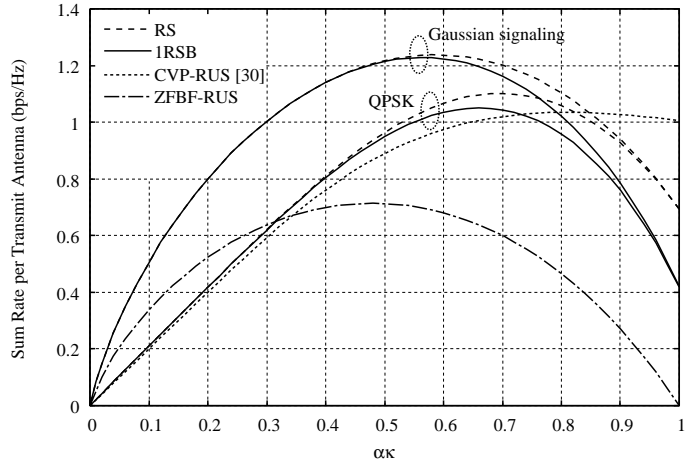}
\end{center}
\caption{
Upper bound~(\ref{sum_rate_bound_DD-US}) versus $\alpha\kappa$ 
for $\alpha=4$, $T=64$, and $P/N_{0}=5$~dB. 
}
\label{fig6} 
\end{figure}

Figure~\ref{fig6} shows the upper bound~(\ref{sum_rate_bound_DD-US}) on the 
achievable sum rate per transmit antenna of the DD-US. There is optimal 
$\alpha\kappa$ or equivalently the optimal number of 
selected users to maximize the sum rate for all schemes. This can be 
understood as follows: Increasing the number of selected users results in a 
degradation of the energy penalty {\em and} in an increase of spatial 
multiplexing gain. The latter effect is dominant for small $\alpha\kappa$, 
whereas the former is for large $\alpha\kappa$. Consequently, the sum rates 
are maximized at an optimal number of selected users.  

\begin{figure}[t]
\begin{center}
\includegraphics[width=0.5\hsize]{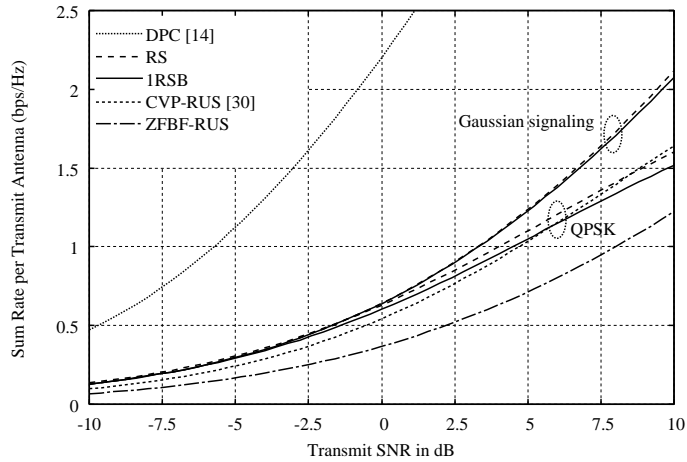}
\end{center}
\caption{
Optimized upper bound versus $P/N_{0}$ for $\alpha=4$ and $T=64$. 
}
\label{fig7} 
\end{figure}

Figure~\ref{fig7} shows the upper bound~(\ref{sum_rate_bound_DD-US}) with 
the optimal number of selected users. The RS and 1RSB solutions for the 
DD-US with Gaussian signaling are close to each other. Furthermore, the upper 
bounds for the DD-US with Gaussian signaling are larger than the achievable 
sum of CVP-RUS~\cite{Zaidel12} for all transmit SNRs, while the DD-US with 
QPSK is comparable to CVP-RUS. Unfortunately, the upper bounds for the DD-US 
with Gaussian signaling are far from the achievable sum rate of DPC. For sum 
rates per transmit antenna of $0.5$~bps/Hz and $1$~bps/Hz, the DD-US with 
Gaussian signaling provides performance gains of $1.2$~dB and $1.4$~dB, 
respectively, compared to CVP-RUS. Note that the SNR loss required for 
detecting whether each user has been selected is ignored for the upper 
bound~(\ref{sum_rate_bound_DD-US}). The upper bound becomes tight as $T$ grows. 
For example, the SNR loss for an iterative detection algorithm proposed in 
\cite{Takeuchi121} is $0.5$~dB for a sum rate per transmit antenna of 
$0.5$~bps/Hz when $T=16$ and QPSK are used. Furthermore, the SNR loss is 
$0.2$~dB for $1$~bps/Hz. These results may imply that the DD-US with 
Gaussian signaling outperforms CVP-RUS in terms of the {\em actual} 
achievable sum rate. 

\section{Conclusions} \label{sec5} 
Joint US-VP has been compared to separate US-VP in the large-system limit, 
where the numbers of transmit antennas, users, and selected users tend 
to infinity while their ratios are kept constant. The analyses under the RS 
and 1RSB assumptions have shown that conventional (data-independent) US 
{\em may} make no sense in the large-system limit: Under the RS and 1RSB 
assumptions, RUS achieves the same performance as optimal data-independent 
US in the large-system limit. Since conventional US is capacity-achieving 
as only the number of users tends to infinity, this implies that 
whether conventional US works well depends on how to take asymptotic limits.  
Joint US-VP can provide a substantial reduction of the energy penalty 
in the large-system limit. Consequently, joint US-VP outperforms separate 
US-VP in terms of the achievable sum rate. In particular, DD-US can be applied 
to general modulation, and implemented easily with a greedy algorithm.  

\appendices
\section{Calculation of (\ref{cdf})} \label{calculation_cdf} 
\subsection{Fourier Representation} 
The cdf~(\ref{cdf}) can be calculated via the characteristic 
function of (\ref{E_k}). Let $G_{T}(\omega)$ denote the characteristic 
function of (\ref{E_k}),  
\begin{equation} \label{characteristic_function}
G_{T}(\omega) 
= \mathbb{E}\left[
 \mathrm{e}^{\mathrm{i}\omega E_{k}(q)}
\right]. 
\end{equation}
Since (\ref{E_k}) is the sum of i.i.d.\ random variables, 
(\ref{characteristic_function}) is decomposed into 
\begin{equation} \label{characteristic_function1} 
G_{T}(\omega) 
= G\left(
 \frac{\omega}{2T} 
\right)^{2T}, 
\end{equation} 
with 
\begin{equation} \label{each_characteristic_function}
G(\omega) 
= \mathbb{E}\left[
 \mathrm{e}^{
  \mathrm{i}\omega\min_{\Re[\tilde{x}_{k,t}]\in\tilde{\mathcal{M}}_{\Re[x_{k,t}]}}
  (\sqrt{2}\Re[\tilde{x}_{k,t}] - \sqrt{2q}\Re[z_{k,t}])^{2}
 }
\right]. 
\end{equation}
In (\ref{each_characteristic_function}), 
$\tilde{\mathcal{M}}_{x}$ is given by $\{x\}$ for the DD-US and 
(\ref{each_CVP}) for the US-CVP, respectively. 

It is well-known that the pdf of (\ref{E_k}) is given by 
the inverse Fourier transform 
\begin{equation} \label{E_k_pdf} 
p(E_{k}(q)=E) 
= \frac{1}{2\pi}\int_{-\infty}^{\infty}
G_{T}(\omega)\mathrm{e}^{-\mathrm{i}\omega E}d\omega. 
\end{equation} 
Integrating the pdf~(\ref{E_k_pdf}) from $0$ to $x$, we obtain 
\begin{equation} \label{cdf1} 
F_{T}(x;q) 
=\int_{-\infty}^{\infty}
\frac{1-\mathrm{e}^{-\mathrm{i}\omega x}}{2\pi\mathrm{i}\omega} 
G_{T}(\omega)d\omega. 
\end{equation}
Since 
\begin{equation}
\int_{-\infty}^{\infty}
\frac{\mathrm{e}^{\mathrm{i}\omega E_{k}(q)}}{2\pi\mathrm{i}\omega}d\omega 
= \int_{-\infty}^{\infty}\frac{\sin(\omega E_{k}(q))}{2\pi\omega}d\omega
= \frac{1}{2}, 
\end{equation}
(\ref{cdf1}) reduces to 
\begin{equation} \label{cdf2} 
F_{T}(x;q) 
=\frac{1}{2} - \frac{1}{2\pi\mathrm{i}}\int_{-\infty}^{\infty}
\frac{1}{\omega}G\left(
 \frac{\omega}{2T} 
\right)^{2T}\mathrm{e}^{-\mathrm{i}\omega x}d\omega, 
\end{equation}
where we have used (\ref{characteristic_function1}). 
It is possible to calculate (\ref{cdf2}) numerically when the characteristic 
function~(\ref{each_characteristic_function}) is given.  

\subsection{DD-US} 
Let us calculate the characteristic 
function~(\ref{each_characteristic_function}) for the DD-US. 
Since $\sqrt{2q}\Re[z_{k,t}]\sim\mathcal{N}(0,q)$ and 
$\tilde{\mathcal{M}}_{x}=\{x\}$ for the DD-US, we obtain 
\begin{equation} \label{each_characteristic_function1} 
G(\omega) 
= \mathbb{E}\left[
 \frac{1}{\sqrt{1-2\mathrm{i}q\omega}}
 \exp\left(
  \frac{2\mathrm{i}\omega\Re[x_{k,t}]^{2}}{1-2\mathrm{i}q\omega}
 \right)
\right]. 
\end{equation}
For QPSK $\Re[x_{k,t}]=\pm1/\sqrt{2}$, 
\begin{equation}
G(\omega) 
= \frac{1}{\sqrt{1-2\mathrm{i}q\omega}}
\exp\left(
 \frac{\mathrm{i}\omega}{1-2\mathrm{i}q\omega}
\right). 
\end{equation}

For Gaussian signaling $\Re[x_{k,t}]\sim\mathcal{N}(0,1/2)$, 
(\ref{each_characteristic_function1}) reduces to 
\begin{equation} \label{each_characteristic_function_Gauss} 
G(\omega) 
= \frac{1}{\sqrt{1 - 2\mathrm{i}(1+q)\omega}}, 
\end{equation}
which is associated with the characteristic function for the chi-square 
distribution with one degree of freedom. In this case, the cdf~(\ref{cdf}) 
is associated with that for the chi-square distribution with $2T$ degrees 
of freedom: 
\begin{equation} \label{cdf_gauss} 
F_{T}(x;q) 
= \gamma\left(
 T,\frac{Tx}{1+q}
\right), 
\end{equation}
where $\gamma(a,x)$ is the incomplete gamma function 
\begin{equation}
\gamma(a,x) 
= \frac{1}{\Gamma(a)}\int_{0}^{x}y^{a-1}\mathrm{e}^{-y}dy,  
\end{equation}
with $\Gamma(x)$ denoting the gamma function. 

\subsection{US-CVP} 
Let us calculate the characteristic 
function~(\ref{each_characteristic_function}) for the US-CVP. 
From (\ref{each_CVP}), we obtain 
\begin{equation}
G(\omega) 
= \mathbb{E}\left[
 \exp\left(
  \mathrm{i}\omega\min_{\tilde{x}\in[1,\infty)}(\tilde{x}-z)^{2}
 \right)
\right], 
\end{equation}
where the expectations are taken with respect to $z\sim\mathcal{N}(0,q)$. 
Calculating the expectation yields 
\begin{equation} \label{each_characteristic_function_US-CVP} 
G(\omega) 
= \int_{-\infty}^{1/\sqrt{q}}\mathrm{e}^{\mathrm{i}\omega(1-\sqrt{q}u)^{2}}Du
+ Q\left(
 \frac{1}{\sqrt{q}}
\right). 
\end{equation}
In (\ref{each_characteristic_function_US-CVP}), $Du$ denotes the 
standard Gaussian measure~(\ref{Gaussian_measure}). Furthermore, 
$Q(x)$ is given by (\ref{Q_function}).

\section{Sum Rate for DD-US} 
\label{calculation_mutual_inf_DD-US} 
\subsection{Calculation of (\ref{selection_probability})} 
\label{calculation_distribution} 
The conditional probability~(\ref{selection_probability}) for the DD-US 
can be calculated in the same manner as in Appendix~\ref{calculation_cdf}. 
Let $G_{T}(\omega,\{\omega_{t}\};\mathcal{X}_{k})$ denote the conditional 
characteristic function of (\ref{E_k}),  
\begin{equation}
G_{T}(\omega,\{\omega_{t}\};\mathcal{X}_{k}) 
= \mathbb{E}\left[
 \left. 
  \mathrm{e}^{\mathrm{i}\omega E_{k}(q)
   + \mathrm{i}\sum_{t=0}^{T-1}
   \Re[\omega_{t}^{*}\tilde{x}_{k,t}^{(\mathrm{opt})}(q)] 
  }
 \right| \mathcal{X}_{k}
\right], \label{conditional_characteristic_function}
\end{equation} 
where $\tilde{x}_{k,t}^{(\mathrm{opt})}(q)$ is given by (\ref{modified_symbol}). 
From (\ref{E_k}), 
the characteristic function~(\ref{conditional_characteristic_function}) is 
decomposed into 
\begin{equation}
G_{T}(\omega,\{\omega_{t}\};\mathcal{X}_{k}) 
= \prod_{t=0}^{T-1}G\left(
 \frac{\omega}{2T},\omega_{t};x_{k,t}  
\right), 
\end{equation}
where $G(\omega,\omega_{t};x_{k,t})$ is given by 
\begin{IEEEeqnarray}{r}
G(\omega,\omega_{t};x_{k,t}) 
= \mathbb{E}\left[
 \exp\left\{
  2\mathrm{i}\omega
  \left|
   \tilde{x}_{k,t}^{(\mathrm{opt})}(q) - \sqrt{q}z_{k,t}
  \right|^{2}
 \right. 
\right. \nonumber \\ 
\left. 
 \left. 
  \left. 
   + \mathrm{i}\Re[\omega_{t}^{*}\tilde{x}_{k,t}^{(\mathrm{opt})}(q)]
  \right\}
 \right| x_{k,t} 
\right]. \label{each_conditional_characteristic_function}
\end{IEEEeqnarray}
Then, the conditional probability~(\ref{selection_probability}) is given by 
\begin{IEEEeqnarray}{rl} 
&\mathrm{Pr}(s_{k}=1,\tilde{x}_{k,t}\in\tilde{\mathcal{A}}|x_{k,t}) 
\nonumber \\ 
=& \mathbb{E}\left[
 \left. 
  \int_{\tilde{\mathcal{A}}\times\mathbb{C}^{T-1}} 
  p(s_{k}=1,\tilde{\mathcal{X}}_{k}|\mathcal{X}_{k})
  d\tilde{\mathcal{X}}_{k}
 \right| x_{k,t}
\right], \label{selection_probability0} 
\end{IEEEeqnarray}
with 
\begin{IEEEeqnarray}{rl} 
&p(s_{k}=1,\tilde{\mathcal{X}}_{k}|\mathcal{X}_{k}) \nonumber \\ 
=& \int_{-\infty}^{\infty}\frac{1-\mathrm{e}^{-\mathrm{i}\omega
\xi_{\kappa,T}(q)}}{2\pi\mathrm{i}\omega}\prod_{t=0}^{T-1}f\left(
 \frac{\omega}{2T},\tilde{x}_{k,t};x_{k,t}
\right)d\omega. \label{selection_probability1} 
\end{IEEEeqnarray}
In (\ref{selection_probability1}), $f(\omega,\tilde{x}_{k,t};x_{k,t})$ 
is defined as   
\begin{equation} \label{function_f} 
f(\omega,\tilde{x}_{k,t};x_{k,t})  
= \frac{1}{(2\pi)^{2}}\int_{\mathbb{C}}
G(\omega,\omega_{t};x_{k,t})
\mathrm{e}^{-\mathrm{i}\Re[\omega_{t}^{*}\tilde{x}_{k,t}]}d\omega_{t}, 
\end{equation}
with (\ref{each_conditional_characteristic_function}). 

Let us calculate (\ref{each_conditional_characteristic_function}) for 
the DD-US to calculate (\ref{selection_probability0}). In the same manner 
as in the derivation of (\ref{each_characteristic_function1}), we obtain 
\begin{equation} \label{each_conditional_characteristic_function1} 
G(\omega,\omega_{t};x_{k,t}) 
= G(\omega;x_{k,t}) 
\mathrm{e}^{\mathrm{i}\Re[\omega_{t}^{*}x_{k,t}]}, 
\end{equation}
with 
\begin{equation} \label{each_conditional_characteristic_function_DD-US}
G(\omega;x_{k,t}) 
= \frac{1}{1-2\mathrm{i}q\omega}
\exp\left(
 \frac{2\mathrm{i}\omega |x_{k,t}|^{2}}{1-2\mathrm{i}q\omega}
\right). 
\end{equation}
Substituting (\ref{each_conditional_characteristic_function1}) into 
(\ref{function_f}) yields 
\begin{equation}
f(\omega,\tilde{x}_{k,t};x_{k,t})  
= G(\omega;x_{k,t})\delta(\tilde{x}_{k,t}-x_{k,t}),  
\end{equation} 
which implies that (\ref{selection_probability1}) reduces to 
\begin{IEEEeqnarray}{l} 
p(s_{k}=1,\tilde{\mathcal{X}}_{k}|\mathcal{X}_{k})
= \prod_{t=0}^{T-1}\delta(\tilde{x}_{k,t}-x_{k,t}) \nonumber \\ 
\cdot\left[
 \frac{1}{2} - \frac{1}{2\pi\mathrm{i}}\int_{-\infty}^{\infty}
  \prod_{t=0}^{T-1}G\left(
   \frac{\omega}{2T};x_{k,t}  
  \right)
 \frac{\mathrm{e}^{-\mathrm{i}\omega \xi_{\kappa,T}(q)}}{\omega}d\omega
\right], \label{selection_probability2} 
\end{IEEEeqnarray}
with (\ref{each_conditional_characteristic_function_DD-US}).
The expressions~(\ref{each_conditional_characteristic_function_DD-US}) and 
(\ref{selection_probability2}) imply that $s_{k}$ is independent of 
$\mathcal{X}_{k}$ for the DD-US when QPSK $|x_{k,t}|^{2}=1$ is used.  
Substituting (\ref{selection_probability2}) into 
(\ref{selection_probability0}), we find that (\ref{selection_probability0}) 
for the DD-US is given by 
\begin{IEEEeqnarray}{rl} 
&\mathrm{Pr}(s_{k}=1,\tilde{x}_{k,t}\in\tilde{\mathcal{A}}|x_{k,t}) 
\nonumber \\ 
=& 1(x_{k,t}\in\tilde{\mathcal{A}})\left[
 \frac{1}{2} 
 - \frac{1}{2\pi\mathrm{i}}\int_{-\infty}^{\infty}
 G\left(
  \frac{\omega}{2T};x_{k,t}  
 \right)
\right. \nonumber \\ 
&\left. 
 \cdot G\left(
  \frac{\omega}{2T} 
 \right)^{2(T-1)}
 \frac{\mathrm{e}^{-\mathrm{i}\omega \xi_{\kappa,T}(q)}}{\omega}d\omega
\right], \label{selection_probability3} 
\end{IEEEeqnarray}
where $G(\omega)$ and $G(\omega;x_{k,t})$ are given by 
(\ref{each_characteristic_function1}) and 
(\ref{each_conditional_characteristic_function_DD-US}), respectively. 

\subsection{Calculation of (\ref{mutual_information_DD-US})} 
The conditional pdf~(\ref{conditional_pdf_yx}) for the DD-US reduces to 
\begin{equation} \label{conditional_pdf_yx1}
p(y_{k,t}|x_{k,t},s_{k}=1) = p(y_{k,t}|\tilde{x}_{k,t}=x_{k,t},s_{k}=1), 
\end{equation} 
with (\ref{channel_pdf}). 
We shall evaluate the conditional pdf~(\ref{conditional_pdf_y}) for 
Gaussian signaling $x_{k,t}\sim\mathcal{CN}(0,1)$. 
Substituting (\ref{posterior_pdf}) with (\ref{selection_probability3}) 
into (\ref{conditional_pdf_y}) and then calculating the integration 
with respect to $x_{k,t}$, we obtain 
\begin{IEEEeqnarray}{rl}  
&p(y_{k,t}|s_{k}=1) \nonumber \\ 
=& \frac{1}{\kappa}\left[
 \frac{1}{2}p_{\mathrm{CG}}\left(
  y_{k,t}; \frac{P}{q} + N_{0} 
 \right)
 - \frac{1}{2\pi\mathrm{i}}\int_{-\infty}^{\infty}
 G\left(
  \frac{\omega}{2T}
 \right)^{2T}
\right. \nonumber \\ 
& \left.
 \cdot p_{\mathrm{CG}}\left(
  y_{k,t}; \frac{P}{q}\sigma^{2}\left(\frac{\omega}{T}\right)+N_{0}
 \right)
 \frac{\mathrm{e}^{-\mathrm{i}\omega \xi_{\kappa,T}(q)}}{\omega}d\omega
\right], \label{conditional_pdf_y1}
\end{IEEEeqnarray}   
with 
\begin{equation}
\sigma^{2}(\omega) 
= \frac{1 - \mathrm{i}q\omega}{1 - \mathrm{i}(1+q)\omega}. 
\end{equation}
In (\ref{conditional_pdf_y1}), $p_{\mathrm{CG}}(z;\sigma^{2})$ and $G(\omega)$ 
are given by (\ref{complex_Gauss}) and 
(\ref{each_characteristic_function_Gauss}), respectively. 
It is possible to calculate the mutual 
information~(\ref{mutual_information_DD-US}) with 
(\ref{conditional_pdf_yx1}) and (\ref{conditional_pdf_y1}).  

\section{Derivation of Proposition~\ref{proposition1}} 
\label{proposition1_proof}  
\subsection{Statistical Physics} \label{statistical_physics}
Before deriving Proposition~\ref{proposition1}, we shall present a brief 
introduction on statistical physics. Statistical physics elucidates 
macroscopic properties of many-body systems that consist of many 
microscopic elements with interaction. Let $s_{i}$ denote a variable 
that represents the state of the $i$th microscopic element 
for $i=1,\ldots,N$. Suppose 
that the interactions between the microscopic elements are characterized 
by Hamiltonian $H(\boldsymbol{s})$, which is a real-valued function of 
the configuration $\boldsymbol{s}=(s_{i},\ldots,s_{N})^{\mathrm{T}}$. 
It is known that the distribution of $\boldsymbol{s}$ is given by the 
so-called Gibbs-Boltzmann distribution with a positive parameter $\beta>0$, 
\begin{equation} \label{GB_distribution} 
\mathrm{Pr}(\boldsymbol{s};\beta) = Z(\beta)^{-1}
\mathrm{e}^{-\beta H(\boldsymbol{s})}, 
\end{equation}
with 
\begin{equation} \label{normalization} 
Z(\beta) = \sum_{\{\boldsymbol{s}\}}\mathrm{e}^{-\beta H(\boldsymbol{s})}. 
\end{equation}
The parameter $\beta$ is called ``inverse temperature.'' Let 
$\mathcal{S}_{\mathrm{g}}$ denote the set of ground states $\{\boldsymbol{s}\}$ 
to minimize the Hamiltonian $H(\boldsymbol{s})$. Only the ground states 
contribute to the Gibbs-Boltzmann distribution in the low-temperature 
limit $\beta\rightarrow\infty$: The Gibbs-Boltzmann 
distribution~(\ref{GB_distribution}) converges to 
\begin{equation} \label{ground_state_distribution} 
\mathrm{Pr}(\boldsymbol{s};\beta) 
\rightarrow \frac{1}{|\mathcal{S}_{\mathrm{g}}|}
1(\boldsymbol{s}\in\mathcal{S}_{\mathrm{g}}), 
\end{equation} 
in the low-temperature limit $\beta\rightarrow\infty$~\cite{Mezard09}.  

The normalization constant~(\ref{normalization}) is called ``partition 
function,'' and is utilized to calculate several macroscopic quantities.  
As an example, let us calculate the internal energy 
$\langle H(\boldsymbol{s}) \rangle_{\beta}$, with $\langle\cdots\rangle_{\beta}$ 
denoting the expectation with respect to the Gibbs-Boltzmann 
distribution~(\ref{GB_distribution}). We define the free energy as 
\begin{equation} \label{general_free_energy} 
f(\beta) = - \frac{1}{\beta}\ln Z(\beta), 
\end{equation}
with (\ref{normalization}). It is straightforward to find that the internal 
energy is given by 
\begin{equation}
\langle H(\boldsymbol{s}) \rangle_{\beta}  
= \frac{\partial}{\partial\beta}\left(
 \beta f(\beta)
\right). 
\end{equation}
This implies that calculating the internal energy reduces to calculating 
the free energy. 

Since the Gibbs-Boltzmann distribution~(\ref{GB_distribution}) converges 
to (\ref{ground_state_distribution}) in the low-temperature limit, 
the internal energy tends to the ground state energy, which is the minimum of 
the Hamiltonian $H(\boldsymbol{s})$, in the low-temperature limit. 
The ground state energy is possible to calculate 
from (\ref{general_free_energy}) directly: 
\begin{equation} \label{ground_state_energy} 
\langle H(\boldsymbol{s}) \rangle_{\infty} 
= \lim_{\beta\rightarrow\infty}f(\beta). 
\end{equation}
We will use the formula~(\ref{ground_state_energy}) to calculate the 
energy penalty. 

\subsection{Formulation} 
The average energy penalty $\bar{\mathcal{E}}=\mathbb{E}[
\mathcal{E}(\{\boldsymbol{H}_{\mathcal{K}_{i}}\},
\{\tilde{\boldsymbol{x}}_{\mathcal{K}_{i},t}\}) ]$ for US-VP~(\ref{US-VP}), 
given via (\ref{time_averaged_energy_penalty}), is 
equal to the average $\bar{\mathcal{E}}_{i}$ of the energy 
penalty~(\ref{energy_penalty_US-VP}) for any block~$i$. Without loss of 
generality, we focus on the first block $i=0$ and  
drop the subscripts $i$ from $\bar{\mathcal{E}}_{i}$, $\mathcal{K}_{i}$, and 
$s_{k,i}$.  

The asymptotic energy penalty for VP was analyzed with the R-transform for 
the empirical eigenvalue distribution of 
$(\boldsymbol{H}\boldsymbol{H}^{\mathrm{H}})^{-1}$~\cite{Mueller08,Zaidel12}. 
Unfortunately, it is difficult to apply this method to our 
case, since the empirical eigenvalue distribution of 
$(\boldsymbol{H}_{\mathcal{K}}\boldsymbol{H}_{\mathcal{K}}^{\mathrm{H}})^{-1}$ 
is hard to calculate. Instead, we use the following lemma to calculate 
the average energy penalty $\bar{\mathcal{E}}$ without using the R-transform. 

\begin{lemma} \label{lemma4} 
Let us define $\boldsymbol{S}=\mathrm{diag}\{s_{1},\ldots,s_{K}\}$ and 
$\tilde{\boldsymbol{x}}_{t}
=(\tilde{x}_{1,t},\ldots,\tilde{x}_{K,t})^{\mathrm{T}}
\in\prod_{k=1}^{K}\mathcal{M}_{x_{k,t}}$, with $s_{k}$ given by 
(\ref{indicator}). The energy penalty~(\ref{energy_penalty_US-VP}) 
for the first block $i=0$ is equal to  
\begin{equation} \label{energy_penalty_appen} 
\mathcal{E}_{\mathrm{min}} 
=\lim_{\lambda\rightarrow\infty}
\min_{\{s_{k}\}}\min_{\{\tilde{\boldsymbol{x}}_{t}\}}
\min_{\{\boldsymbol{u}_{t}\}}\frac{1}{T}
\mathcal{H}_{\lambda}
(\boldsymbol{S},\{\tilde{\boldsymbol{x}}_{t}\},\{\boldsymbol{u}_{t}\}), 
\end{equation}
where the minimizations with respect to 
$\{s_{k}\}$, $\{\tilde{\boldsymbol{x}}_{t}\}$, and $\{\boldsymbol{u}_{t}\}$ are 
over $\{0,1\}^{K}$, $\prod_{t=0}^{T-1}\prod_{k=1}^{K}\mathcal{M}_{x_{k,t}}$, and 
$\mathbb{C}^{NT}$, respectively. In (\ref{energy_penalty_appen}),  
$\mathcal{H}_{\lambda}
(\boldsymbol{S},\{\tilde{\boldsymbol{x}}_{t}\},\{\boldsymbol{u}_{t}\})$, 
is given by  
\begin{equation} \label{Hamiltonian} 
\mathcal{H}_{\lambda}
(\boldsymbol{S},\{\tilde{\boldsymbol{x}}_{t}\},\{\boldsymbol{u}_{t}\})
= \sum_{t=0}^{T-1}\|\boldsymbol{u}_{t}\|^{2}
+ \lambda g(\boldsymbol{S},\{\tilde{\boldsymbol{x}}_{t}\},
\{\boldsymbol{u}_{t}\}),  
\end{equation}
with 
\begin{equation} \label{constraint} 
g(\boldsymbol{S},\{\tilde{\boldsymbol{x}}_{t}\},\{\boldsymbol{u}_{t}\}) 
= \sum_{t=0}^{T-1}\|\boldsymbol{S}
(\boldsymbol{H}\boldsymbol{u}_{t}  - \tilde{\boldsymbol{x}}_{t})\|^{2} 
+ \left(
 \mathrm{Tr}\boldsymbol{S}-\tilde{K}
\right)^{2}. 
\end{equation}
\end{lemma}
\begin{IEEEproof}[Proof of Lemma~\ref{lemma4}]
Since the function~(\ref{constraint}) is non-negative, the 
function~(\ref{Hamiltonian}) is bounded in $\lambda\rightarrow\infty$ only when 
$g(\boldsymbol{S},\{\tilde{\boldsymbol{x}}_{t}\},\{\boldsymbol{u}_{t}\})=0$. 
This implies 
\begin{equation}
\sum_{k=1}^{K}s_{k}=\tilde{K},
\end{equation}
\begin{equation}
\boldsymbol{u}_{t} = 
\boldsymbol{H}_{\mathcal{K}}^{\mathrm{H}}\left(
 \boldsymbol{H}_{\mathcal{K}}\boldsymbol{H}_{\mathcal{K}}^{\mathrm{H}}
\right)^{-1}\tilde{\boldsymbol{x}}_{\mathcal{K},t}
= \boldsymbol{u}_{t}^{(\mathrm{ZF})}(\boldsymbol{H}_{\mathcal{K}},
\tilde{\boldsymbol{x}}_{\mathcal{K},t}),  
\end{equation}
where we have used $\mathcal{K}=\{k\in\mathcal{K}_{\mathrm{all}}:s_{k}=1\}$, 
obtained from (\ref{indicator}). Thus, (\ref{energy_penalty_appen}) 
reduces to 
\begin{equation}
\mathcal{E}_{\mathrm{min}}
= \min_{\mathcal{K}\subset\mathcal{K}_{\mathrm{all}}:|\mathcal{K}|=\tilde{K}}
\min_{\{\tilde{\boldsymbol{x}}_{\mathcal{K},t}\}}
\frac{1}{T}\sum_{t=0}^{T-1}\left\|
 \boldsymbol{u}_{t}^{(\mathrm{ZF})}(\boldsymbol{H}_{\mathcal{K}},
 \tilde{\boldsymbol{x}}_{\mathcal{K},t})
\right\|^{2}, 
\end{equation} 
which is equal to the energy penalty~(\ref{energy_penalty_US-VP}) 
with US-VP~(\ref{US-VP}). 
\end{IEEEproof}

We start with defining the free energy as
\begin{equation} \label{free_energy} 
f = -\frac{1}{\beta\tilde{K}T}\mathbb{E}\left[
 \ln Z(\beta,\lambda)
\right], 
\end{equation}
where the so-called partition function $Z(\beta,\lambda)$ is given by   
\begin{IEEEeqnarray}{l}  
Z(\beta,\lambda)   
= \sum_{\{s_{k}\in\{0,1\}\}}
\prod_{t=0}^{T-1}\left(
 \int_{\prod_{k=1}^{K}\mathcal{M}_{x_{k,t}}}
\right) \nonumber \\ 
\int_{\mathbb{C}^{NT}}\mathrm{e}^{
 -\beta\mathcal{H}_{\lambda}
 (\boldsymbol{S},\{\tilde{\boldsymbol{x}}_{t}\},\{\boldsymbol{u}_{t}\})
}\prod_{t=0}^{T-1}d\tilde{\boldsymbol{x}}_{t}
\prod_{t=0}^{T-1}d\boldsymbol{u}_{t}, \label{partition_function}
\end{IEEEeqnarray}
with (\ref{Hamiltonian}). Only the minimums of (\ref{Hamiltonian}) contribute 
to the free energy~(\ref{free_energy}) in $\beta\rightarrow\infty$, so that 
taking the limit $\beta\rightarrow\infty$ in 
(\ref{free_energy}) before $\lambda\rightarrow\infty$ yields 
\begin{equation}
\lim_{\lambda\rightarrow\infty}\lim_{\beta\rightarrow\infty}f 
= \frac{1}{\tilde{K}}\mathbb{E}\left[
 \mathcal{E}_{\mathrm{min}}
\right], 
\end{equation}
which is equal to the average energy penalty per selected user 
$\bar{\mathcal{E}}/\tilde{K}$ for US-VP~(\ref{US-VP}) from 
Lemma~\ref{lemma4}. 
Thus, calculating the average energy penalty is equivalent to 
evaluating the free energy~(\ref{free_energy}).  

We use the replica method to calculate the free energy~(\ref{free_energy}) 
in the large-system limit. The replica method is based on the identity 
\begin{equation} \label{free_energy1} 
f = -\lim_{u\rightarrow0}\frac{1}{\beta u\tilde{K}T}
\ln\mathbb{E}\left[
 Z(\beta,\lambda)^{u}
\right]. 
\end{equation}
Since the RHS is difficult to calculate for real $u>0$, 
we regard $u$ as a natural number to obtain a special expression 
for (\ref{free_energy1}) with (\ref{partition_function}),  
\begin{equation} \label{free_energy2} 
f = - \lim_{u\rightarrow0}\frac{1}{\beta u\tilde{K}T}
\ln Z_{u}(\beta,\lambda), 
\end{equation}
with 
\begin{IEEEeqnarray}{l}  
Z_{u}(\beta,\lambda)  
= \mathbb{E}\left[
 \prod_{a=0}^{u-1}\left\{\sum_{\{s_{k,a}\in\{0,1\}\}}
  \prod_{t=0}^{T-1}\left(
   \int_{\prod_{k=1}^{K}\mathcal{M}_{x_{k,t}}}
  \right)
 \right.
\right. \nonumber \\ 
\left. 
 \left.  
  \int_{\mathbb{C}^{NT}}
  \mathrm{e}^{
   -\beta\mathcal{H}_{\lambda}
   (\boldsymbol{S}_{a},\{\tilde{\boldsymbol{x}}_{t,a}\},\{\boldsymbol{u}_{t,a}\})
  }\prod_{t=0}^{T-1}\left(
   d\tilde{\boldsymbol{x}}_{t,a}d\boldsymbol{u}_{t,a} 
  \right)
 \right\}
\right]. \label{replica_partition_function}
\end{IEEEeqnarray}
In (\ref{replica_partition_function}), 
$\boldsymbol{u}_{t,a}\in\mathbb{C}^{N}$, 
$\tilde{\boldsymbol{x}}_{t,a}\in\prod_{k=1}^{K}\mathcal{M}_{x_{k,t}}$, 
and $s_{k,a}\in\{0,1\}$ denote replicas of the transmit vector 
$\boldsymbol{u}_{t}$, the modified data symbol vector 
$\tilde{\boldsymbol{x}}_{t}$, and the indicator variable $s_{k}$, 
respectively. Furthermore, the diagonal matrix $\boldsymbol{S}_{a}$ is given 
by $\boldsymbol{S}_{a}=\mathrm{diag}\{s_{1,a},\ldots,s_{K,a}\}$.  

\subsection{Average over Quenched Randomness} 
We first evaluate the expectation in (\ref{replica_partition_function}) with 
respect to the channel matrix $\boldsymbol{H}$. 
Using (\ref{Hamiltonian}) yields 
\begin{equation} \label{replica_partition_function1} 
Z_{u}(\beta,\lambda) 
= \int_{\mathbb{C}^{uNT}}\Xi_{\beta\lambda}^{(u)}(\{\boldsymbol{u}_{t,a}\})
\prod_{a=0}^{u-1}\prod_{t=0}^{T-1}\left\{
 \mathrm{e}^{
  -\beta\|\boldsymbol{u}_{t,a}\|^{2} 
 }d\boldsymbol{u}_{t,a}
\right\},  
\end{equation}
with 
\begin{IEEEeqnarray}{rl}  
\Xi_{\beta\lambda}^{(u)}(\{\boldsymbol{u}_{t,a}\}) 
=& \mathbb{E}\left[ 
 \prod_{a=0}^{u-1}\left\{
  \sum_{\{s_{k,a}\in\{0,1\}\}}
  \prod_{t=0}^{T-1}\left(
   \int_{\prod_{k=1}^{K}\mathcal{M}_{x_{k,t}}}
  \right)
 \right. 
\right. \nonumber \\ 
&\left.
 \left. 
  \mathrm{e}^{
   - \beta\lambda g
   (\boldsymbol{S}_{a},\{\tilde{\boldsymbol{x}}_{t,a}\},\{\boldsymbol{u}_{t,a}\})
  }\prod_{t=0}^{T-1}d\tilde{\boldsymbol{x}}_{t,a}
 \right\} 
\right], \label{Xi_function} 
\end{IEEEeqnarray}
where $g(\{s_{k,a}\},\{\tilde{\boldsymbol{x}}_{t,a}\},\{\boldsymbol{u}_{t,a}\})$ 
is given by (\ref{constraint}). Let us define a random vector 
$\boldsymbol{v}_{a}(k)\in\mathbb{C}^{T}$ as 
\begin{equation} \label{v} 
\boldsymbol{v}_{a}(k) = \sum_{n=1}^{N}(\boldsymbol{H})_{k,n}
\boldsymbol{u}_{a}(n),
\end{equation}
with $\boldsymbol{u}_{a}(n)=(u_{n,0,a},\ldots,
u_{n,T-1,a})^{\mathrm{T}}$, in which $u_{n,t,a}=(\boldsymbol{u}_{t,a})_{n}$ 
denotes the $n$th element of $\boldsymbol{u}_{t,a}$. 
Since we have assumed that $\boldsymbol{H}$ has independent circularly 
symmetric complex Gaussian elements with variance $1/N$,  
$\boldsymbol{v}(k)=(\boldsymbol{v}_{0}(k)^{\mathrm{T}},\ldots,
\boldsymbol{v}_{u-1}(k)^{\mathrm{T}})^{\mathrm{T}}$ conditioned on 
$\{\boldsymbol{u}_{t,a}\}$ is a circularly symmetric complex Gaussian random 
vector with the covariance matrix 
\begin{equation} \label{Q}
\boldsymbol{Q} = \frac{1}{N}\sum_{n=1}^{N}\boldsymbol{u}(n)
\boldsymbol{u}(n)^{\mathrm{H}}, 
\end{equation}
with $\boldsymbol{u}(n)=(\boldsymbol{u}_{0}(n)^{\mathrm{T}},\ldots,
\boldsymbol{u}_{u-1}(n)^{\mathrm{T}})^{\mathrm{T}}$. 
The function (\ref{constraint}) in (\ref{Xi_function}) depends on 
$\{\boldsymbol{u}_{t,a}\}$ only through the covariance matrix~(\ref{Q}), 
so that we can re-write (\ref{Xi_function}) as 
$\Xi_{\beta\lambda}^{(u)}(\boldsymbol{Q})$ to find that 
(\ref{replica_partition_function1}) reduces to 
\begin{IEEEeqnarray}{rl}  
Z_{u}(\beta,\lambda) 
=& \left(
 \frac{\pi}{\beta}
\right)^{uNT}\int_{\mathbb{C}^{uNT}}\Xi_{\beta\lambda}^{(u)}(\boldsymbol{Q})
\nonumber \\ 
&\cdot\prod_{n=1}^{N}\left\{
 \left(
  \frac{\beta}{\pi}
 \right)^{uT}\mathrm{e}^{-\beta
 \|\boldsymbol{u}(n)\|^{2}}
 d\boldsymbol{u}(n)
\right\}. \label{replica_partition_function2}
\end{IEEEeqnarray}
In (\ref{replica_partition_function2}), 
$\Xi_{\beta\lambda}^{(u)}(\boldsymbol{Q})$ is given by 
\begin{IEEEeqnarray}{l} 
\Xi_{\beta\lambda}^{(u)}(\boldsymbol{Q}) 
= \mathbb{E}\left[ 
 \prod_{a=0}^{u-1}\left\{
  \sum_{\{s_{k,a}\in\{0,1\}\}}
  \prod_{k=1}^{K}\left(
   \int_{\prod_{t=0}^{T-1}\mathcal{M}_{x_{k,t}}}
  \right) 
 \right. 
\right. \nonumber \\ 
\left. 
 \left. 
  \mathrm{e}^{
   - \beta\lambda\tilde{g}
   (\{s_{k,a}\},\{\tilde{\boldsymbol{x}}_{a}(k)\},\{\boldsymbol{v}_{a}(k)\})
  }\prod_{k=1}^{K}d\tilde{\boldsymbol{x}}_{a}(k)
 \right\} 
\right], \label{Xi_function1}
\end{IEEEeqnarray}
\begin{IEEEeqnarray}{rl}  
&\tilde{g}(\{s_{k,a}\},\{\tilde{\boldsymbol{x}}_{a}(k)\},
\{\boldsymbol{v}_{a}(k)\}) \nonumber \\ 
=& \sum_{k=1}^{K}s_{k,a}
\|\boldsymbol{v}_{a}(k) - \tilde{\boldsymbol{x}}_{a}(k)\|^{2} 
+ \left(
 \sum_{k=1}^{K}s_{k,a} - \tilde{K} 
\right)^{2}, \label{tilde_g}
\end{IEEEeqnarray}
with $\tilde{\boldsymbol{x}}_{a}(k)
=((\tilde{\boldsymbol{x}}_{0,a})_{k},\ldots,
(\tilde{\boldsymbol{x}}_{T-1,a})_{k})^{\mathrm{T}}$. 

\subsection{Average over Spin Variables}
We next calculate the integration in (\ref{replica_partition_function2}) 
with respect to $\{\boldsymbol{u}(n)\}$. 
The expression~(\ref{replica_partition_function2}) implies that  
$\{\boldsymbol{u}(n)\}$ can be regarded as independent circularly symmetric 
complex Gaussian random vectors with the covariance matrix 
$\beta^{-1}\boldsymbol{I}_{uT}$. Thus, the covariance matrix~(\ref{Q}) is 
regarded as a complex Wishart matrix~\cite{Tulino04} with $N$ degrees of 
freedom, so that the pdf of (\ref{Q}) is 
given by 
\begin{equation} \label{pdf_Q} 
p(\boldsymbol{Q}) 
= C_{u}\mathrm{e}^{-\beta N\mathrm{Tr}\boldsymbol{Q}}
\det\boldsymbol{Q}^{N-uT}, 
\end{equation}
with 
\begin{equation} \label{constant} 
C_{u} = \frac{(\beta N)^{uTN}}{\pi^{uT(uT-1)/2}\prod_{i=1}^{uT}(N-i)!}. 
\end{equation}
Replacing the integration in (\ref{replica_partition_function2}) with respect 
to $\{\boldsymbol{u}(n)\}$ by the average over $\boldsymbol{Q}$ after 
substituting (\ref{replica_partition_function2}) into the free 
energy~(\ref{free_energy2}), we obtain  
\begin{IEEEeqnarray}{rl}  
f =& -\lim_{u\rightarrow0}
\frac{1}{\beta u\tilde{K}T}\ln 
\int C_{u}\det\boldsymbol{Q}^{-uT} \nonumber \\ 
&\cdot\exp\left\{
 \beta N\left(
  \frac{1}{\beta N}\ln\Xi_{\beta\lambda}^{(u)}(\boldsymbol{Q})
  - I_{u}(\boldsymbol{Q})
 \right)
\right\}d\boldsymbol{Q} \nonumber \\ 
&- \frac{1}{\beta\alpha\kappa}\ln\left(
 \frac{\pi}{\beta}
\right),  
\end{IEEEeqnarray} 
with 
\begin{equation} \label{rate_function} 
I_{u}(\boldsymbol{Q}) 
= \mathrm{Tr}\boldsymbol{Q}
- \frac{1}{\beta}\ln\det\boldsymbol{Q}. 
\end{equation}
Assuming that the large-system limit and the limit $u\rightarrow0$ are 
commutative, we use the saddle-point method to arrive at 
\begin{equation} \label{free_energy3} 
\lim_{K\rightarrow\infty}f  
= \lim_{u\rightarrow0}\frac{1}{u\alpha\kappa T}
\Phi_{u}(\boldsymbol{Q}_{\mathrm{s}})
- \frac{1}{\beta\alpha\kappa}\ln(\pi\mathrm{e}), 
\end{equation}
with 
\begin{equation} \label{Phi} 
\Phi_{u}(\boldsymbol{Q}) 
= I_{u}(\boldsymbol{Q}) 
- \lim_{K\rightarrow\infty}\frac{\alpha}{\beta K}
\ln\Xi_{\beta\lambda}^{(u)}(\boldsymbol{Q}), 
\end{equation}
where we have used the asymptotic formula for (\ref{constant})  
\begin{equation}
\lim_{K\rightarrow\infty}\frac{1}{\beta u\tilde{K}T}\ln C_{u} 
= \frac{1}{\beta\alpha\kappa}\ln(\beta\mathrm{e}) + o(1) 
\end{equation}
in the large-system limit. In (\ref{free_energy3}), 
the limit $\lim_{K\rightarrow\infty}$ denotes the large-system limit. 
Furthermore, $\boldsymbol{Q}_{\mathrm{s}}$ is the solution 
to minimize (\ref{Phi}): 
\begin{equation} \label{solution} 
\boldsymbol{Q}_{\mathrm{s}} 
= \argmin_{\boldsymbol{Q}}\Phi_{u}(\boldsymbol{Q}). 
\end{equation}

\subsection{Replica Symmetry Solution}
Let us assume RS for the solution~(\ref{solution}). 
\begin{assumption}[Replica Symmetry]
\begin{equation} \label{RS} 
\boldsymbol{Q}_{\mathrm{s}} = 
(\chi\boldsymbol{I}_{u} + q_{0}\boldsymbol{1}_{u}\boldsymbol{1}_{u}^{\mathrm{T}})
\otimes\boldsymbol{I}_{T}. 
\end{equation}
\end{assumption}

We first calculate (\ref{rate_function}) to obtain 
\begin{IEEEeqnarray}{rl}
&\frac{1}{uT}I_{u}(\boldsymbol{Q}_{\mathrm{s}}) \nonumber \\ 
=& \frac{1}{u}\left[
 u(\chi+q_{0}) - \frac{1}{\beta}\ln(\chi + uq_{0}) - \frac{u-1}{\beta}\ln\chi 
\right] \label{rate_function_rs_marginal} \\ 
\rightarrow& \chi + q_{0} - \frac{q_{0}}{\beta \chi} 
- \frac{1}{\beta}\ln\chi, \label{rate_function_rs} 
\end{IEEEeqnarray} 
in $u\rightarrow0$. 

We next evaluate (\ref{Xi_function1}). 
The RS assumption~(\ref{RS}) implies that $(\ref{v})$ is represented as 
\begin{equation} \label{v_RS} 
\boldsymbol{v}_{a}(k) 
= \sqrt{\chi}\boldsymbol{w}_{a}(k) 
+ \sqrt{q_{0}}\boldsymbol{z}(k), 
\end{equation}
where $\{\boldsymbol{w}_{a}(k)\in\mathbb{C}^{T}\}$ and  
$\{\boldsymbol{z}(k)\in\mathbb{C}^{T}\}$ are 
independent circularly symmetric complex Gaussian random vectors with 
covariance $\boldsymbol{I}_{T}$, respectively. We calculate the 
expectation with respect to $\{\boldsymbol{w}_{a}(k)\}$ to obtain  
\begin{equation}  
\Xi_{\beta\lambda}^{(u)}(\boldsymbol{Q}_{\mathrm{s}}) 
= \mathbb{E}\left[
 \left(
  \Xi_{\beta\lambda}^{(\mathrm{RS})}(\{\boldsymbol{z}(k)\})
 \right)^{u}
\right], 
\end{equation}
with
\begin{IEEEeqnarray}{r}  
\Xi_{\beta\lambda}^{(\mathrm{RS})}(\{\boldsymbol{z}(k)\})
= \sum_{\{s_{k}\in\{0,1\}\}}
\prod_{k=1}^{K}\left(
 \int_{\prod_{t=0}^{T-1}\mathcal{M}_{x_{k,t}}}
\right) \nonumber \\ 
\frac{\mathrm{e}^{
 - H_{\beta\lambda}^{(\mathrm{RS})}(\{s_{k}\},\{\tilde{\boldsymbol{x}}(k)\},
 \{\boldsymbol{z}(k)\})    
}}{(1+\beta\lambda\chi)^{T\sum_{k=1}^{K}s_{k}}}
\prod_{k=1}^{K}d\tilde{\boldsymbol{x}}(k), \label{Xi_function_RS}
\end{IEEEeqnarray}
where $H_{\beta\lambda}^{(\mathrm{RS})}
(\{s_{k}\},\{\tilde{\boldsymbol{x}}(k)\},\{\boldsymbol{z}(k)\})$ is 
given by   
\begin{IEEEeqnarray}{l} 
H_{\beta\lambda}^{(\mathrm{RS})}
(\{s_{k}\},\{\tilde{\boldsymbol{x}}(k)\},\{\boldsymbol{z}(k)\}) 
= \beta\lambda\left(
 \sum_{k=1}^{K}s_{k} - \tilde{K}
\right)^{2}\nonumber \\  
+ \sum_{k=1}^{K}\frac{\beta\lambda s_{k}}{1+\beta\lambda s_{k}\chi}
\|\tilde{\boldsymbol{x}}(k) - \sqrt{q_{0}}\boldsymbol{z}(k)\|^{2}.  
\label{H_rs}
\end{IEEEeqnarray}
Taking $u\rightarrow0$ yields 
\begin{IEEEeqnarray}{rl} 
&\lim_{u\rightarrow0}\lim_{K\rightarrow\infty}
\frac{1}{\beta uKT}\ln\Xi_{\beta\lambda}^{(u)}(\boldsymbol{Q}_{\mathrm{s}}) 
\nonumber \\ 
=& \lim_{K\rightarrow\infty}\frac{1}{\beta KT}
\mathbb{E}\left[
 \ln\Xi_{\beta\lambda}^{(\mathrm{RS})}(\{\boldsymbol{z}(k)\})
\right], 
\end{IEEEeqnarray}
with (\ref{Xi_function_RS}). 
Since (\ref{H_rs}) should be $O(\beta)$ in $\beta\rightarrow\infty$, 
$\chi$ must be $O(\beta^{-1})$ in $\beta\rightarrow\infty$. 
Taking $\beta\rightarrow\infty$ with $\hat{\chi}=\beta\chi$ fixed before 
$\lambda\rightarrow\infty$ yields 
\begin{IEEEeqnarray}{rl} 
&\lim_{\lambda\rightarrow\infty}\lim_{\beta\rightarrow\infty}\lim_{u\rightarrow0}
\lim_{K\rightarrow\infty}
\frac{1}{\beta uKT}\ln\Xi_{\beta\lambda}^{(u)}(\boldsymbol{Q}_{\mathrm{s}})  
\nonumber \\ 
=& -\frac{1}{\hat{\chi}}\lim_{K\rightarrow\infty}
\mathbb{E}[E_{\mathrm{RS}}(q_{0})], \label{Xi_function_rs}
\end{IEEEeqnarray}
where $E_{\mathrm{RS}}(q_{0})$ is given by 
\begin{equation} \label{cost_RS}
E_{\mathrm{RS}}(q_{0}) = \frac{1}{K}
\min_{\{s_{k}\in\{0,1\}\}:\sum_{k=1}^{K}s_{k}=\tilde{K}}
\sum_{k=1}^{K}s_{k}E_{k}^{(\mathrm{RS})}(q_{0}), 
\end{equation}
with 
\begin{equation} \label{E_k_RS} 
E_{k}^{(\mathrm{RS})}(q_{0}) = 
\frac{1}{T}\sum_{t=0}^{T-1}\min_{\tilde{x}_{k,t}\in\mathcal{M}_{x_{k,t}}}\left| 
 \tilde{x}_{k,t} - \sqrt{q_{0}}(\boldsymbol{z}(k))_{t}
\right|^{2}. 
\end{equation}

In order to evaluate the expectation of (\ref{cost_RS}), 
we write the order statistics of $\{E_{k}^{(\mathrm{RS})}(q_{0})\}$ as 
$\{E_{(k)}^{(\mathrm{RS})}(q_{0})\}$, i.e.\ 
$E_{(1)}^{(\mathrm{RS})}(q_{0})\leq E_{(2)}^{(\mathrm{RS})}(q_{0}) \leq \cdots \leq 
E_{(K)}^{(\mathrm{RS})}(q_{0})$~\cite{David03}. 
Since (\ref{cost_RS}) can be represented as 
\begin{equation} \label{cost_RS_tmp1} 
E_{\mathrm{RS}}(q_{0}) = \frac{1}{K}
\sum_{k=1}^{\tilde{K}}E_{(k)}^{(\mathrm{RS})}(q_{0}), 
\end{equation}
Lemma~\ref{lemma2} implies 
\begin{equation} \label{Stigler_mean}
\lim_{K\rightarrow\infty}\mathbb{E}[E_{\mathrm{RS}}(q_{0})]
= \mu_{\kappa,T}(q_{0}), 
\end{equation}
with (\ref{mean}). 

Finally, we substitute (\ref{rate_function_rs}) and (\ref{Xi_function_rs}) 
with (\ref{Stigler_mean}) into the free energy~(\ref{free_energy3}) 
to arrive at 
\begin{equation} \label{free_energy_RS_appen}
\lim_{\lambda\rightarrow\infty}\lim_{\beta\rightarrow\infty}
\lim_{K\rightarrow\infty}f 
= \frac{1}{\alpha\kappa}\left(
 q_{0} - \frac{q_{0}-\alpha\mu_{\kappa,T}(q_{0})}{\hat{\chi}} 
\right),  
\end{equation}
where $\lim_{\beta\rightarrow\infty}$ denotes the limit in which 
$\beta\rightarrow\infty$ and $\chi\rightarrow0$ with 
$\hat{\chi}=\beta\chi$ fixed. In (\ref{free_energy_RS_appen}),
$\hat{\chi}$ and $q_{0}$ are chosen so as to extremize the free 
energy~(\ref{free_energy_RS_appen}). 
The stationarity condition for $\hat{\chi}$ implies that $q_{0}$ is the 
solution to the fixed-point equation 
\begin{equation} \label{deriv_chi_RS} 
q_{0} = \alpha\mu_{\kappa,T}(q_{0}). 
\end{equation}
Substituting (\ref{deriv_chi_RS}) into the free 
energy~(\ref{free_energy_RS_appen}) yields $f=q_{0}/(\alpha\kappa)$. 

If the fixed-point equation~(\ref{deriv_chi_RS}) has multiple solutions, 
the solution $q_{0}$ to minimize (\ref{Phi}) or equivalently the free energy 
is selected. Since the free energy is given by $q_{0}/(\alpha\kappa)$, 
this criterion is equivalent to selecting the smallest solution to 
the fixed-point equation~(\ref{deriv_chi_RS}). 

\section{Derivation of Proposition~\ref{proposition2}} 
\label{proposition2_proof} 
We start with (\ref{free_energy3}). Let us assume 1RSB for the 
solution~(\ref{solution}).  
\begin{assumption}[1-step Replica Symmetry Breaking]
\begin{equation} \label{1RSB} 
\boldsymbol{Q}_{\mathrm{s}} = \left[
 \chi\boldsymbol{I}_{u} 
 + q_{0}\boldsymbol{1}_{u}\boldsymbol{1}_{u}^{\mathrm{T}}
 + q_{1}\boldsymbol{I}_{u/m_{1}}\otimes(
 \boldsymbol{1}_{m_{1}}\boldsymbol{1}_{m_{1}}^{\mathrm{T}})
\right]\otimes\boldsymbol{I}_{T},  
\end{equation}
for a positive integer $m_{1}$ satisfying $u/m_{1}\in\mathbb{N}$. 
\end{assumption}

We first calculate (\ref{rate_function}) to obtain 
\begin{IEEEeqnarray}{rl}
&\frac{1}{uT}I_{u}(\boldsymbol{Q}_{\mathrm{s}}) \nonumber \\ 
=& \frac{1}{\beta u}\left[
 \beta u(\chi + q_{0} + q_{1}) 
 - \frac{u(m_{1}-1)}{m_{1}}\ln\chi 
\right. \nonumber \\ 
-&\left. 
  \left(
  \frac{u}{m_{1}} - 1
 \right)\ln(\chi+m_{1}q_{1})
 - \ln(\chi+uq_{0}+m_{1}q_{1}) 
\right] \label{rate_function_1RSB_marginal} \\ 
\rightarrow& \chi+q_{0}+q_{1} 
- \frac{q_{0}}{\beta(\chi+m_{1}q_{1})} 
- \frac{1}{\beta m_{1}}\ln\left(
 1 + \frac{m_{1}q_{1}}{\chi}
\right) \nonumber \\ 
&- \frac{1}{\beta}\ln\chi, \label{rate_function_1RSB} 
\end{IEEEeqnarray} 
in $u\rightarrow0$. 

We next evaluate (\ref{Xi_function1}). The 1RSB assumption~(\ref{1RSB}) 
implies that $(\ref{v})$ is represented as 
\begin{equation} \label{v_1RSB} 
\boldsymbol{v}_{a}(k) 
= \sqrt{\chi}\boldsymbol{w}_{a}(k) 
+ \sqrt{q_{0}}\boldsymbol{z}(k) 
+ \sqrt{q_{1}}\boldsymbol{z}_{\lfloor a/m_{1} \rfloor}(k), 
\end{equation}
where $\{\boldsymbol{w}_{a}(k)\in\mathbb{C}^{T}\}$, 
$\{\boldsymbol{z}(k)\in\mathbb{C}^{T}\}$, and 
$\{\boldsymbol{z}_{c}(k)\in\mathbb{C}^{T}\}$ are 
independent circularly symmetric complex Gaussian random vectors with 
covariance $\boldsymbol{I}_{T}$, respectively. We calculate the 
expectation with respect to $\{\boldsymbol{w}_{a}(k)\}$ to obtain  
\begin{IEEEeqnarray}{rl} 
&\Xi_{\beta\lambda}^{(u)}(\boldsymbol{Q}_{\mathrm{s}}) 
\nonumber \\ 
=& \mathbb{E}\left[
 \mathbb{E}_{\{\boldsymbol{z}_{0}(k)\}}\left\{
  \Xi_{\beta\lambda}^{(\mathrm{1RSB})}
  (\{\boldsymbol{z}(k)\},\{\boldsymbol{z}_{0}(k)\})^{m_{1}}
 \right\}^{u/m_{1}} 
\right],
\end{IEEEeqnarray}
with
\begin{IEEEeqnarray}{l} \label{Xi_function_1RSB_tmp} 
\Xi_{\beta\lambda}^{(\mathrm{1RSB})}
(\{\boldsymbol{z}(k)\},\{\boldsymbol{z}_{0}(k)\})
= \sum_{\{s_{k}\in\{0,1\}\}}
\prod_{k\in\mathcal{K}}\left(
 \int_{\prod_{t=0}^{T-1}\mathcal{M}_{x_{k,t}}}
\right) \nonumber \\ 
\frac{
 \mathrm{e}^{-H_{\beta\lambda}^{(\mathrm{1RSB})}(\{s_{k}\},
 \{\tilde{\boldsymbol{x}}(k)\},\{\boldsymbol{z}(k)\},
 \{\boldsymbol{z}_{0}(k)\})}
}{(1+\beta\lambda\chi)^{T\sum_{k=1}^{K}s_{k}}}
\prod_{k=1}^{K}d\tilde{\boldsymbol{x}}(k), 
\end{IEEEeqnarray}
where $H_{\beta\lambda}^{(\mathrm{1RSB})}(\{s_{k}\},\{\tilde{\boldsymbol{x}}(k)\},
\{\boldsymbol{z}(k)\},\{\boldsymbol{z}_{0}(k)\})$ is given by   
\begin{IEEEeqnarray}{rl} 
&H_{\beta\lambda}^{(\mathrm{1RSB})}(\{s_{k}\},\{\tilde{\boldsymbol{x}}(k)\},
\{\boldsymbol{z}(k)\},\{\boldsymbol{z}_{0}(k)\})  
\nonumber \\ 
=& \sum_{k=1}^{K}\frac{\beta\lambda s_{k}}{1+\beta\lambda s_{k}\chi}
\left\|
 \tilde{\boldsymbol{x}}(k) - \sqrt{q_{0}}\boldsymbol{z}(k) 
 - \sqrt{q_{1}}\boldsymbol{z}_{0}(k) 
\right\|^{2} \nonumber \\ 
&+ \beta\lambda\left(
 \sum_{k=1}^{K}s_{k} - \tilde{K}
\right)^{2}. \label{H_1RSB}
\end{IEEEeqnarray}
Taking $u\rightarrow0$ yields 
\begin{IEEEeqnarray}{r} 
\lim_{u\rightarrow0}\lim_{K\rightarrow\infty}
\frac{1}{\beta uKT}\ln\Xi_{\beta\lambda}^{(u)}(\boldsymbol{Q}_{\mathrm{s}}) 
= \lim_{K\rightarrow\infty}
\frac{1}{\beta m_{1}KT} \nonumber \\ 
\cdot\mathbb{E}\left[
 \ln\mathbb{E}_{\{\boldsymbol{z}_{0}(k)\}}\left\{
  \Xi_{\beta\lambda}^{(\mathrm{1RSB})}
  (\{\boldsymbol{z}(k)\},\{\boldsymbol{z}_{0}(k)\})^{m_{1}}
 \right\} 
\right], \label{Xi_function_1RSB}
\end{IEEEeqnarray}
with (\ref{Xi_function_1RSB_tmp}). 
The function~(\ref{Xi_function_1RSB}) converges in the limit  
$\beta\rightarrow\infty$, $m_{1}\rightarrow0$, and 
$\chi\rightarrow0$ with $\mu_{1}=\beta m_{1}$ 
and $\hat{\chi}=\beta\chi$ fixed. 
Taking this limit before $\lambda\rightarrow\infty$ yields 
\begin{IEEEeqnarray}{l}   
\lim_{\lambda\rightarrow\infty}\lim_{\beta\rightarrow\infty}
\lim_{u\rightarrow0}\lim_{K\rightarrow\infty}
\frac{1}{\beta uKT}\ln\Xi_{\beta\lambda}^{(u)}(\boldsymbol{Q}_{\mathrm{s}}) 
= \lim_{K\rightarrow\infty}\frac{1}{\mu_{1}KT} \nonumber \\ 
\cdot\mathbb{E}\left[
 \ln\mathbb{E}_{\{\boldsymbol{z}_{0}(k)\}}\left\{
  \exp\left[
   -\frac{\mu_{1}KT}{\hat{\chi}}E_{1\mathrm{RSB}}(q_{0},q_{1}) 
  \right]
 \right\}
\right], \label{Xi_function_1RSB1} 
\end{IEEEeqnarray}
where $E_{1\mathrm{RSB}}(q_{0},q_{1})$ is given by 
\begin{IEEEeqnarray}{rl} 
&E_{1\mathrm{RSB}}(q_{0},q_{1}) \nonumber \\ 
=& \frac{1}{K}
\min_{\{s_{k}\in\{0,1\}\}:\sum_{k=1}^{K}s_{k}=\tilde{K}}
\sum_{k=1}^{K}s_{k}E_{k}^{(\mathrm{1RSB})}(q_{0},q_{1}), \label{cost_1RSB} 
\end{IEEEeqnarray}
with 
\begin{equation} \label{E_k_appen} 
E_{k}^{(\mathrm{1RSB})}(q_{0},q_{1}) 
= \frac{1}{T}\sum_{t=0}^{T-1}
|\tilde{x}_{k,t} - \sqrt{q_{0}}(\boldsymbol{z}(k))_{t} 
- \sqrt{q_{1}}(\boldsymbol{z}_{0}(k))_{t}|^{2}. 
\end{equation}

In order to evaluate the distribution of (\ref{cost_1RSB}), 
we write the order statistics of $\{E_{k}^{(\mathrm{1RSB})}(q_{0},q_{1})\}$ as 
$\{E_{(k)}^{(\mathrm{1RSB})}(q_{0},q_{1})\}$, i.e.\ 
$E_{(1)}^{(\mathrm{1RSB})}(q_{0},q_{1})\leq E_{(2)}^{(\mathrm{1RSB})}(q_{0},q_{1}) 
\leq \cdots \leq E_{(K)}^{(\mathrm{1RSB})}(q_{0},q_{1})$~\cite{David03}. 
Expression~(\ref{cost_1RSB}) can be represented as 
\begin{equation}
E_{1\mathrm{RSB}}(q_{0},q_{1})  
= \frac{1}{K}\sum_{k=1}^{\tilde{K}}E_{(k)}^{(\mathrm{1RSB})}(q_{0},q_{1}).  
\end{equation}
Since $E_{1\mathrm{RSB}}(q_{0},q_{1})$ conditioned on 
$\{\tilde{\boldsymbol{x}}(k)\}$ and $\{\boldsymbol{z}(k)\}$ converges in law 
to a Gaussian random variable in the large-system 
limit~\cite[Theorem~6]{Stigler74}, (\ref{Xi_function_1RSB1}) reduces to 
\begin{IEEEeqnarray}{rl} 
&\lim_{\lambda\rightarrow\infty}\lim_{\beta\rightarrow\infty}
\lim_{u\rightarrow0}\lim_{K\rightarrow\infty}
\frac{1}{\beta uKT}\ln\Xi_{\beta\lambda}^{(u)}(\boldsymbol{Q}_{\mathrm{s}}) 
\nonumber \\ 
=& \lim_{K\rightarrow\infty}\left\{
 \frac{\mathbb{E}[E_{1\mathrm{RSB}}(q_{0},q_{1})]}{\hat{\chi}}
 - \frac{\mu_{1}TK\mathbb{V}[E_{1\mathrm{RSB}}(q_{0},q_{1})]}{2\hat{\chi}^{2}}
\right\}. \label{Xi_function_1RSB2}
\end{IEEEeqnarray}
Lemma~\ref{lemma2} implies  
\begin{equation} \label{mean_1RSB}
\lim_{K\rightarrow\infty}\mathbb{E}[E_{1\mathrm{RSB}}(q_{0},q_{1})]
= \mu_{\kappa,T}(q_{0}+q_{1}), 
\end{equation}
\begin{equation} \label{variance_1RSB} 
\lim_{K\rightarrow\infty}K\mathbb{V}[E_{1\mathrm{RSB}}(q_{0},q_{1})]
= \sigma_{\kappa,T}^{2}(q_{0}+q_{1}), 
\end{equation}
with (\ref{mean}) and (\ref{variance}). 
 
Finally, we substitute (\ref{rate_function_1RSB}) and 
(\ref{Xi_function_1RSB2}) with (\ref{mean_1RSB}) and 
(\ref{variance_1RSB}) into the free 
energy~(\ref{free_energy3}) to arrive at 
\begin{IEEEeqnarray}{rl} 
&\lim_{\lambda\rightarrow\infty}\lim_{\beta\rightarrow\infty}
\lim_{K\rightarrow\infty}f \nonumber \\ 
=& \frac{1}{\alpha\kappa}\left[
 q_{0} + q_{1}  
 - \frac{1}{\mu_{1}}\left\{
  \frac{\mu_{1}q_{0}}{\hat{\chi}+\mu_{1}q_{1}} 
  + \ln\left(
   \frac{\hat{\chi} + \mu_{1}q_{1}}{\hat{\chi}}
  \right)
 \right. 
\right. \nonumber \\ 
-&\left. 
 \left. 
  \alpha\left[
   \frac{\mu_{1}\mu_{\kappa,T}(q_{0}+q_{1})}{\hat{\chi}}
   - \frac{T\mu_{1}^{2}\sigma_{\kappa,T}^{2}(q_{0}+q_{1})}{2\hat{\chi}^{2}}
  \right]
 \right\}
\right], \label{free_energy_1RSB_appen}
\end{IEEEeqnarray}
where $\lim_{\beta\rightarrow\infty}$ denotes the limit 
$\beta\rightarrow\infty$, $m_{1}\rightarrow0$, and 
$\chi\rightarrow0$ with $\mu_{1}=\beta m_{1}$ 
and $\hat{\chi}=\beta\chi$ fixed. 
In (\ref{free_energy_1RSB_appen}), $\mu_{1}$, $\hat{\chi}$, $q_{0}$, and 
$q_{1}$ or equivalently $\mu_{1}$, $\bar{\chi}=\hat{\chi}/\mu_{1}$, $q_{0}$, 
and $q_{1}$ are chosen so as to extremize the free 
energy~(\ref{free_energy_1RSB_appen}). 
The stationarity conditions for $\mu_{1}$ and $\bar{\chi}$ are given by
\begin{IEEEeqnarray}{rl} 
&\frac{q_{0}}{\bar{\chi}+q_{1}} 
+ \ln\left(
 1 + \frac{q_{1}}{\bar{\chi}}
\right) \nonumber \\ 
=& \alpha\left[
 \frac{\mu_{\kappa,T}(q_{0}+q_{1})}{\bar{\chi}}
 - \frac{T\sigma_{\kappa,T}^{2}(q_{0}+q_{1})}{2\bar{\chi}^{2}}
\right], \label{deriv_mu}
\end{IEEEeqnarray} 
\begin{IEEEeqnarray}{rl} 
&- \frac{q_{0}}{(\bar{\chi}+q_{1})^{2}} 
-\frac{q_{1}}{\bar{\chi}(\bar{\chi}+q_{1})} \nonumber \\ 
=& \alpha\left[
 - \frac{1}{\bar{\chi}^{2}}\mu_{\kappa,T}(q_{0}+q_{1})
 + \frac{T}{\bar{\chi}^{3}}
 \sigma_{\kappa,T}^{2}(q_{0}+q_{1})
\right], \label{deriv_chi} 
\end{IEEEeqnarray}
respectively. From the stationarity conditions for $q_{0}$ and $q_{1}$, 
we obtain 
\begin{equation}
\frac{q_{0}}{(\bar{\chi}+q_{1})^{2}} = 0, 
\end{equation}
which implies $q_{0}\rightarrow0$, $\bar{\chi}\rightarrow\infty$, 
or $q_{1}\rightarrow\infty$. The free energy~(\ref{free_energy_1RSB_appen}) 
diverges in $q_{1}\rightarrow\infty$. Furthermore, the limit 
$\bar{\chi}\rightarrow\infty$ corresponds to the RS solution. 
Taking $q_{0}\rightarrow0$ yields 
\begin{equation}
\lim_{\lambda\rightarrow\infty}\lim_{\beta\rightarrow\infty}
\lim_{K\rightarrow\infty}f 
= \frac{q_{1}}{\alpha\kappa},   
\end{equation}
where $q_{1}$ satisfies the coupled fixed-point equations, 
\begin{equation} \label{fixed_point_1RSB_appen1}
\ln\left(
 1 + \frac{q_{1}}{\bar{\chi}}
\right)
= \alpha\left[
 \frac{1}{\bar{\chi}}\mu_{\kappa,T}(q_{1})
 - \frac{T}{2\bar{\chi}^{2}}
 \sigma_{\kappa,T}^{2}(q_{1})
\right], 
\end{equation}
\begin{equation} \label{fixed_point_1RSB_appen2}
\frac{q_{1}}{\bar{\chi}+q_{1}}  
= \alpha\left[
 \frac{1}{\bar{\chi}}\mu_{\kappa,T}(q_{1})
 - \frac{T}{\bar{\chi}^{2}}
 \sigma_{\kappa,T}^{2}(q_{1})
\right]. 
\end{equation}
If the coupled fixed-point equations~(\ref{fixed_point_1RSB_appen1}) 
and (\ref{fixed_point_1RSB_appen2}) have multiple solutions, the 
solution $q_{1}$ to minimize the free energy or equivalently the smallest 
solution $q_{1}$ to the coupled fixed-point equations is selected. 

\section{Proof of Corollary~\ref{corollary1}}
\label{proof_corollary1} 
We prove that the fixed-point equation~(\ref{fixed_point_RS}) reduces to 
(\ref{fixed_point_inf}) in $T\rightarrow\infty$. We first show that 
the $\kappa$-quantile (\ref{quantile}) converges to the expectation 
\begin{equation} \label{expectation} 
\mathbb{E}[E_{k}(q)] 
= \mathbb{E}\left[
 \min_{\tilde{x}_{k,t}\in\mathcal{M}_{x_{k,t}}}
 |\tilde{x}_{k,t} - \sqrt{q}z_{k,t}|^{2}
\right] 
\end{equation}
in $T\rightarrow\infty$. Let $\overline{\xi}$ denote a variable that 
satisfies 
\begin{equation} \label{upper_bound_quantile} 
1 = F_{T}\left(
 \overline{\xi};q
\right) 
\end{equation}
in $T\rightarrow\infty$. 
Since the cdf~(\ref{cdf}) is monotonically increasing, 
we find $\overline{\xi}\geq\xi_{\kappa,T}(q)$, with (\ref{quantile}). 
The weak law of large numbers implies that the random variable~(\ref{E_k}) 
converges in probability to (\ref{expectation}) in $T\rightarrow\infty$, 
so that the cdf~(\ref{cdf}) converges to  
\begin{equation} \label{inf_cdf}
\lim_{T\rightarrow\infty}F_{T}(x;q) = 1(x\geq \mathbb{E}[E_{k}(q)]) 
\end{equation}
in $T\rightarrow\infty$. Thus, (\ref{upper_bound_quantile}) reduces to 
\begin{equation} \label{inf_upper_bound_quantile} 
1 = 1(\overline{\xi}\geq \mathbb{E}[E_{k}(q)]) 
\end{equation}
in $T\rightarrow\infty$. The smallest variable $\overline{\xi}$ satisfying 
(\ref{inf_upper_bound_quantile}) is given by (\ref{expectation}).  
This implies that (\ref{quantile}) is bounded from above by 
(\ref{expectation}) in $T\rightarrow\infty$: 
\begin{equation} \label{limsup} 
\limsup_{T\rightarrow\infty}\xi_{\kappa,T}(q) 
\leq \mathbb{E}[E_{k}(q)]. 
\end{equation}
Similarly, considering a variable $\underline{\xi}$ that satisfies 
\begin{equation}
0 = F_{T}(\underline{\xi};q) 
\end{equation}
in $T\rightarrow\infty$, we obtain the lower bound on (\ref{quantile}) 
\begin{equation} \label{liminf} 
\liminf_{T\rightarrow\infty}\xi_{\kappa,T}(q) 
\geq \mathbb{E}[E_{k}(q)]. 
\end{equation}
Combining the two bounds (\ref{limsup}) and (\ref{liminf}) yields 
\begin{equation} \label{limit}
\lim_{T\rightarrow\infty}\xi_{\kappa,T}(q) 
= \mathbb{E}[E_{k}(q)], 
\end{equation} 
with (\ref{expectation}). 

We next calculate (\ref{mean}) in $T\rightarrow\infty$. 
Integrating (\ref{mean}) by parts after the transformation $y=F_{T}^{-1}(x;q)$, 
we obtain 
\begin{IEEEeqnarray}{rl} 
\mu_{\kappa,T}(q) 
=& \int_{0}^{\xi_{\kappa,T}(q)}yF_{T}'(y;q)dy \nonumber \\ 
=& \kappa\xi_{\kappa,T}(q) - \int_{0}^{\xi_{\kappa,T}(q)}F_{T}(x;q)dx, 
\label{inf_mean} 
\end{IEEEeqnarray}
with (\ref{quantile}). 
Applying (\ref{inf_cdf}) and (\ref{limit}) to (\ref{inf_mean}) yields 
\begin{equation}
\lim_{T\rightarrow\infty}\mu_{\kappa,T}(q)  
= \kappa\mathbb{E}[E_{k}(q)], 
\end{equation} 
with (\ref{expectation}). This implies that 
(\ref{fixed_point_RS}) reduces to (\ref{fixed_point_inf}) in 
$T\rightarrow\infty$.

\section{Derivation of Proposition~\ref{proposition3}}
\label{proposition3_proof} 
\subsection{Replica Method} 
As shown in Appendix~\ref{statistical_physics}, the Gibbs-Boltzmann 
distribution~(\ref{GB_distribution}) converges to 
(\ref{ground_state_distribution}) in the low-temperature limit 
$\beta\rightarrow\infty$. This implies that the marginal distribution 
$\mathrm{Pr}(s_{i};\beta)=\sum_{\backslash s_{i}}
\mathrm{Pr}(\boldsymbol{s};\beta)$ tends to 
\begin{equation}
\mathrm{Pr}(s_{i};\beta) 
\rightarrow \frac{1}{|\mathcal{S}_{\mathrm{g}}(i)|}
1(s_{i}\in\mathcal{S}_{\mathrm{g}}(i))  
\end{equation}
in the low-temperature limit, where $\mathcal{S}_{\mathrm{g}}(i)$ is the 
set of the $i$th element $s_{i}$ included in the ground states 
$\mathcal{S}_{\mathrm{g}}$. 
Thus, evaluating the conditional joint distribution 
$\mathrm{Pr}(s_{k}=1, \tilde{\mathcal{X}}_{k}\in\prod_{t=0}^{T-1}
\mathcal{A}_{t} | \mathcal{X}_{k})$  
reduces to calculating a marginal distribution of 
the Gibbs-Boltzmann distribution associated with the 
Hamiltonian~(\ref{Hamiltonian}).  

We start with the identity 
\begin{IEEEeqnarray}{rl}  
&\mathrm{Pr}\left(
 \left. 
  s_{k}=1, \tilde{\mathcal{X}}_{k}\in\prod_{t=0}^{T-1}
  \mathcal{A}_{t} 
 \right| \mathcal{X}_{k}
\right) \nonumber \\ 
=& \int_{\mathcal{A}_{0}\times\cdots\times\mathcal{A}_{T-1}} 
p(s_{k}=1,\tilde{\mathcal{X}}_{k} | \mathcal{X}_{k})
d\tilde{\mathcal{X}}_{k}, \label{target_distribution} 
\end{IEEEeqnarray} 
with 
\begin{IEEEeqnarray}{l} 
p(s_{k},\tilde{\mathcal{X}}_{k} | \mathcal{X}_{k}) 
= \lim_{\lambda\rightarrow\infty}\lim_{\beta\rightarrow\infty}
\lim_{u\rightarrow0}Z(\beta,\lambda)^{u-1} \nonumber \\ 
\cdot\mathbb{E}\left[
 \left. 
  \sum_{\backslash s_{k}}\int
  \mathrm{e}^{
   -\beta\mathcal{H}_{\lambda}
   (\boldsymbol{S},\{\tilde{\boldsymbol{x}}_{t}\},\{\boldsymbol{u}_{t}\})
  }d\backslash\tilde{\mathcal{X}}_{k}\prod_{t=0}^{T-1}d\boldsymbol{u}_{t} 
 \right| \mathcal{X}_{k}
\right], \label{original_marginal} 
\end{IEEEeqnarray}
where $\mathcal{H}_{\lambda}(\boldsymbol{S},\{\tilde{\boldsymbol{x}}_{t}\},
\{\boldsymbol{u}_{t}\})$ and $Z(\beta,\lambda)$ are given by 
(\ref{Hamiltonian}) and (\ref{partition_function}), respectively. 
In (\ref{original_marginal}), $\sum_{\backslash s_{k}}$ denotes the 
marginalization over $\{s_{k'}\in\{0,1\}:k'\neq k\}$. Furthermore, 
$\int d\backslash\tilde{\mathcal{X}}_{k}$ represents the marginalization over 
$\{\tilde{x}_{k',t}\in\mathcal{M}_{x_{k',t}}: \hbox{for all $t$ and $k'\neq k$}
\}$. Regarding $u$ in (\ref{original_marginal}) as a non-negative integer gives 
\begin{equation} \label{marginal} 
p(s_{k},\tilde{\mathcal{X}}_{k} | \mathcal{X}_{k})
= \lim_{\lambda\rightarrow\infty}\lim_{\beta\rightarrow\infty}
\lim_{u\rightarrow0}Z_{u}(s_{k},\tilde{\mathcal{X}}_{k},\mathcal{X}_{k}
;\beta,\lambda), 
\end{equation}
with
\begin{IEEEeqnarray}{rl} 
&Z_{u}(s_{k},\tilde{\mathcal{X}}_{k},\mathcal{X}_{k};\beta,\lambda) 
\nonumber \\ 
=& \left(
  \frac{\pi}{\beta}
 \right)^{uNT} 
\int\Xi_{\beta\lambda}^{(u)}(\boldsymbol{Q},s_{k},\tilde{\mathcal{X}}_{k},
\mathcal{X}_{k}) 
p(\boldsymbol{Q})d\boldsymbol{Q}. \label{marginal_partition_function}
\end{IEEEeqnarray}
In (\ref{marginal_partition_function}), the pdf $p(\boldsymbol{Q})$ is 
given by (\ref{pdf_Q}). Furthermore, 
$\Xi_{\beta\lambda}^{(u)}(\boldsymbol{Q},s_{k},\tilde{\mathcal{X}}_{k},
\mathcal{X}_{k})$ is 
defined as 
\begin{IEEEeqnarray}{r} 
\Xi_{\beta\lambda}^{(u)}(\boldsymbol{Q},s_{k,0},\tilde{\mathcal{X}}_{k,0},
\mathcal{X}_{k}) 
= \mathbb{E}\left[
 \left. 
  \sum_{\backslash s_{k,0}}\int 
  \prod_{a=0}^{u-1}\exp\{
 \right. 
\right. \nonumber \\ 
\left. 
 \left. 
   - \beta\lambda\tilde{g}
   (\{s_{k,a}\},\{\tilde{\boldsymbol{x}}_{a}(k)\},\{\boldsymbol{v}_{a}(k)\})
  \}d\backslash\tilde{\mathcal{X}}_{k,0}
 \right| \mathcal{X}_{k} 
\right], \label{marginal_Xi_function} 
\end{IEEEeqnarray}
with (\ref{tilde_g}), where $\tilde{\mathcal{X}}_{k,0}$ is given by 
$\tilde{\mathcal{X}}_{k,0}=\{\tilde{x}_{k,t,0}:t=0,\ldots,T-1\}$. 
Substituting (\ref{pdf_Q}) into (\ref{marginal_partition_function}) yields 
\begin{IEEEeqnarray}{l} 
Z_{u}(s_{k},\tilde{\mathcal{X}}_{k},\mathcal{X}_{k};\beta,\lambda)
= \left(
 \frac{\pi}{\beta}
\right)^{uNT} C_{u}\int d\boldsymbol{Q}\det\boldsymbol{Q}^{-uT} \nonumber \\ 
\cdot\exp\left\{
 \beta N\left(
  \frac{1}{\beta N}\ln\Xi_{\beta\lambda}^{(u)}
  (\boldsymbol{Q},s_{k},\tilde{\mathcal{X}}_{k},\mathcal{X}_{k}) 
  - I_{u}(\boldsymbol{Q})
 \right)
\right\}, 
\end{IEEEeqnarray}
with (\ref{rate_function}). 
Assuming that the large-system limit and the limits in (\ref{marginal})  
are commutative, we use the saddle-point method to obtain 
\begin{IEEEeqnarray}{rl}  
p(s_{k},\tilde{\mathcal{X}}_{k} | \mathcal{X}_{k}) 
=& \lim_{\lambda\rightarrow\infty}\lim_{\beta\rightarrow\infty}
\lim_{u\rightarrow0}\lim_{K\rightarrow\infty}\Bigl\{ \nonumber \\ 
&\left.  
 \mathrm{e}^{
  -\beta NI_{u}(\boldsymbol{Q}_{\mathrm{s}})
 }\Xi_{\beta\lambda}^{(u)}
 (\boldsymbol{Q}_{\mathrm{s}},s_{k},\tilde{\mathcal{X}}_{k},\mathcal{X}_{k})  
\right\}, \label{marginal1}
\end{IEEEeqnarray}
where $\boldsymbol{Q}_{\mathrm{s}}$ is the solution to minimize 
(\ref{Phi}), given by (\ref{solution}). 
In the derivation of (\ref{marginal1}), we have used the fact that the 
difference between 
$\ln\Xi_{\beta\lambda}^{(u)}(\boldsymbol{Q},s_{k},\tilde{\mathcal{X}}_{k},
\mathcal{X}_{k})$ 
and $\ln\Xi_{\beta\lambda}(\boldsymbol{Q})$ should be $O(1)$. 

\subsection{Replica Symmetry Solution} \label{proposition3_proof_RS} 
We evaluate (\ref{marginal1}) under the RB assumption~(\ref{RS}). 
The order parameter $q_{0}$ satisfies the fixed-point 
equation~(\ref{deriv_chi_RS}). Furthermore, from 
(\ref{rate_function_rs_marginal}), it is straightforward to find that 
$I_{u}(\boldsymbol{Q}_{\mathrm{s}})$ tends to zero in $u\rightarrow0$. 

We next calculate (\ref{marginal_Xi_function}) with (\ref{v_RS}) to obtain 
\begin{IEEEeqnarray}{rl}
&\lim_{u\rightarrow0}\lim_{K\rightarrow\infty} \Xi_{\beta\lambda}^{(u)}
(\boldsymbol{Q}_{\mathrm{s}},s_{k,0},\tilde{\mathcal{X}}_{k,0},
\mathcal{X}_{k}) \nonumber \\ 
=& \lim_{K\rightarrow\infty}\mathbb{E}\left[
 \left\{
  \Xi_{\beta\lambda}^{(\mathrm{RS})}(\{\boldsymbol{z}(k)\})
 \right\}^{-1}
\right. \nonumber \\ 
\cdot&\left. 
 \left. 
  \sum_{\backslash s_{k,0}}
  \int\frac{\mathrm{e}^{
   - H_{\beta\lambda}^{(\mathrm{RS})}(\{s_{k,0}\},\{\tilde{\boldsymbol{x}}_{0}(k)\},
   \{\boldsymbol{z}(k)\})    
  }}{(1+\beta\lambda\chi)^{T\sum_{k'=1}^{K}s_{k',0}}}
  d\backslash\tilde{\mathcal{X}}_{k,0} 
 \right| \mathcal{X}_{k}
\right], \label{marginal_Xi_function_rs} 
\end{IEEEeqnarray}
where $\Xi_{\beta\lambda}^{(\mathrm{RS})}(\{\boldsymbol{z}(k)\})$ 
and $H_{\beta\lambda}^{(\mathrm{RS})}(\{s_{k,0}\},
\{\tilde{\boldsymbol{x}}_{0}(k)\},\{\boldsymbol{z}(k)\})$ are given by 
(\ref{Xi_function_RS}) and (\ref{H_rs}), respectively.  
Substituting (\ref{marginal_Xi_function_rs}) into (\ref{marginal1}) and then 
taking $\beta\rightarrow\infty$ with $\hat{\chi}=\beta\chi$ fixed before 
$\lambda\rightarrow\infty$, we have 
\begin{IEEEeqnarray}{rl}  
&p(s_{k},\tilde{\mathcal{X}}_{k} | \mathcal{X}_{k}) \nonumber \\ 
=& \lim_{K\rightarrow\infty}\lim_{\beta\rightarrow\infty}\mathbb{E}\left[
 \left. 
  \exp\left\{ 
   -\frac{\beta}{\hat{\chi}}H^{(\mathrm{RS})}(s_{k},\tilde{\mathcal{X}}_{k})  
  \right\} 
 \right| \mathcal{X}_{k} 
\right], \label{marginal2}
\end{IEEEeqnarray}
with 
\begin{IEEEeqnarray}{r}
H^{(\mathrm{RS})}(s_{k},\tilde{\mathcal{X}}_{k}) 
= \frac{s_{k}}{T}\sum_{t=0}^{T-1}\left| 
 \tilde{x}_{k,t} - \sqrt{q_{0}}(\boldsymbol{z}(k))_{t}
\right|^{2} \nonumber \\ 
+ \min_{\backslash s_{k}:\sum_{k'=1}^{K}s_{k'}=\tilde{K}}
\sum_{k'\neq k}s_{k'}E_{k'}^{(\mathrm{RS})}(q_{0}) \nonumber \\ 
- \min_{\{s_{k'}\in\{0,1\}\}:\sum_{k'=1}^{K}s_{k'}=\tilde{K}}
\sum_{k'=1}^{K}s_{k'}E_{k'}^{(\mathrm{RS})}(q_{0}), \label{marginal_Hamiltonian_RS} 
\end{IEEEeqnarray}
where $E_{k}^{(\mathrm{RS})}(q_{0})$ is given by (\ref{E_k_RS}). The 
quantity~(\ref{marginal_Hamiltonian_RS}) is non-negative for any $s_{k}$ and 
$\tilde{\mathcal{X}}_{k}$, and zero if and only if 
$(s_{k}, \tilde{\mathcal{X}}_{k})$ is equal to the optimal solution 
$(s_{k}^{(\mathrm{opt})}(q_{0}),\{\tilde{x}_{k,t}^{(\mathrm{opt})}(q_{0})\})$, 
given by 
\begin{equation} \label{tilde_x_k_RS} 
\tilde{x}_{k,t}^{(\mathrm{opt})}(q_{0})   
= \argmin_{\tilde{x}_{k,t}\in\mathcal{M}_{x_{k,t}}}\left| 
 \tilde{x}_{k,t} - \sqrt{q_{0}}(\boldsymbol{z}(k))_{t}
\right|^{2}, 
\end{equation}
\begin{IEEEeqnarray}{rl}  
s_{k}^{(\mathrm{opt})}(q_{0})   
=& \argmin_{s_{k}\in\{0,1\}}\Biggl\{
s_{k}E_{k}^{(\mathrm{RS})}(q_{0}) \nonumber \\ 
+&\left.
  \min_{\backslash s_{k}:\sum_{k'=1}^{K}s_{k'}=\tilde{K}}
 \sum_{k'\neq k}s_{k'}E_{k'}^{(\mathrm{RS})}(q_{0})
\right\}, \label{s_k_RS}
\end{IEEEeqnarray}
with (\ref{E_k_RS}). Substituting (\ref{marginal2}) into 
(\ref{target_distribution}), we arrive at 
\begin{IEEEeqnarray}{l} 
\mathrm{Pr}\left(
 \left. 
  s_{k}=1, \tilde{\mathcal{X}}_{k}\in\prod_{t=0}^{T-1}
  \mathcal{A}_{t} 
 \right| \mathcal{X}_{k}
\right)  
= \lim_{K\rightarrow\infty}\mathbb{E}\Biggl[ \nonumber \\
\left. 
 \left. 
  1\left(
   s_{k}^{(\mathrm{opt})}(q_{0})=1
  \right)\prod_{t=0}^{T-1}
  1\left(
   \tilde{x}_{k,t}^{(\mathrm{opt})}(q_{0})\in\mathcal{A}_{t}
  \right) 
 \right| \mathcal{X}_{k}
\right], \label{marginal_rs}
\end{IEEEeqnarray}
where $\tilde{x}_{k,t}^{(\mathrm{opt})}(q_{0})$ and 
$s_{k}^{(\mathrm{opt})}(q_{0})$ are given by (\ref{tilde_x_k_RS}) and 
(\ref{s_k_RS}), respectively.  

In order to complete the derivation of Proposition~\ref{proposition3}, 
we prove that (\ref{marginal_rs}) reduces to (\ref{joint_distribution}) 
with $q=q_{0}$. The solution~(\ref{s_k_RS}) takes $1$ if and only if 
$E_{k}^{(\mathrm{RS})}(q_{0})$ is smaller than the $\tilde{K}$th order 
statistic, i.e.\ 
$E_{k}^{(\mathrm{RS})}(q_{0})\leq E_{(\tilde{K})}^{(\mathrm{RS})}(q_{0})$. 
Lemma~\ref{lemma1} implies that the $\tilde{K}$th order statistic 
$E_{(\tilde{K})}^{(\mathrm{RS})}(q_{0})$ converges in probability to the 
$\kappa$-quantile $\xi_{\kappa,T}(q_{0})$, given by (\ref{quantile}), 
in the large-system limit. This observation implies that 
(\ref{marginal_rs}) reduces to (\ref{joint_distribution}) with $q=q_{0}$. 

\section{Derivation of Proposition~\ref{proposition4}}
\label{proposition4_proof} 
We start with (\ref{marginal1}). Let us calculate (\ref{marginal1}) under the 
1RSB assumption~(\ref{1RSB}). As shown in 
Appendix~\ref{proposition2_proof}, $q_{0}$ tends to zero, and $q_{1}$ 
satisfies the coupled fixed-point equations~(\ref{fixed_point_1RSB_appen1}) 
and (\ref{fixed_point_1RSB_appen2}). Furthermore, from 
(\ref{rate_function_1RSB_marginal}), it is straightforward to find that 
$I_{u}(\boldsymbol{Q}_{\mathrm{s}})$ tends to zero in $u\rightarrow0$. 

We next calculate (\ref{marginal_Xi_function}) with (\ref{v_1RSB}) to obtain 
\begin{IEEEeqnarray}{rl}
&\lim_{u\rightarrow0}\lim_{K\rightarrow\infty} \Xi_{\beta\lambda}^{(u)}
(\boldsymbol{Q}_{\mathrm{s}},s_{k,0},\tilde{\mathcal{X}}_{k,0},
\mathcal{X}_{k}) 
\nonumber \\ 
=& \lim_{K\rightarrow\infty}\mathbb{E}\left[
 \left\{
  \Xi_{\beta\lambda}^{(\mathrm{1RSB})}(\{\boldsymbol{z}(k)\},
  \{\boldsymbol{z}_{0}(k)\})
 \right\}^{-1}\sum_{\backslash s_{k,0}}\int d\backslash\tilde{\mathcal{X}}_{k,0} 
\right. \nonumber \\ 
&\cdot\left. 
 \left. 
  \frac{\mathrm{e}^{
   - H_{\beta\lambda}^{(\mathrm{1RSB})}(\{s_{k,0}\},
   \{\tilde{\boldsymbol{x}}_{0}(k)\},\{\boldsymbol{z}(k)\},
   \{\boldsymbol{z}_{0}(k)\}) 
  }}{(1+\beta\lambda\chi)^{T\sum_{k'=1}^{K}s_{k',0}}}
 \right| \mathcal{X}_{k}
\right], \label{marginal_Xi_function_1RSB} 
\end{IEEEeqnarray}
where $\Xi_{\beta\lambda}^{(\mathrm{1RSB})}(\{\boldsymbol{z}(k)\},
\{\boldsymbol{z}_{0}(k)\})$ and 
$H_{\beta\lambda}^{(\mathrm{1RSB})}(\{s_{k,0}\},\{\tilde{\boldsymbol{x}}_{0}(k)\},
\{\boldsymbol{z}(k)\},\{\boldsymbol{z}_{0}(k)\})$ are given by 
(\ref{Xi_function_1RSB_tmp}) and (\ref{H_1RSB}), respectively.  
Substituting (\ref{marginal_Xi_function_1RSB}) into (\ref{marginal1}) and then 
taking the limit $\beta\rightarrow\infty$, $m_{1}\rightarrow0$, and 
$\chi\rightarrow0$ with $\mu_{1}=\beta m_{1}$ and $\hat{\chi}=\beta\chi$ 
fixed before taking $\lambda\rightarrow\infty$, we have 
\begin{IEEEeqnarray}{rl}  
&p(s_{k},\tilde{\mathcal{X}}_{k} | \mathcal{X}_{k}) \nonumber \\ 
=& \lim_{K\rightarrow\infty}\lim_{\beta\rightarrow\infty}\mathbb{E}\left[
 \left. 
  \exp\left\{ 
   -\frac{\beta}{\hat{\chi}}H^{(\mathrm{1RSB})}(s_{k},\tilde{\mathcal{X}}_{k})  
  \right\} 
 \right| \mathcal{X}_{k} 
\right], \label{marginal2_1RSB}
\end{IEEEeqnarray}
with 
\begin{IEEEeqnarray}{r}
H^{(\mathrm{1RSB})}(s_{k},\tilde{\mathcal{X}}_{k}) 
= \frac{s_{k}}{T}\sum_{t=0}^{T-1}
|\tilde{x}_{k,t} - \sqrt{q_{1}}(\boldsymbol{z}_{0}(k))_{t}|^{2} \nonumber \\ 
+ \min_{\backslash s_{k}:\sum_{k'=1}^{K}s_{k'}=\tilde{K}}
\sum_{k'\neq k}s_{k'}E_{k'}^{(\mathrm{1RSB})}(0,q_{1}) \nonumber \\ 
- \min_{\{s_{k'}\in\{0,1\}\}:\sum_{k'=1}^{K}s_{k'}=\tilde{K}}
\sum_{k'=1}^{K}s_{k'}E_{k'}^{(\mathrm{1RSB})}(0,q_{1}), 
\label{marginal_Hamiltonian_1RSB} 
\end{IEEEeqnarray}
where $E_{k}^{(\mathrm{1RSB})}(q_{0},q_{1})$ is given by (\ref{E_k_appen}). The 
quantity~(\ref{marginal_Hamiltonian_1RSB}) is non-negative for any $s_{k}$ and 
$\tilde{\mathcal{X}}_{k}$, and zero if and only if 
$(s_{k}, \tilde{\mathcal{X}}_{k})$ is equal to the optimal solution 
$(s_{k}^{(\mathrm{opt})}(0,q_{1}),
\{\tilde{x}_{k,t}^{(\mathrm{opt})}(0,q_{1})\})$, 
given by 
\begin{equation} \label{tilde_x_k_1RSB} 
\tilde{x}_{k,t}^{(\mathrm{opt})}(0,q_{1})   
= \argmin_{\tilde{x}_{k,t}\in\mathcal{M}_{x_{k,t}}}\left| 
 \tilde{x}_{k,t} - \sqrt{q_{1}}(\boldsymbol{z}_{0}(k))_{t}
\right|^{2}, 
\end{equation}
\begin{IEEEeqnarray}{rl} 
s_{k}^{(\mathrm{opt})}(0,q_{1})   
&= \argmin_{s_{k}\in\{0,1\}}\Biggl\{
s_{k}E_{k}^{(\mathrm{1RSB})}(0,q_{1}) 
\nonumber \\ 
+&\left.
 \min_{\backslash s_{k}:\sum_{k'=1}^{K}s_{k'}=\tilde{K}}
 \sum_{k'\neq k}s_{k'}E_{k'}^{(\mathrm{1RSB})}(0,q_{1})
\right\}, \label{s_k_1RSB} 
\end{IEEEeqnarray}
with (\ref{E_k_appen}). Substituting (\ref{marginal2_1RSB}) into 
(\ref{target_distribution}), we arrive at 
\begin{IEEEeqnarray}{l} 
\mathrm{Pr}\left(
 \left. 
  s_{k}=1, \tilde{\mathcal{X}}_{k}\in\prod_{t=0}^{T-1}
  \mathcal{A}_{t} 
 \right| \mathcal{X}_{k}
\right) 
= \lim_{K\rightarrow\infty}\mathbb{E}\Biggl[ \nonumber \\ 
\left. 
 \left. 
  1\left(
   s_{k}^{(\mathrm{opt})}(0,q_{1})=1
  \right)\prod_{t=0}^{T-1}
  1\left(
   \tilde{x}_{k,t}^{(\mathrm{opt})}(0,q_{1})\in\mathcal{A}_{t}
  \right) 
 \right| \mathcal{X}_{k}
\right], \label{marginal_1RSB}
\end{IEEEeqnarray}
where $\tilde{x}_{k,t}^{(\mathrm{opt})}(0,q_{1})$ and 
$s_{k}^{(\mathrm{opt})}(0,q_{1})$ are given by (\ref{tilde_x_k_1RSB}) and 
(\ref{s_k_1RSB}), respectively. Repeating the argument in the end of 
Appendix~\ref{proposition3_proof_RS}, we find that (\ref{marginal_1RSB}) 
reduces to (\ref{joint_distribution}) with $q=q_{1}$.  

\ifCLASSOPTIONcaptionsoff
  \newpage
\fi



\bibliographystyle{IEEEtran}
\bibliography{IEEEabrv,kt-it2012}

\begin{thebibliography}{10}
\providecommand{\url}[1]{#1}
\csname url@samestyle\endcsname
\providecommand{\newblock}{\relax}
\providecommand{\bibinfo}[2]{#2}
\providecommand{\BIBentrySTDinterwordspacing}{\spaceskip=0pt\relax}
\providecommand{\BIBentryALTinterwordstretchfactor}{4}
\providecommand{\BIBentryALTinterwordspacing}{\spaceskip=\fontdimen2\font plus
\BIBentryALTinterwordstretchfactor\fontdimen3\font minus
  \fontdimen4\font\relax}
\providecommand{\BIBforeignlanguage}[2]{{%
\expandafter\ifx\csname l@#1\endcsname\relax
\typeout{** WARNING: IEEEtran.bst: No hyphenation pattern has been}%
\typeout{** loaded for the language `#1'. Using the pattern for}%
\typeout{** the default language instead.}%
\else
\language=\csname l@#1\endcsname
\fi
#2}}
\providecommand{\BIBdecl}{\relax}
\BIBdecl

\bibitem{Foschini98}
G.~J. Foschini and M.~J. Gans, ``On limits of wireless communications in a
  fading environment when using multiple antennas,'' \emph{Wireless Pers.
  Commun.}, vol.~6, pp. 311--335, 1998.

\bibitem{Telatar99}
E.~Telatar, ``Capacity of multi-antenna {Gaussian} channels,'' \emph{Euro.
  Trans. Telecommun.}, vol.~10, no.~6, pp. 585--595, Nov.--Dec. 1999.

\bibitem{Marzetta99}
T.~L. Marzetta and B.~M. Hochwald, ``Capacity of a mobile multiple-antenna
  communication link in {Rayleigh} flat fading,'' \emph{{IEEE} Trans. Inf.
  Theory}, vol.~45, no.~1, pp. 139--157, Jan. 1999.

\bibitem{Zheng02}
L.~Zheng and D.~N.~C. Tse, ``Communication on the {Grassmann} manifold: A
  geometric approach to the noncoherent multiple-antenna channel,''
  \emph{{IEEE} Trans. Inf. Theory}, vol.~48, no.~2, pp. 359--383, Feb. 2002.

\bibitem{Biglieri02}
E.~Biglieri, G.~Taricco, and A.~Tulino, ``Performance of space-time codes for a
  large number of antennas,'' \emph{{IEEE} Trans. Inf. Theory}, vol.~48, no.~7,
  pp. 1794--1803, Jul. 2002.

\bibitem{Tse99}
D.~N.~C. Tse and S.~V. Hanly, ``Linear multiuser receivers: effective
  interference, effective bandwidth and user capacity,'' \emph{{IEEE} Trans.
  Inf. Theory}, vol.~45, no.~2, pp. 641--657, Mar. 1999.

\bibitem{Verdu99}
S.~Verd\'u and S.~{Shamai (Shitz)}, ``Spectral efficiency of {CDMA} with random
  spreading,'' \emph{{IEEE} Trans. Inf. Theory}, vol.~45, no.~2, pp. 622--640,
  Mar. 1999.

\bibitem{Mueller03}
R.~R. M\"uller, ``Channel capacity and minimum probability of error in large
  dual antenna array systems with binary modulation,'' \emph{{IEEE} Trans.
  Signal Process.}, vol.~51, no.~11, pp. 2821--2828, Nov. 2003.

\bibitem{Viswanath02}
P.~Viswanath, D.~N.~C. Tse, and R.~Laroia, ``Opportunistic beamforming using
  dumb antennas,'' \emph{{IEEE} Trans. Inf. Theory}, vol.~48, no.~6, pp.
  1277--1294, Jun. 2002.

\bibitem{Chung03}
J.~Chung, C.-S. Hwang, K.~Kim, and Y.~K. Kim, ``A random beamforming technique
  in {MIMO} systems exploiting multiuser diversity,'' \emph{{IEEE} J. Sel.
  Areas Commun.}, vol.~21, no.~5, pp. 848--855, Jun. 2003.

\bibitem{Marzetta06}
T.~L. Marzetta, ``How much training is required for multiuser {MIMO}?'' in
  \emph{Proc. 40th Asilomar Conf. Signals, Systems, \& Computers}, Pacific
  Grove, CA, USA, Oct.--Nov. 2006, pp. 359--363.

\bibitem{Marzetta10}
------, ``Noncooperative cellular wireless with unlimited numbers of base
  station antennas,'' \emph{{IEEE} Trans. Wireless Commun.}, vol.~9, no.~11,
  pp. 3590--3600, Nov. 2010.

\bibitem{Caire03}
G.~Caire and S.~{Shamai (Shitz)}, ``On the achievable throughput of a
  multiantenna {Gaussian} broadcast channel,'' \emph{{IEEE} Trans. Inf.
  Theory}, vol.~49, no.~7, pp. 1691--1706, Jul. 2003.

\bibitem{Viswanath03}
P.~Viswanath and D.~N.~C. Tse, ``Sum capacity of the vector {Gaussian}
  broadcast channel and uplink-downlink duality,'' \emph{{IEEE} Trans. Inf.
  Theory}, vol.~49, no.~8, pp. 1912--1921, Aug. 2003.

\bibitem{Yu04}
W.~Yu and J.~M. Cioffi, ``Sum capacity of {Gaussian} vector broadcast
  channels,'' \emph{{IEEE} Trans. Inf. Theory}, vol.~50, no.~9, pp. 1875--1892,
  Sep. 2004.

\bibitem{Weingarten06}
H.~Weingarten, Y.~Steinberg, and S.~{Shamai (Shitz)}, ``The capacity region of
  the {Gaussian} multiple-input multiple-output broadcast channel,''
  \emph{{IEEE} Trans. Inf. Theory}, vol.~52, no.~9, pp. 3936--3964, Sep. 2006.

\bibitem{Costa83}
M.~H.~M. Costa, ``Writing on dirty paper,'' \emph{{IEEE} Trans. Inf. Theory},
  vol.~29, no.~3, pp. 439--441, May 1983.

\bibitem{Spencer04}
Q.~H. Spencer, L.~Swindlehurst, and M.~Haardt, ``Zero-forcing methods for
  downlink spatial multiplexing in multiuser {MIMO} channels,'' \emph{{IEEE}
  Trans. Signal Process.}, vol.~52, no.~2, pp. 461--471, Feb. 2004.

\bibitem{Choi04}
L.-U. Choi and R.~D. Murch, ``A transmit preprocessing technique for multiuser
  {MIMO} systems using a decomposition approach,'' \emph{{IEEE} Trans. Wireless
  Commun.}, vol.~3, no.~1, pp. 20--24, Jan. 2004.

\bibitem{Wiesel08}
A.~Wiesel, Y.~C. Eldar, and S.~{Shamai (Shitz)}, ``Zero-forcing precoding and
  generalized inverses,'' \emph{{IEEE} Trans. Signal Process.}, vol.~56, no.~9,
  pp. 4409--4418, Sep. 2008.

\bibitem{Tu03}
Z.~Tu and R.~S. Blum, ``Multiuser diversity for a dirty paper approach,''
  \emph{{IEEE} Commun. Lett.}, vol.~7, no.~8, pp. 370--372, Aug. 2003.

\bibitem{Dimic05}
G.~Dimi\'c and N.~D. Sidiropoulos, ``On downlink beamforming with greedy user
  selection: Performance analysis and a simple new algorithm,'' \emph{{IEEE}
  Trans. Signal Process.}, vol.~53, no.~10, pp. 3857--3868, Oct. 2005.

\bibitem{Yoo06}
T.~Yoo and A.~Goldsmith, ``On the optimality of multiantenna broadcast
  scheduling using zero-forcing beamforming,'' \emph{{IEEE} J. Sel. Areas
  Commun.}, vol.~24, no.~3, pp. 528--541, Mar. 2006.

\bibitem{Shen06}
Z.~Shen, R.~Chen, J.~G. Andrews, R.~W. {Heath, Jr.}, and B.~L. Evans, ``Low
  complexity user selection algorithms for multiuser {MIMO} systems with block
  diagonalization,'' \emph{{IEEE} Trans. Signal Process.}, vol.~54, no.~9, pp.
  3658--3663, Sep. 2006.

\bibitem{Wang08}
J.~Wang, D.~J. Love, and M.~D. Zoltowski, ``User selection with zero-forcing
  beamforming achieves the asymptotically optimal sum rate,'' \emph{{IEEE}
  Trans. Signal Process.}, vol.~56, no.~8, pp. 3713--3726, Aug. 2008.

\bibitem{Hochwald05}
B.~M. Hochwald, C.~B. Peel, and A.~L. Swindlehurst, ``A vector-perturbation
  technique for near-capacity multiantenna multiuser communication---part~{II}:
  Perturbation,'' \emph{{IEEE} Trans. Commun.}, vol.~53, no.~3, pp. 537--544,
  Mar. 2005.

\bibitem{Mueller08}
R.~R. M\"uller, D.~Guo, and A.~L. Moustakas, ``Vector precoding for wireless
  {MIMO} systems and its replica analysis,'' \emph{{IEEE} J. Sel. Areas
  Commun.}, vol.~26, no.~3, pp. 530--540, Apr. 2008.

\bibitem{Razi10}
A.~Razi, D.~J. Ryan, I.~B. Collings, and J.~Yuan, ``Sum rates, rate allocation,
  and user selection for multi-user {MIMO} vector perturbation precoding,''
  \emph{{IEEE} Trans. Wireless Commun.}, vol.~9, no.~1, pp. 356--365, Jan.
  2010.

\bibitem{Boyd04}
S.~P. Boyd and L.~Vandenberghe, \emph{Convex Optimization}.\hskip 1em plus
  0.5em minus 0.4em\relax Cambridge University Press, 2004.

\bibitem{Zaidel12}
B.~M. Zaidel, R.~R. M\"uller, A.~L. Moustakas, and R.~de~Miguel, ``Vector
  precoding for {Gaussian} {MIMO} broadcast channels: Impact of replica
  symmetry breaking,'' \emph{IEEE Trans. Inf. Theory}, vol.~58, no.~3, pp.
  1413--1440, Mar. 2012.

\bibitem{Takeuchi121}
K.~Takeuchi and T.~Kawabata, ``A greedy algorithm of data-dependent user
  selection for fast fading {Gaussian} vector broadcast channels,'' \emph{{\rm
  submitted to} IEICE Trans. Fundamentals}, 2012, [Online]. Available:
  http://arxiv.org/abs/1201.6453.

\bibitem{Sherrington75}
D.~Sherrington and S.~Kirkpatrick, ``Solvable model of a spin-glass,''
  \emph{Phys. Rev. Lett.}, vol.~35, no.~26, pp. 1792--1796, Dec. 1975.

\bibitem{Mezard87}
M\'ezard, G.~Parisi, and M.~A. Virasoro, \emph{Spin Glass Theory and
  Beyond}.\hskip 1em plus 0.5em minus 0.4em\relax Singapore: World Scientific,
  1987.

\bibitem{Nishimori01}
H.~Nishimori, \emph{Statistical Physics of Spin Glasses and Information
  Processing}.\hskip 1em plus 0.5em minus 0.4em\relax New York: Oxford
  University Press, 2001.

\bibitem{Moustakas03}
A.~L. Moustakas, S.~H. Simon, and A.~M. Sengupta, ``{MIMO} capacity through
  correlated channels in the presence of correlated interferers and noise: A
  (not so) large ${N}$ analysis,'' \emph{{IEEE} Trans. Inf. Theory}, vol.~49,
  no.~10, pp. 2545--2561, Oct. 2003.

\bibitem{Mueller04}
R.~R. M\"{u}ller and W.~H. Gerstacker, ``On the capacity loss due to separation
  of detection and decoding,'' \emph{{IEEE} Trans. Inf. Theory}, vol.~50,
  no.~8, pp. 1769--1778, Aug. 2004.

\bibitem{Guo05}
D.~Guo and S.~Verd\'u, ``Randomly spread {CDMA}: Asymptotics via statistical
  physics,'' \emph{{IEEE} Trans. Inf. Theory}, vol.~51, no.~6, pp. 1983--2010,
  Jun. 2005.

\bibitem{Takeda06}
K.~Takeda, S.~Uda, and Y.~Kabashima, ``Analysis of {CDMA} systems that are
  characterized by eigenvalue spectrum,'' \emph{Europhys. Lett.}, vol.~76,
  no.~6, pp. 1193--1199, 2006.

\bibitem{Wen07}
C.~K. Wen and K.~K. Wong, ``Asymptotic analysis of spatially correlated {MIMO}
  multiple-access channels with arbitrary signaling inputs for joint and
  separate decoding,'' \emph{{IEEE} Trans. Inf. Theory}, vol.~53, no.~1, pp.
  252--268, Jan. 2007.

\bibitem{Takeuchi08}
K.~Takeuchi, T.~Tanaka, and T.~Yano, ``Asymptotic analysis of general multiuser
  detectors in {MIMO DS-CDMA} channels,'' \emph{{IEEE} J. Sel. Areas Commun.},
  vol.~26, no.~3, pp. 486--496, Apr. 2008.

\bibitem{Takeuchi122}
K.~Takeuchi, M.~Vehkaper\"a, T.~Tanaka, and R.~R. M\"uller, ``Large-system
  analysis of joint channel and data estimation for {MIMO} {DS-CDMA} systems,''
  \emph{{IEEE} Trans. Inf. Theory}, vol.~58, no.~3, pp. 1385--1412, Mar. 2012.

\bibitem{Tanaka02}
T.~Tanaka, ``A statistical-mechanics approach to large-system analysis of
  {CDMA} multiuser detectors,'' \emph{{IEEE} Trans. Inf. Theory}, vol.~48,
  no.~11, pp. 2888--2910, Nov. 2002.

\bibitem{Hemmen79}
J.~L. van Hemmen and R.~G. Palmer, ``The replica method and a solvable spin
  glass model,'' \emph{J. Phys. A: Math. Gen.}, vol.~12, no.~4, pp. 563--580,
  1979.

\bibitem{Parisi80}
G.~Parisi, ``A sequence of approximate solutions to the {S-K} model for spin
  glasses,'' \emph{J. Phys. A: Math. Gen.}, vol.~13, no.~4, pp. L115--L121,
  Apr. 1980.

\bibitem{Korada11}
S.~B. Korada and A.~Montanari, ``Applications of the {Lindeberg} principle in
  communicaitons and statistical learning,'' \emph{{IEEE} Trans. Inf. Theory},
  vol.~57, no.~4, pp. 2440--2450, Apr. 2011.

\bibitem{Guerra03}
F.~Guerra, ``Broken replica symmetry bounds in the mean field spin glass
  model,'' \emph{Commun. Math. Phys.}, vol. 233, pp. 1--12, 2003.

\bibitem{Talagrand06}
M.~Talagrand, ``The {Parisi} formula,'' \emph{Annals of Mathematics}, vol. 163,
  pp. 221--263, 2006.

\bibitem{David03}
H.~A. David and H.~N. Nagaraja, \emph{Order Statistics}, 3rd~ed.\hskip 1em plus
  0.5em minus 0.4em\relax New Jersey, USA: Wiley, 2003.

\bibitem{Neeser93}
F.~D. Neeser and J.~L. Massey, ``Proper complex random processes with
  applications to information theory,'' \emph{{IEEE} Trans. Inf. Theory},
  vol.~39, no.~4, pp. 1293--1302, Jul. 1993.

\bibitem{Tulino04}
A.~M. Tulino and S.~Verd\'{u}, \emph{Random Matrix Theory and Wireless
  Communications}.\hskip 1em plus 0.5em minus 0.4em\relax Hanover, MA USA: Now
  Publishers, 2004.

\bibitem{Bahadur66}
R.~R. Bahadur, ``A note on quantiles in large samples,'' \emph{Ann. Math.
  Statist.}, vol.~37, no.~3, pp. 577--580, 1966.

\bibitem{Ghosh71}
J.~K. Ghosh, ``A new proof of the {Bahadur} representation of quantiles and an
  application,'' \emph{Ann. Math. Statist.}, vol.~42, no.~6, pp. 1957--1961,
  1971.

\bibitem{Stigler74}
S.~M. Stigler, ``Linear functions of order statistics with smooth weight
  functions,'' \emph{Ann. Statist.}, vol.~2, no.~4, pp. 676--693, Jul. 1974.

\bibitem{Cover06}
T.~M. Cover and J.~A. Thomas, \emph{Elements of Information Theory},
  2nd~ed.\hskip 1em plus 0.5em minus 0.4em\relax New Jersey: Wiley, 2006.

\bibitem{Tse05}
D.~N.~C. Tse and P.~Viswanath, \emph{Fundamentals of Wireless
  Communication}.\hskip 1em plus 0.5em minus 0.4em\relax Cambridge, UK:
  Cambridge University Press, 2005.

\bibitem{Pastur91}
L.~A. Pastur and M.~V. Shcherbina, ``Absence of self-averaging of the order
  parameter in the {Sherrington}-{Kirkpatrick} model,'' \emph{J. Stat. Phys.},
  vol.~62, no. 1/2, pp. 1--19, 1991.

\bibitem{Guerra02}
F.~Guerra and F.~L. Toninelli, ``The thermodynamic limit in mean field spin
  glass models,'' \emph{Commun. Math. Phys.}, vol. 230, pp. 71--79, 2002.

\bibitem{Guerra032}
------, ``The infinite volume limit in generalized mean field disordered
  models,'' \emph{Markov. Proc. Rel. Fields}, vol.~9, pp. 195--207, 2003.

\bibitem{Korada10}
S.~B. Korada and N.~Macris, ``Tight bounds on the capacity of binary input
  random {CDMA} systems,'' \emph{{IEEE} Trans. Inf. Theory}, vol.~56, no.~11,
  pp. 5590--5613, Nov. 2010.

\bibitem{Mezard09}
M.~M\'ezard and A.~Montanari, \emph{Information, Physics, and
  Computation}.\hskip 1em plus 0.5em minus 0.4em\relax New York, USA: Oxford
  University Press, 2009.

\end{thebibliography}
\end{document}